\documentclass{jfm}

\usepackage{graphicx}
\usepackage{natbib}
\usepackage{hyperref}
\usepackage{amsmath}
\usepackage{ar}
\usepackage{tikz}
\usepackage{pgfplots}
\usepackage{tikz-3dplot}
\usepackage{bm}
\usepackage{multirow}

\newcommand{\aver}[1]{ \! \left\langle {#1} \right \rangle \!}

\title[Finite-size inertial spherical particles in turbulence]{Finite-size inertial spherical particles in turbulence}

\author[A.~Chiarini, M.E.~Rosti]{Alessandro Chiarini\corresp{\email{alessandro.chiarini@oist.jp}} and Marco Edoardo Rosti\corresp{\email{marco.rosti@oist.jp}}}
\affiliation{Complex Fluids and Flows Unit, Okinawa Institute of Science and Technology Graduate University, 1919-1 Tancha, Onna-son, Okinawa 904-0495, Japan}

\begin{document}
\maketitle

\begin{abstract}
We investigate by direct numerical simulations the fluid-solid interaction of non-dilute suspensions of spherical particles moving in triperiodic turbulence, at the relatively large Reynolds number of $Re_\lambda \approx 400$. The solid-to-fluid density ratio is varied between $1.3$ and $100$, the particle diameter $D$ ranges between $16 \le D/\eta \le 123$ ($\eta$ is the Kolmogorov scale), and the volume fraction of the suspension is $0.079$. Turbulence is sustained using the Arnold-Beltrami-Childress cellular-flow forcing.  
The influence of the solid phase on the largest and energetic scales of the flow changes with the size and density of the particles. Light and large particles modulate all scales in a isotropic way, while heavier and smaller particles modulate the largest scales of the flow towards an anisotropic state. Smaller scales are isotropic and homogeneous for all cases. The mechanism driving the energy transfer across scales changes with the size and the density of the particles. For large and light particles the energy transfer is only marginally influenced by the fluid-solid interaction. For small and heavy particles, instead, the classical energy cascade is subdominant at all scales, and the energy transfer is essentially driven by the fluid-solid coupling. The influence of the solid phase on the flow intermittency is also discussed. Besides, the collective motion of the particles and their preferential location in relation with properties of the carrier flow are analysed. The solid phase exhibits moderate clustering; for large particles the level of clustering decreases with their density, while for small particles it is maximum for intermediate values. 
\end{abstract}

\begin{keywords}
\end{keywords}

\section{Introduction}
\label{sec:introduction}

Particle-laden turbulent flows have attracted the attention of many scholars over the last decades because of their relevance that goes beyond a fundamental interest, and encompasses several applications in natural, industrial and geophysical fields
\citep{delillo-etal-2014,breard-etal-2016,sengupta-etal-2017,falkinhoff-etal-2020}.

The inclusion of solid particles in the flow introduces additional momentum in the suspension that may result in modulation of the flow structures. When the suspension is dilute enough, the carrier flow is almost not altered by the presence of the particles. When the suspension is non-dilute, instead, the carrier flow undergoes macroscopic changes, and the particle-fluid interaction can not be neglected \citep{balachandar-eaton-2010,brandt-coletti-2022}.

The rheological properties of suspensions have been first studied in the limit of viscous Stokesian regime, where the inertial effects are negligible. The presence of the particles affects the deformation of the surrounding fluid, and the effective viscosity $\mu_e$ of the particle-fluid mixture depends on the dynamics of the dispersed phase. For a dilute suspension with negligible inter-particle interactions, \cite{einstein-1906} derived an analytical linear relation for the effective viscosity, i.e. $\mu_e = \mu(1+2.5\Phi_V)$, with $\Phi_V = V_p/(V_p + V_f)$ being the volume fraction of the dispersed phase; $V_p$ and $V_f$ are the volumes of the solid and fluid phases. Later, a quadratic correction was proposed by \cite{batchelor-1970} and \cite{batchelor-green-1972} to account for mutual particle interaction for slightly higher volume fractions. For higher concentrations, where the inter-particle interactions play a crucial role, only semi-empirical formulas exist \citep{krieger-dougherty-1959}.

In the turbulent regime the particle size has been found to play a crucial role in determining the effect of the dispersed phase on the carrier flow. \cite{gore-crowe-1989} considered together data from particulate turbulent pipe flows and jets, and found that the turbulence modulation depends on the ratio of the particle diameter $D$ and the flow integral scale $\mathcal{L}$. When $D/\mathcal{L}>0.1$ the turbulent intensity of the carrier phase increases with respect to the single phase case, while when $D/\mathcal{L}<0.1$ it decreases. Large particles produce fluctuations in their wake and enhance the turbulent activity, while small ones drain energy from the large-scale turbulent eddies of the flow. Indeed, \cite{tsuji-morikawa-1982} experimentally considered an air-solid two-phase flow in a horizontal pipe with diameter of $30\,\mathrm{mm}$, and investigated the influence of small and large particles with diameter of $0.2\,\mathrm{mm}$ and $3.4\,\mathrm{mm}$ on the carrier flow, varying the section average air velocity between $6$ and $20\,\mathrm{m/s}$. They observed that large particles markedly increase the turbulent activity, while small ones reduce it. These results were later confirmed by the same authors, with experiments of an air-solid two-phase flow in a vertical pipe \citep{tsuji-etal-1984}. More recently, \cite{hoque-etal-2016} studied the turbulence modulation of a nearly isotropic flow field due to the presence of single glass particles with varying diameter, and observed that the critical ratio of the particle size to integral scale of $ D/\mathcal{L} = 0.41$ separates the regimes of attenuation and enhancement of turbulent intensity.

Large steps towards a complete description of the interaction between rigid particles of different size and density and the fluid phase have been performed over the last years thanks to the increase in the supercomputers' power, which enables the study of such problem via direct numerical simulations (DNS) \citep{crowe-etal-1996,burton-etal-2005,maxey-2017}. One of the most common approaches used today to model many particle-laden flows is based on the point-particle approximation. In general, each particle is treated as a mathematical point source of mass, momentum and energy, with the particles assumed to be much smaller than any structure of the flow. The point-particle approximation is valid in the limit in which the volume of each particle is negligible, i.e. the case with vanishing disperse-phase volume fraction. Indeed, the point-particle assumption requires that the fluid velocity field does not display a turbulent behaviour at the scale of the particle, meaning that $D \le \eta$ ($\eta$ being the Kolmogorov scale).  
According to this method, the fluid-solid interaction is completely described by a forcing term that should be accurately modelled; see for example \cite{maxey-riley-1983}. The theoretical developments of the point-particle approximation, however, are limited to a relatively narrow range of parameters. For example, investigating the motion of a single particle in homogeneous isotropic turbulence, \cite{homann-bec-2010} found that the particle velocity variance reflects what predicted by the point-particle approximation for $D/ \eta \le 4$ only. Within the point-particle approximation, \cite{elghobashi-truesdell-1993} and \cite{elghobashi-1994} showed the importance of the particle Stokes number in determining the turbulence modulation. \cite{costa-brandt-picano-2020} tested the point-particle approximation in a turbulent channel flow laden with small inertial particle, with high particle-to-fluid density ratio. They considered a volume fraction of $\Phi_V \approx 10^{-5}$ to ensure that the particle feedback on the flow is negligible, and discussed the validity of models that approximate the particle dynamics considering only the inertial and nonlinear drag forces. They observed that in the bulk of the channel these models predict pretty well the statistics of the particle velocity. Close to the wall, however, the models fail as they are not able to capture the shear-induced lift force acting on the particles in this region, which instead is well predicted by the lift force model introduced by \cite{saffman-1964}.

A complete understanding of the fluid-solid interaction problem over a wider range of parameters \citep[$D/\eta \gtrsim 1$, $\Phi_V \gtrsim 10^{-4}$ and $\rho_p/\rho_f \gtrsim 10$; see][]{brandt-coletti-2022} requires to properly resolve the flow around each particle, without the need of relying on models. In this respect, over the last years, several numerical methods based on the coupling between direct numerical simulations and the immerse boundary method (IBM) have appeared, and have been used to study particulate turbulence; see for example the numerical methods described in \cite{kajishima-etal-2001,uhlmann-2005,huang-etal-2007,breugem_2012a,kempe-etal-2012,hori-rosti-takagi-2022}. Based on these developments, several works have carried out DNS of particulate turbulence, but mainly considering small Reynolds numbers and/or large particles, due to the extremely large computational cost; see for example \cite{lucci-etal-2010,lucci-etal-2011,oka-goto-2022} for homogeneous isotropic turbulence, \cite{uhlmann-2008,shao-etal-2012,wang-abbas-climent-2018,peng-etal-2019,yousefi-etal-2020,rosti_brandt_2020a, costa_brandt_picano_2021a,gao-etal-2023} for channel flow, \cite{lin-etal-2017} for duct flow and \cite{wang-etal-2015,zahtila-etal-2023} for pipe flow.

In this work, we consider non-dilute suspensions of finite-size spherical particles in periodic turbulence, and investigate the fluid-solid interaction in the two-dimensional parameter space of particle size and density. This flow configuration has been investigated by several authors over the years. \cite{tenCate-etal-2004} investigated the influence of suspensions of finite-size particles with a solid-to-fluid density ratio of $ \rho_p/\rho_f \approx 1.5$ and $\Phi_V \approx 0.02-0.1$ on periodic turbulence at a microscale Reynolds number of $Re_\lambda = u' \lambda / \nu = 61$ ($u'$ is the average velocity fluctuation, $\lambda$ is the Taylor length-scale and $\nu$ is the fluid kinematic viscosity). They found that the energy spectrum is enhanced for wavenumbers $\kappa>\kappa_p\approx 0.72 \kappa_d$, where $\kappa_d = 2 \pi /D$ is the wavenumber corresponding to the particle diameter, and that it is attenuated for $\kappa<\kappa_p$. The particles drain energy from the large scales of the flow, and inject it at smaller scales by means of their wake. \cite{hwang-eaton-2006} experimentally studied the influence of heavy particles with a diameter similar to the Kolmogorov scale on homogeneous isotropic turbulence, setting the Reynolds number at $Re_\lambda =230$. They observed that, the fluid turbulent kinetic energy and the viscous dissipation decrease when increasing the mass loading. \cite{yeo-etal-2010} numerically confirmed the results by \cite{tenCate-etal-2004} at $Re_\lambda \approx 60$ with a particle-to-fluid density ratio of $ \rho_p/\rho_f = 1.4$ and a volume fraction of $\Phi_V = 0.06$. \cite{lucci-etal-2010} studied the influence of particles of Taylor length-scale on decaying isotropic turbulence. They found that, in contrast to what happens when using particles with diameter smaller than the Kolmogorov scale, the turbulent kinetic energy is always smaller than that of the single-phase flow, and that the two-way coupling rate of change is always positive. Later, using the same framework, \cite{lucci-etal-2011} found that the Stokes number should not be used as an indicator to estimate at which extent particles of the Taylor length-scale modulate the turbulent carrier flow, as particles with same response time but different diameter or density may have different effect on the flow, unlike what observed for sub-Kolmogorov particles \citep{ferrante-elghobashi-2003,yang-shy-2005}. \cite{uhlmann-chouippe-2017} investigated the influence of spherical particles with diameter of approximately $5$ and $8$ times the Kolmogorov length and particle-to-fluid density ratio of $\rho_p/\rho_f=1.5$ on forced homogeneous isotropic turbulence at $Re_\lambda \approx 130$. They observed that the disperse phase exhibits clustering with moderate intensity, and that this clustering decreases with the particle diameter. They also suggest that small and light finite-size particles follow the expansive directions of fluid acceleration field, as happens also for sub-Kolmogorov particles \citep{chen-goto-vassilicos-2006,goto-vassilicos-2008}. 
More recently, \cite{olivieri-cannon-rosti-2022} considered homogeneous isotropic turbulence with a well developed inertial range of scales (they set the Reynolds number at $Re_{\lambda} \approx 400$), and investigated the turbulence modulation by non-dilute suspensions ($\Phi_V=0.079$) of spherical particles with size that lies in the inertial range ($D/\eta=123$), varying the particle-to-fluid density ratio between $1.3 \le \rho_p/\rho_f \le 100$. Besides the energy depletion at large scales and the energy enhancement at the small ones, they characterise how particles modify the classical energy cascade described by Richardson and Kolmogorov. They observed that the energy transfer is governed by the fluid-solid coupling at scales larger than the particle diameter, and that the classical energy cascade is recovered only at smaller scales.
\cite{oka-goto-2022} performed a parametric DNS study of particulate homogeneous isotropic turbulence, and investigated the turbulence modulation due to spherical solid particles with different Stokes number at a volume fraction of $\Phi_V = 8.1 \times 10^{-3}$ and reference Reynolds number of $Re_\lambda \le 100$. They varied the particle diameter in the $7.8 \le D/\eta \le 64$ range, and found that the turbulent kinetic energy decreases when reducing $D$, due to the additional energy dissipation rate in the wake. 
Similarly, \cite{shen-etal-2022} used a multiple-relaxation-time lattice Boltzman model, and studied the influence of the particle-to-fluid density ratio and particle diameter on the turbulence modulation at $Re_\lambda \approx 75$, varying the parameters between $8 \le D/\eta \le 16$ and $5 \le \rho_p/\rho_f \le 20$. Besides confirming the turbulent kinetic attenuation for smaller and heavier particles, they observed that the region affected by the particles depends on their diameter, and that the dissipation close to their surface increases with the particle density due to the larger slip velocity and particle Reynolds number.

\begin{figure}
\centering
\includegraphics[trim={30 74 60 0},clip,width=1\textwidth]{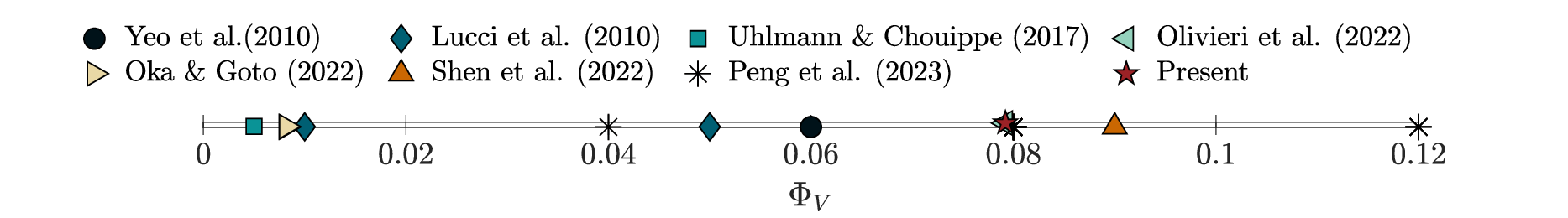}
\includegraphics[width=0.49\textwidth]{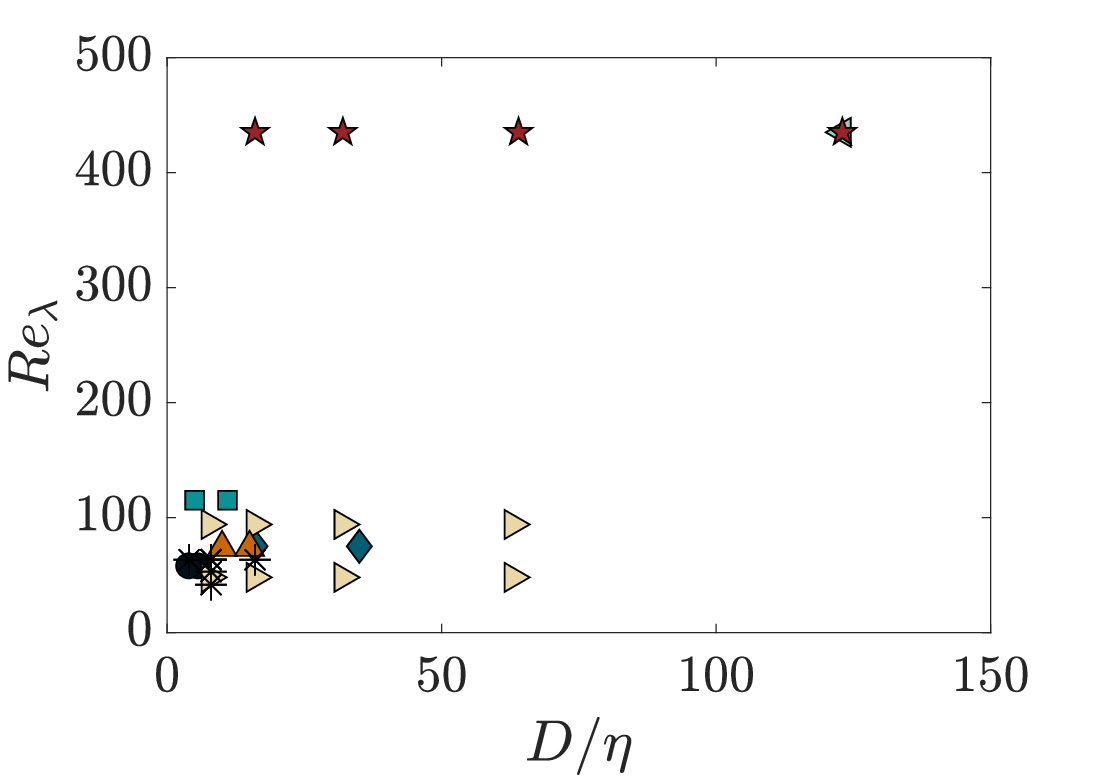}
\includegraphics[width=0.49\textwidth]{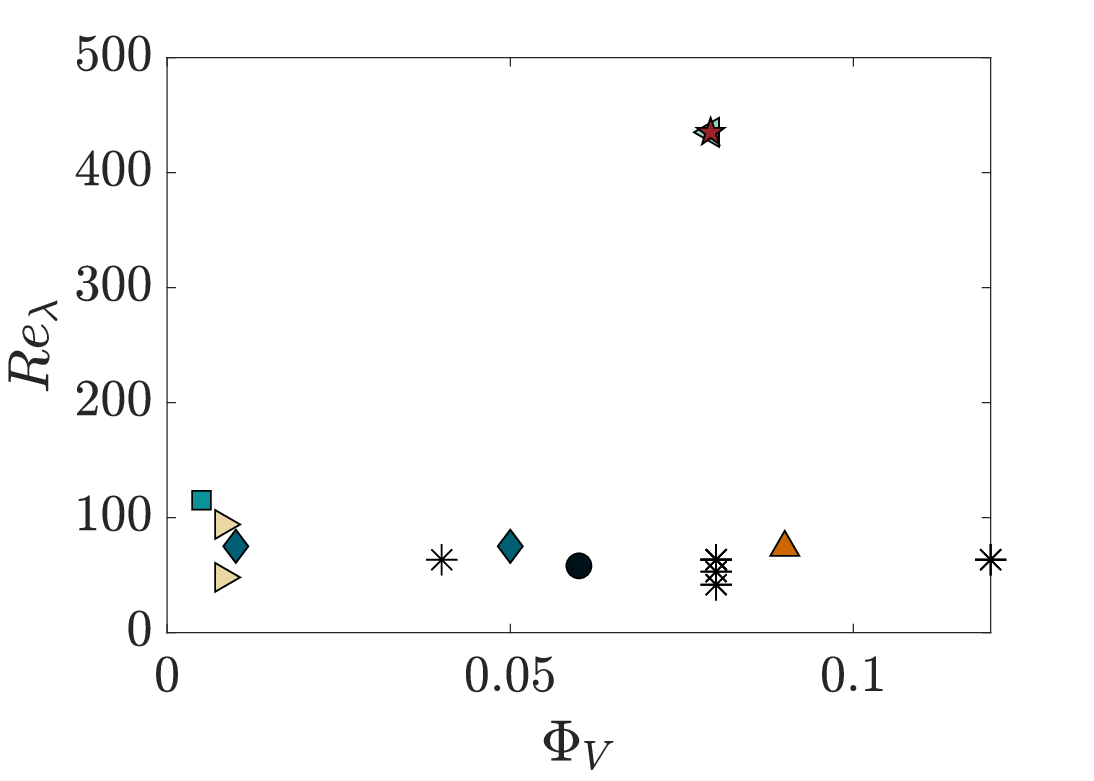}
\caption{Comparison between the values of the Reynolds number $Re_\lambda$, the particle size $D/\eta$, and the volume fraction $\Phi_V$ of the suspension considered in previous works in the literature  and the ones considered in the present work. This list considers only recent numerical works and does not intend to be exhaustive.}
\label{fig:old}
\end{figure}

Despite the large number of studies in the literature, further investigations are necessary for a complete understanding of particle-laden turbulent flows, even in the simple framework of a triperiodic box. Most of the numerical works, indeed, have considered relatively low Reynolds numbers, where the inertial range of turbulence is not completely developed, and relatively low volume fractions, that lead to a rather weak flow modulation (see figure \ref{fig:old}). As reported in \cite{brandt-coletti-2022}, reliable studies at larger Reynolds numbers and at higher concentrations are needed, to bridge the gap between the studies of small, heavy particles and large, weakly buoyant particles, and to promote the development of models that go beyond the point-particle approximation. In this work, we do a step in this direction. By means of a massive parametric study based on DNS and IBM, we investigate the fluid-solid interaction in a triperiodic turbulent flow laden with spherical particles with size that lies within the inertial range. We consider a Reynolds number of $Re_\lambda \approx 400$, which ensures an extensive inertial range of scales, and a volume fraction of $\Phi_V = 0.079$, which is large enough for the suspension to be non dilute, but small enough for the particle-particle interactions to be subdominant; see figure \ref{fig:old}. The fluid-solid interaction is studied in the two-dimensional parameter space of particle size $D$ and particle-to-fluid density ratio $\rho_p/\rho_f$. The particle size is varied between $ 16 \le D/\eta \le 123$, while the particle density is varied between $1.3 \le \rho_p/\rho_f \le 100$. The specific goals of the paper are: (i) to provide an extensive characterisation of the carrier flow modulation in the $D-\rho_p/\rho_f$ space, extending the previous works by \cite{olivieri-cannon-rosti-2022} and \cite{oka-goto-2022} to a wider range of parameters and to a larger Reynolds number; (ii) to describe the near-particle flow modulation and provide insights on how the fluid-solid interaction changes with $D$ and $\rho_p/\rho_f$, which is relevant for modelling purposes; (iii) to investigate the collective motion and preferential location of the particles, extending the work by \cite{uhlmann-chouippe-2017} to larger and heavier particles and to a larger Reynolds number.

The paper is structured as follows. In section \S\ref{sec:methods}, the physical model and the numerical method are briefly presented. Section \S\ref{sec:flow} describes the influence of the solid phase on the carrier flow. In section \S\ref{sec:near-par}, the effect of the particle size and density on the near-particle flow modulation is addressed. The dynamics of the particles and their collective motion are then discussed in section \S\ref{sec:particles}. Eventually, in section \S\ref{sec:conclusions} concluding remarks are provided.

\section{Numerical methods}
\label{sec:methods}

\begin{figure}
\centering
\includegraphics[trim={300 0 300 0},clip,width=0.49\textwidth]{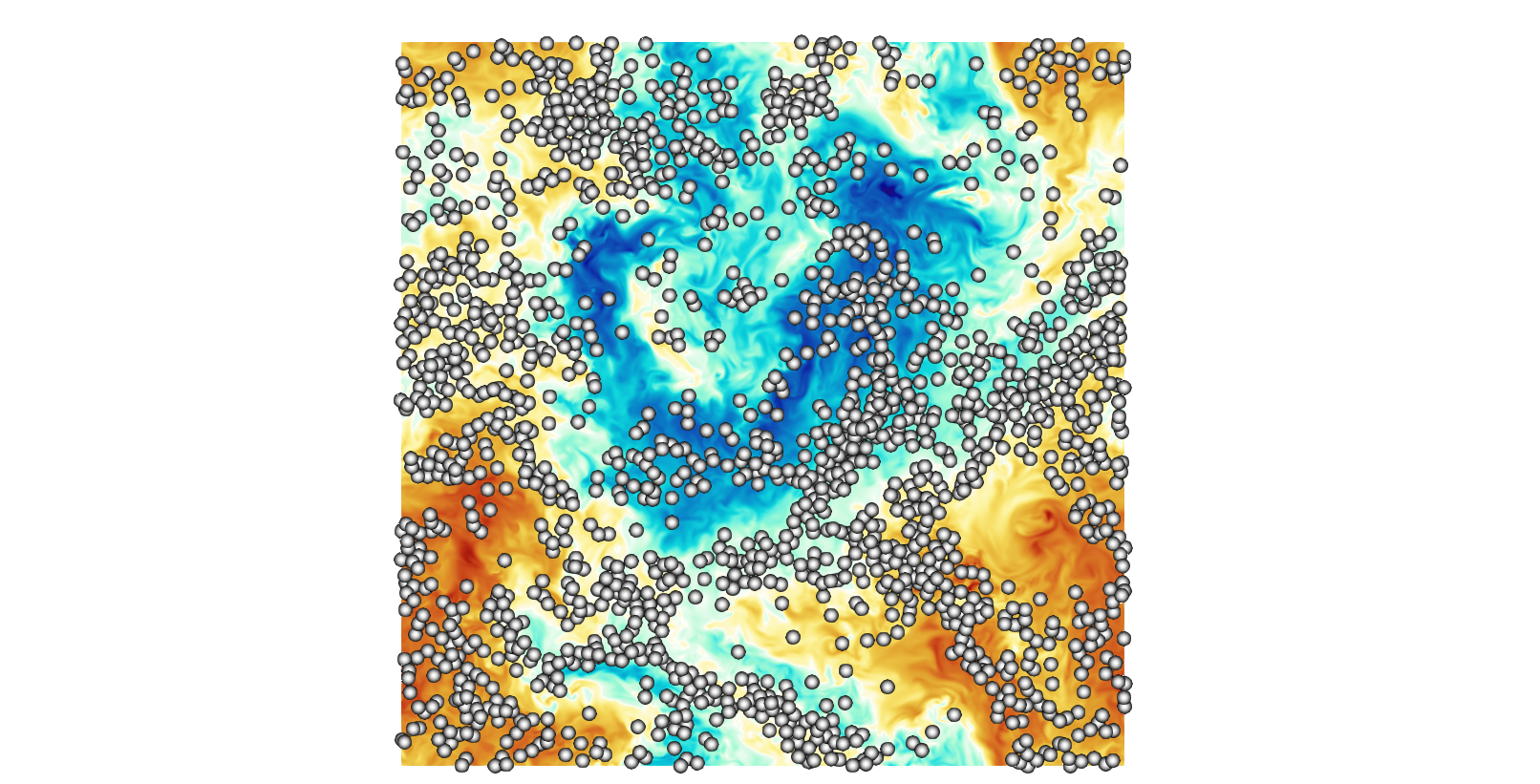}
\includegraphics[trim={300 0 300 0},clip,width=0.49\textwidth]{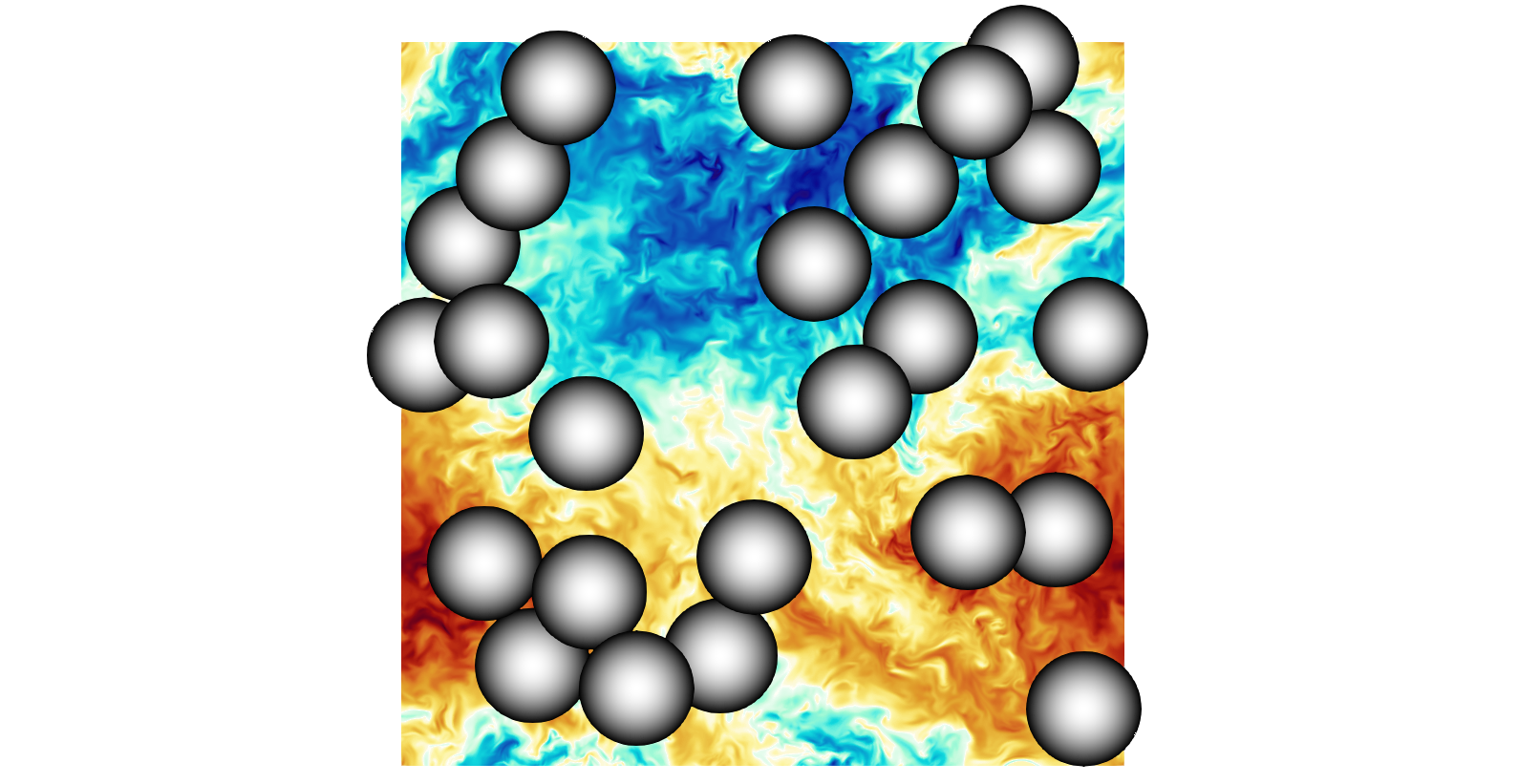}
\caption{Instantaneous velocity field and particles in a two-dimensional plane for the $D/\eta=16$ and $D/\eta=123$ particulate cases with $M=0.3$.}
\label{fig:catchy}
\end{figure}

The fluid-solid interaction of non-dilute suspension of solid spherical particles is investigated by means of a massive parametric study based on direct numerical simulations. We consider an ensemble of $N$ finite-size spherical particles with size $D$ and density $\rho_p$, suspended in a periodic turbulent flow; see figure \ref{fig:catchy}. Particles are released in a cubic domain of size $L=2\pi$, having periodic boundary conditions in all directions, where turbulence is generated and sustained using the Arnold-Beltrami-Childress (ABC) cellular-flow forcing \citep{podvigina-pouquet-1994}. 

The carrier flow is governed by the incompressible Navier--Stokes equations for a Newtonian fluid, i.e.
\begin{equation}
\begin{cases}
\frac{\partial \bm{u}}{\partial t} + \bm{\nabla} \cdot \bm{u} \bm{u} = -\frac{1}{\rho_f} \bm{\nabla} p + \nu \nabla^2 \bm{u}  + \bm{f} + \bm{f}^{\leftarrow p} \\
\bm{\nabla} \cdot \bm{u} = 0.
\end{cases}
\label{eq:Navier-Stokes}
\end{equation}
Here, $\bm{u}=(u,v,w)$ is the velocity vector field, $p$ is the reduced pressure, and $\rho_f$ and $\nu$ are the fluid density and kinematic viscosity. In the momentum equation, the term $\bm{f}^{\leftarrow p}$ is the force due to the solid phase, and $\bm{f}$ is the external ABC body force used to sustain turbulence
\begin{equation}
\bm{f}=\left(\frac{2\pi}{L}\right)^2 F_o
\begin{pmatrix}
 A \sin(\frac{2\pi}{L}z) + C \cos(\frac{2\pi}{L}y) \\
 B \sin(\frac{2\pi}{L}x) + A \cos(\frac{2\pi}{L}z) \\
 C \sin(\frac{2\pi}{L}y) + B \cos(\frac{2\pi}{L}x) \\
\end{pmatrix},
\end{equation}
where $A=B=C=1$, and $F_o$ is a constant parameter. 
The dynamics of the spherical particles is governed by the Euler-Newton equations,
\begin{equation}
\begin{cases}
m \frac{\text{d} \bm{u}_p}{\text{d} t} = \bm{f}^{\leftarrow f} + \bm{f}^{\leftrightarrow p} \\
I \frac{\text{d} \bm{\omega}_p}{\text{d} t} = \bm{L}_p^{\leftarrow f},
\end{cases}
\end{equation}
where $\bm{u}_p$ and $\bm{\omega}_p$ are the velocity and angular velocity of the particles, and $m=\pi \rho_p D^3/6$ and $I = m D^2/10$ are the mass and inertial moment of the particles. Also, $\bm{f}^{\leftarrow f}$ is the force due to the fluid and $\bm{f}^{\leftrightarrow p}$ is the force due to the interaction between particles. In this study gravitational forces are not considered.

We set $\rho_f=1$, $\nu=1/400$ and $F_o=5$ to achieve in the single-phase case a micro-scale Reynolds number of $Re_\lambda = u' \lambda / \nu \approx 435$, ($u'$ is the root mean square of the velocity fluctuations and $\lambda$ is the Taylor length scale), which ensures an extensive inertial range of scales. 
The particle diameter $D$ is varied between $16 \le D/\eta \le 123$ ($\eta$ is the Kolmogorov length scale for the single-phase case). The number $N$ of particles is increased as the diameter decreases, to maintain the volume fraction $\Phi_V =  V_p/(V_p + V_f)$ constant to a value of $\Phi_V = 0.0792$, as in \cite{olivieri-cannon-rosti-2022}. The particle density $\rho_p$ is varied between $1.29 \le \rho_p/\rho_f \le 104.7$ to consider both light and heavy particles, corresponding to a variation of the mass fraction $M = \rho_p V_p / (\rho_f V_f + \rho_p V_p)$ between $0.1 \le M \le 0.9$. The smallest value of $\rho_p/\rho_f=1.29$ and the largest value of $\rho_p/\rho_f = 104.7$ have been chosen based on the results of \cite{olivieri-cannon-rosti-2022}. As shown in figures 2 and 3 of their paper, indeed, the flow modulation only marginally changes when further increasing the density ratio $\rho_p/\rho_f$. 

The governing equations are numerically integrated in time using the in-house solver Fujin (\url{https://groups.oist.jp/cffu/code}), which is based on an incremental pressure-correction scheme. It considers the Navier--Stokes equations written in primitive variables on a staggered grid, and uses second-order finite differences in all the directions. The Adams-Bashforth time scheme is used for advancing the momentum equation in time. The Poisson equation for the pressure enforcing the incompressibility constraint is solved using a fast and efficient approach based on the Fast Fourier Transform. The solver is parallelised and is based on a Cartesian block decomposition of the domain in the $x$ and $y$ directions, and uses the Message Passing Interface (MPI) library for maximum portability. The governing equations for the particles are dealt with the immersed boundary method introduced by \cite{hori-rosti-takagi-2022}. The fluid-solid coupling is achieved in an Eulerian framework \citep{kajishima-etal-2001}, and accounts for the inertia of the fictitious fluid inside the solid phase, to properly reproduce their behaviour in both the neutrally-bouyant case and in presence of density difference between the fluid and solid phases. The soft sphere collision model, first proposed by \cite{tsuji-etal-1993}, is used to prevent the interpenetration between particles. A fixed-radius near neighbours algorithm \citep[see][and references therein]{monti-etal-2021} is used for the particle interaction to avoid an otherwise prohibitive increase of the computational cost when the number of particles increases. 

The fluid domain is discretised using $N_{point}=1024$ points in the three directions, to ensure that all the scales down to the smallest dissipative (Kolmogorov) ones are solved, leading to $\eta/\Delta x= O(1)$, where $\Delta x$ denotes the grid spacing. At the initial time the particles are randomly distributed within the domain. Excluding the initial transient period needed to reach the statistically steady state, all simulations are advanced for approximately $15T$, where $T=L/\bar{u'}$ is the turnover time of the largest eddies. For the particle-laden cases with $D/\eta=32$, the simulations have been advanced for a longer period, i.e. approximately $35T$, to ensure convergence of the temporal averages discussed in section \S\ref{sec:flow}. Note that the particle-laden cases with $D/\eta=123$ are the same considered in \cite{olivieri-cannon-rosti-2022}. Details of the numerical simulations are provided in table \ref{tab:simulations}. The adequacy of the grid resolution is assessed in appendix \ref{sec:resolution}.

\begin{table}
\caption{Details of the numerical simulations considered in the present parametric study. $M$ is the mass fraction; $\rho_p/\rho_f$ is the particle-to-fluid density ratio; $N$ is the number of particles; $D$ is the particle diameter; $\lambda$ is the Taylor length scale; $\eta$ is the Kolomogorov scale; $E$ is the fluid kinetic energy; $\epsilon$ is the dissipation; $Re_\lambda$ is the Reynolds number based on $u'=\sqrt{2E/3}$ and on the Taylor length scale; $St$ is the Stokes number defined as $St=\tau_p/\tau_f$, where $\tau_p=(\rho_p/\rho_f)D^2/(18 \nu)$ is the relaxation time of the particle velocity and $\tau_f=\mathcal{L}/\sqrt{2\overline{\aver{E}}/3}$ is the turnover time of the largest eddies. $\mathcal{L}=\pi/(4\overline{\aver{E}}/3)\int_0^\infty \mathcal{E}(\kappa)/\kappa \text{d} \kappa$ is the fluid integral scale and $\mathcal{E}$ is the energy spectrum.}
\label{tab:simulations}
\centering
\begin{tabular}{cccccccccccccccccccccc}
$M$ & & $\rho_p/\rho_f$ & & $N$ & & $D/L$ & & $D/\eta$ & & $\overline{\aver{\lambda}}$ & & $\overline{\aver{\eta}}$ & & $\overline{\aver{E}}$ & & $\overline{\aver{\epsilon}}$ & & $Re_\lambda$ & & $St$ \\
\hline
$-$   & &  $-$  & &    $-$   & & $-$      & & $-$      & & $0.162$  & & $0.0039$ & & $65.03$ & & $70.13$ & & $435.01$ & &  $-$     \\
      & &       & &          & &          & &          & &          & &         & &         & &          & &  \\
$0.1$ & & $1.29$ & & $300$ & & $0.0796$ & & $123$    & & $0.168$  & & $0.0041$ & & $62.87$ & & $56.80$ & & $434.05$ & &  $10.39$  \\
$0.3$ & & $4.98$   & &$300$ & & $0.0796$ & & $123$    & & $0.163$  & & $0.0042$ & & $55.68$ & & $52.22$ & & $397.47$ & &  $41.34$  \\
$0.6$ & & $17.45$  & & $300$ & & $0.0796$ & & $123$    & & $0.150$  & & $0.0041$ & & $50.31$ & & $55.95$ & & $347.48$ & &  $133.43$ \\
$0.9$ & & $104.7$ & & $300$ & & $0.0796$ & & $123$    & & $0.137$  & & $0.0042$ & & $38.30$ & & $50.51$ & & $278.36$ & &  $712.69$ \\
      & &       & &          & &          & &          & &          & &         & &         & &          & &  \\
$0.1$ & & $1.29$ & & $2129$ & & $0.0414$ & & $64$    & & $0.170$  & & $0.0042$ & & $59.65$ & & $52.14$ & & $428.05$ & &  $4.98$ \\
$0.3$ & & $4.98$ & & $2129$ & & $0.0414$ & & $64$    & & $0.171$  & & $0.0044$ & & $52.01$ & & $45.49$ & & $400.94$ & &  $9.99$ \\
$0.45$& & $9.51$ & & $2129$ & & $0.0414$ & & $64$    & & $0.191$  & & $0.0041$ & & $83.48$ & & $57.31$ & & $569.70$ & &  $21.49$ \\
$0.6$ & & $17.45$ & & $2129$ & & $0.0414$ & & $64$    & & $0.171$  & & $0.0042$ & & $57.30$ & & $48.46$ & & $424.15$ & &  $32.19$ \\
$0.9$ & & $104.7$ & & $2129$ & & $0.0414$ & & $64$    & & $0.132$  & & $0.0045$ & & $27.16$ & & $38.66$ & & $225.29$ & &  $144.65$ \\
      & &        & &          & &         & &          & &          & &         & &         & &          & &  \\
$0.1$  & & $1.29$ & & $17036$& & $0.0207$ & & $32$    & & $0.176$  & & $0.0043$ & & $57.48$ & & $48.87$ & & $435.36$ & &  $0.74$ \\
$0.3$  & & $4.98$ & & $17036$& & $0.0207$ & & $32$    & & $0.167$  & & $0.0045$ & & $44.63$ & & $41.02$ & & $363.32$ & &  $2.69$ \\
$0.45$ & & $9.51$ & & $17036$& & $0.0207$ & & $32$    & & $0.200$  & & $0.0043$ & & $74.40$ & & $46.64$ & & $564.06$ & &  $5.13$ \\
$0.525$& & $12.85$ & &$17036$& & $0.0207$ & & $32$    & & $0.208$  & & $0.0042$ & & $87.30$ & & $50.43$ & & $635.26$ & &  $7.04$ \\
$0.6$  & & $17.45$ & &$17036$& & $0.0207$ & & $32$    & & $0.223$  & & $0.0043$ & & $90.28$ & & $45.41$ & & $691.96$ & &  $9.42$ \\
$0.75$ & & $34.90$ & & $17036$& & $0.0207$ & & $32$    & & $0.177$  & & $0.0046$ & & $43.45$ & & $34.50$ & & $382.00$ & &  $12.53$ \\
$0.9$  & & $104.7$ & & $17036$& & $0.0207$ & & $32$    & & $0.134$  & & $0.0050$ & & $19.04$ & & $25.08$ & & $186.02$ & &  $25.44$ \\
       & &       & &          & &         & &          & &          & &         & &         & &          & &  \\
$0.1$ & &  $1.29$ & & $136293$& & $0.0104$ & & $16$    & & $0.178$  & & $0.0043$ & & $57.80$ & & $46.07$ & & $441.62$ & &  $0.20$ \\
$0.3$ & &  $4.98$ & & $136293$& & $0.0104$ & & $16$    & & $0.177$  & & $0.0046$ & & $43.31$ & & $34.99$ & & $379.86$ & &  $0.62$ \\
$0.6$ & &  $17.45 $ & & $136293$& & $0.0104$ & & $16$    & & $0.164$  & & $0.0050$ & & $28.07$ & & $26.20$ & & $283.98$ & &  $1.50$ \\
$0.9$ & &  $104.7$ & & $136293$& & $0.0104$ & & $16$    & & $0.196$  & & $0.0052$ & & $34.09$ & & $22.07$ & & $374.65$ & &  $8.15$ \\
\end{tabular}
\end{table}

To demonstrate the independence of the results on the external forcing, additional simulations have been carried out using the forcing introduced by \cite{eswaran-pope-1988} to sustain turbulence; see appendix \ref{sec:Pope}. Unlike the ABC forcing, indeed, the Eswaran \& Pope's forcing has a random component, and does not generate an inhomogeneous shear at the largest scales. As shown in section \ref{sec:flow} and in appendix \ref{sec:Pope}, the effect of the solid phase on the largest and energetic scales of the flow changes with the kind of forcing considered. At smaller scales, however, the results do not depend on the external forcing. At small scales the flow modulation does not depend on how energy is injected in the system, and the effect of the solid phase is substantially the same for the two considered forcings.


\section{The carrier flow}
\label{sec:flow}

\subsection{Integral quantities}
\label{sec:integ}

\begin{figure}
\centering
\includegraphics[width=1.0\textwidth]{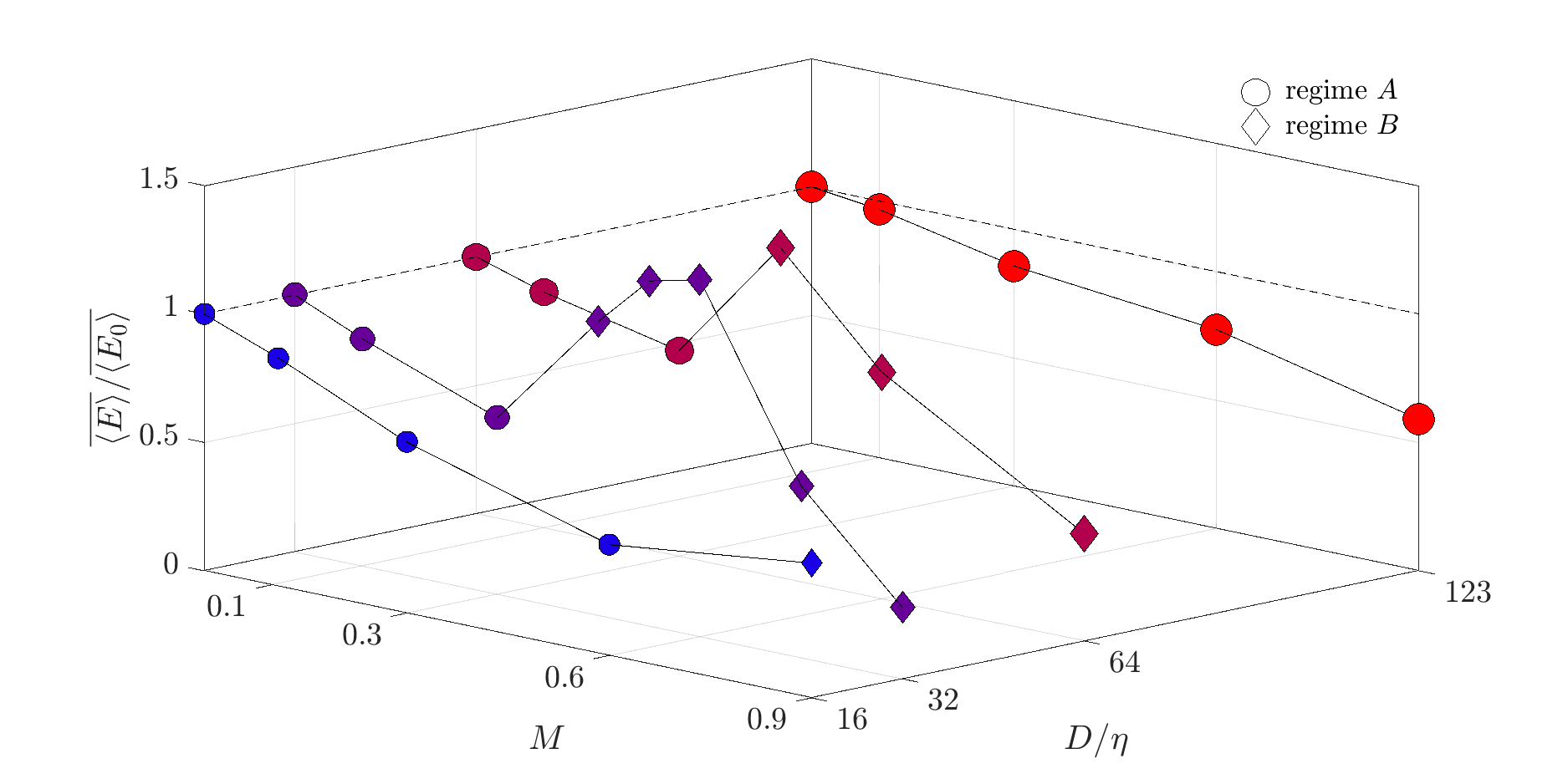}
\caption{Dependence of the average energy of the fluid phase $\overline{\aver{E}}$ on the density and size of the particles. $\overline{\aver{E_0}}$ refers to the single-phase case. Circles refer to regime A, while diamonds to regime B (see text).}
\label{fig:energy}
\end{figure}

In this section we investigate the influence of the particles on the fluid phase. 
Figure \ref{fig:energy} shows the dependence of the fluid average kinetic energy $\overline{\aver{E (\bm{x},t) } }$ on $M$ and $D$. We define the fluid kinetic energy as
\begin{equation}
  E(t) = \frac{1}{2} | \bm{u}(\bm{x},t) - \aver{\bm{U}(\bm{x})}|^2,
\end{equation}
while $\overline{ \cdot }$ and $\aver{ \cdot }$ indicate average in time and along the homogeneous directions respectively; $\bm{U}(\bm{x}) \equiv \overline{ \bm{u}( \bm{x},t) } $ is the three-dimensional fluid velocity field averaged in time. 
As discussed by \cite{chiarini-etal-2023}, due to the presence of the inhomogeneous mean shear induced by the external ABC forcing, the dispersed phase modulates the largest scales of the carrier flow in a way that substantially changes with the size and density of the particles. Similarly to what found by previous authors \citep{oka-goto-2022,peng-etal-2023}, for large and/or light particles, i.e. $D/\eta > 64$ and/or $M < 0.45$, the kinetic energy $\aver{\overline{E}}$ of the carrier flow monotonically decreases when the mass fraction increases and/or the particle size decreases; we call this regime A in figure \ref{fig:energy}. When fixing $D$, indeed, an increase of the density of the particles leads to an increase of the inertia of the fluid-particle system, and the same large-scale external forcing generates weaker fluctuations. When fixing $M$, instead, smaller particles are more effective in attenuating the fluid velocity fluctuations. A decrease of $D$ corresponds to an increase of the total solid surface area and, therefore, to an increase of the high dissipation rate regions that form around the particles (see section \ref{sec:near-par}). On the other hand, when particles are smaller and heavier, i.e. $D/\eta \le 64$ and $M \gtrapprox 0.45$, the scenario changes: the kinetic energy $\aver{\overline{E}}$ sharply increases, and does not have a monotonous dependence on $D$ and $M$. This is what we call regime B in figure \ref{fig:energy}. The value of $M$ that delimits regimes A and B increases when the particle size decreases, suggesting that the occurrence of regime B is mainly driven by the inertia of the single particles. Note that for $D/\eta=32$ and $0.45 \le M \le 0.6$ and $D/\eta=64$ and $M=0.45$ the presence of the solid phase enhances the total fluid energy compared to the single-phase case. 

\begin{figure}
\centering
\includegraphics[width=0.49\textwidth]{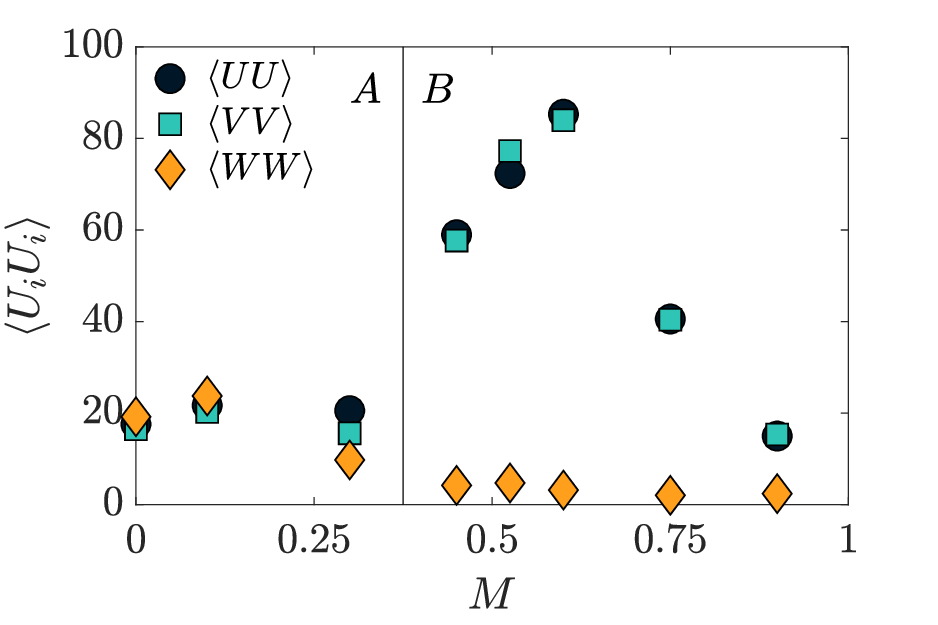}
\includegraphics[width=0.49\textwidth]{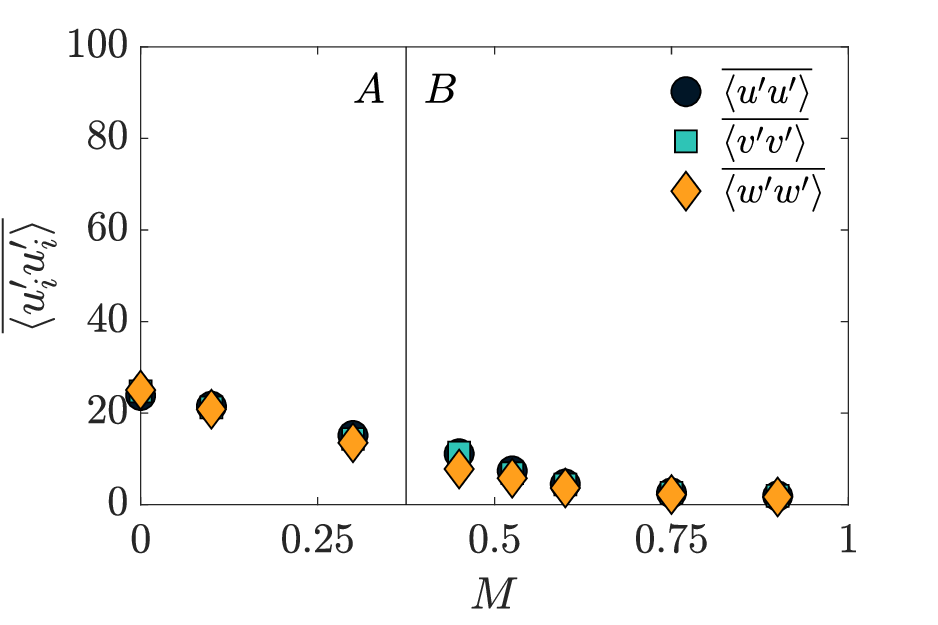}  
\caption{Dependence of the (left) mean and (right) fluctuating  energy on the mass fraction for the $D/\eta=32$ particle-laden case.}
\label{fig:energyM}
\end{figure}

In this work, regimes A (circles in figure \ref{fig:energy}) and B (squares in figure \ref{fig:energy}) are only briefly described; we refer the interested reader to \cite{chiarini-etal-2023} for more details. To characterise the two regimes, we use the temporal average operator and isolate the influence of the dispersed phase on the largest and smaller scales of the flow. We focus on $D/\eta=32$, being the case that shows the strongest flow modulation, and for which $\aver{\overline{E}}$ undergoes the maximum enhancement. We decompose the complete velocity field $\bm{u}(\bm{x},t)$ into its temporal mean field $\bm{U}(\bm{x})$ and the fluctuating field $\bm{u}'(\bm{x},t) = \bm{u}(\bm{x},t) - \bm{U}(\bm{x})$, and plot the variances of their three components in figure \ref{fig:energyM}.
Light particles (regime A) influence the flow without introducing a preferential direction, modulating both the mean and fluctuating fields in a isotropic way: here $\aver{UU} \approx \aver{VV} \approx \aver{WW}$ and $\aver{u'u'} \approx \aver{v'v'} \approx \aver{w'w'}$. Heavy particles (regime B), instead, modulate differently the largest and smaller scales of the flow. They attenuate the fluctuating field (i.e. the smaller scales of the flow) in a isotropic way, but modulate the mean flow field (i.e. the largest scales of the flow) towards a more energetic and anisotropic state. In this case, the presence of the dispersed phase enhances two components of the mean flow ($U$ and $V$) and attenuates the third one ($W$). Here $\aver{UU} \approx \aver{VV} \gg \aver{WW}$ and $\aver{u'u'} \approx \aver{v'v'} \approx \aver{w'w'}$. The increase of the total fluid energy $\aver{\overline{E}}$ observed in figure \ref{fig:energy} for $0.45 \le M \le 0.6$ is entirely due to the enhanced mean-flow contribution. In regime B the scenario is the following. Due to their large inertia, particles are not able to follow the inhomogeneous mean shear generated by the ABC forcing, and deviate towards almost straight trajectories that lay in $x-y$ plane (see section \ref{sec:traj}), showing anomalous transport. In doing this, they modulate the flow towards an anisotropic and almost two-dimensional state. Here the mean-flow velocity components that lay in the plane of the trajectories of the particles ($U$ and $V$) are enhanced, while the out-of-plane component ($W$) is attenuated. As detailed in \cite{chiarini-etal-2023}, this anisotropic modulation of the largest scales is due to the interaction of the particle with the inhomogeneous mean shear induced by the ABC forcing. When the inhomogeneous mean shear is not present the anisotropic modulation of the largest scales is not observed; see appendix \ref{sec:Pope}. 

Note that when presenting the results, we deliberately choose a reference system such that for all cases the $z$ axis is aligned with the mostly attenuated mean-flow velocity component, similarly to what done in \cite{chiarini-etal-2023}. In the reference system of the simulation, however, the direction aligned with the attenuated mean-flow velocity component changes with the initial condition.

\subsection{Mean flow modulation}
\label{sec:meanflow}

\begin{figure}
\centering
\includegraphics[width=0.8\textwidth]{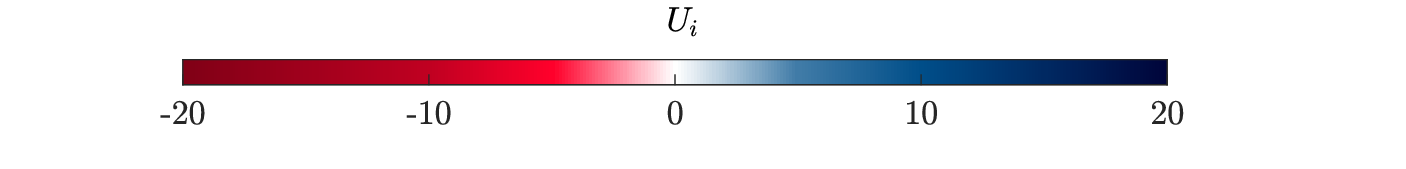}\\
\includegraphics[width=0.32\textwidth]{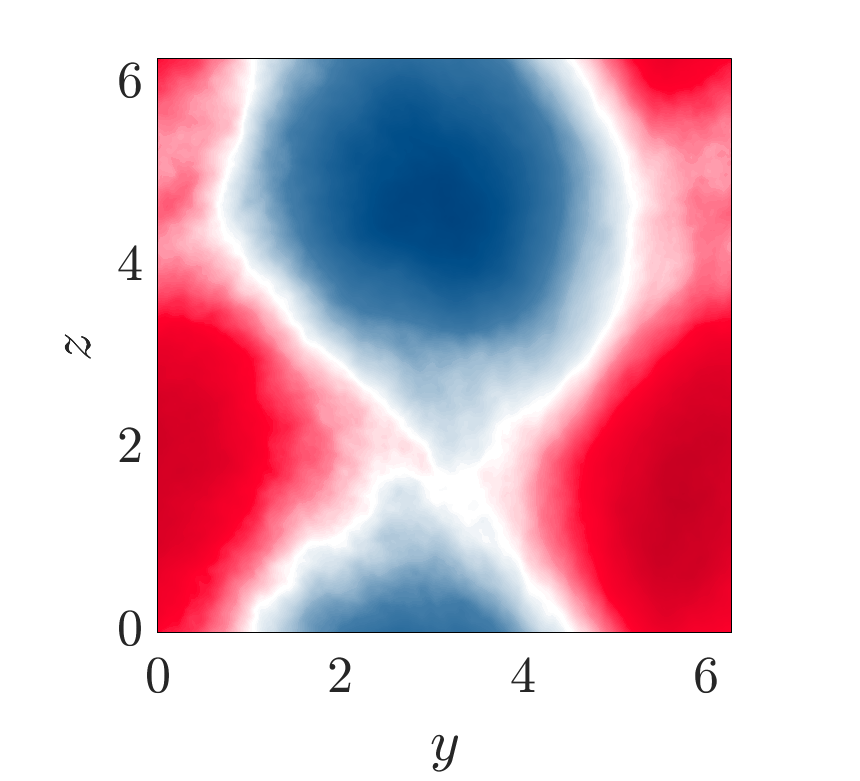}
\includegraphics[width=0.32\textwidth]{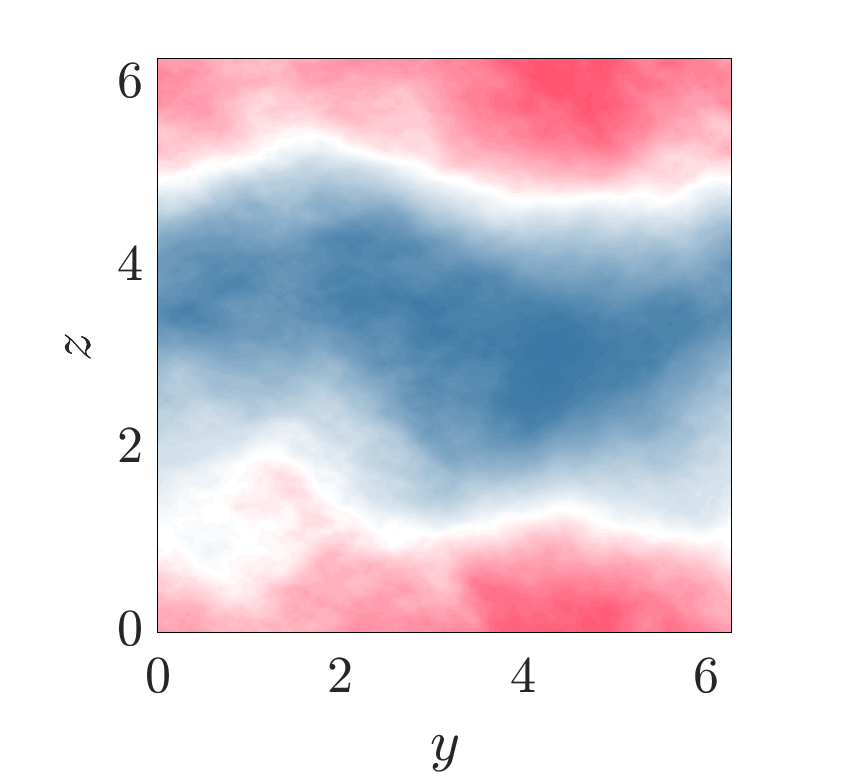}
\includegraphics[width=0.32\textwidth]{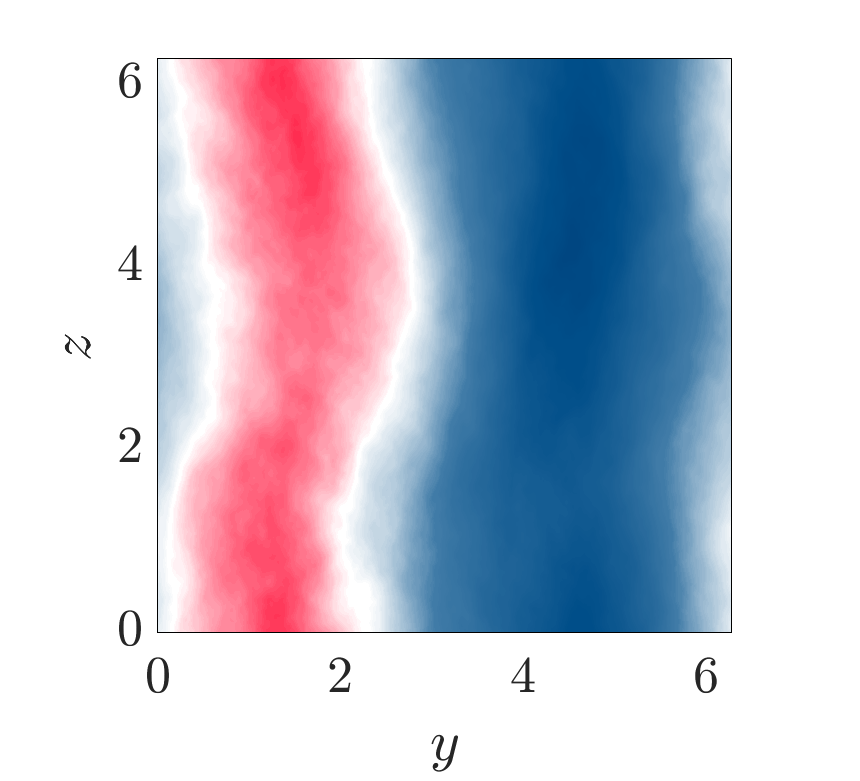}
\includegraphics[width=0.32\textwidth]{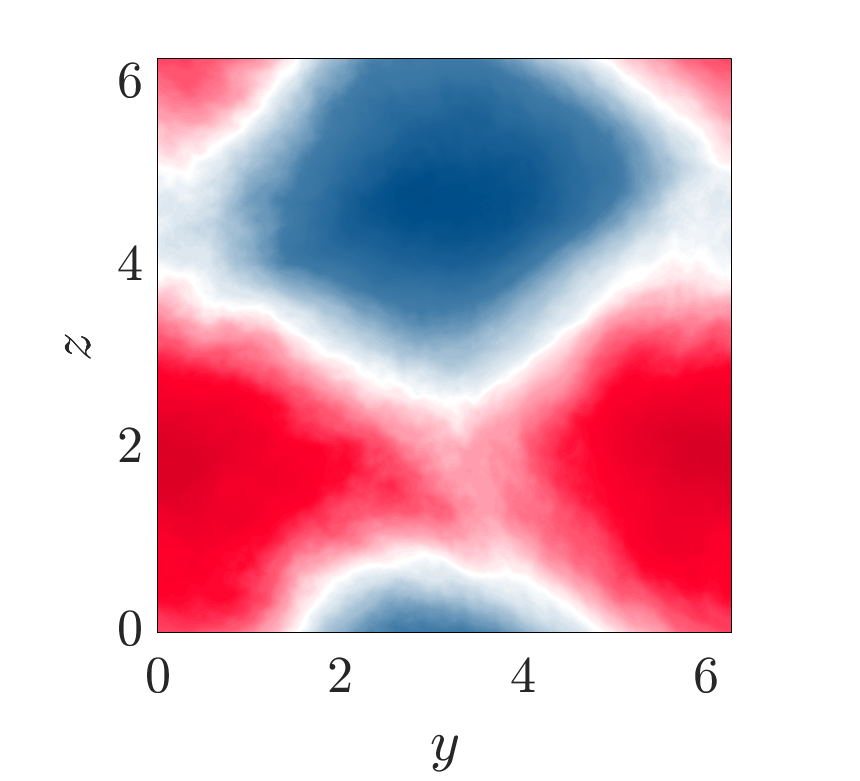}
\includegraphics[width=0.32\textwidth]{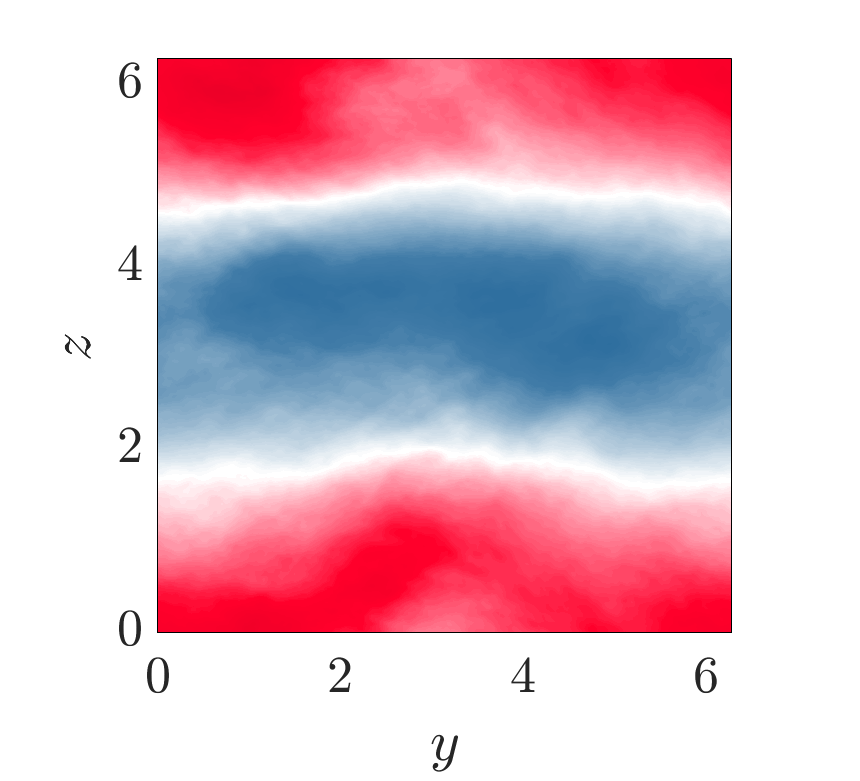}
\includegraphics[width=0.32\textwidth]{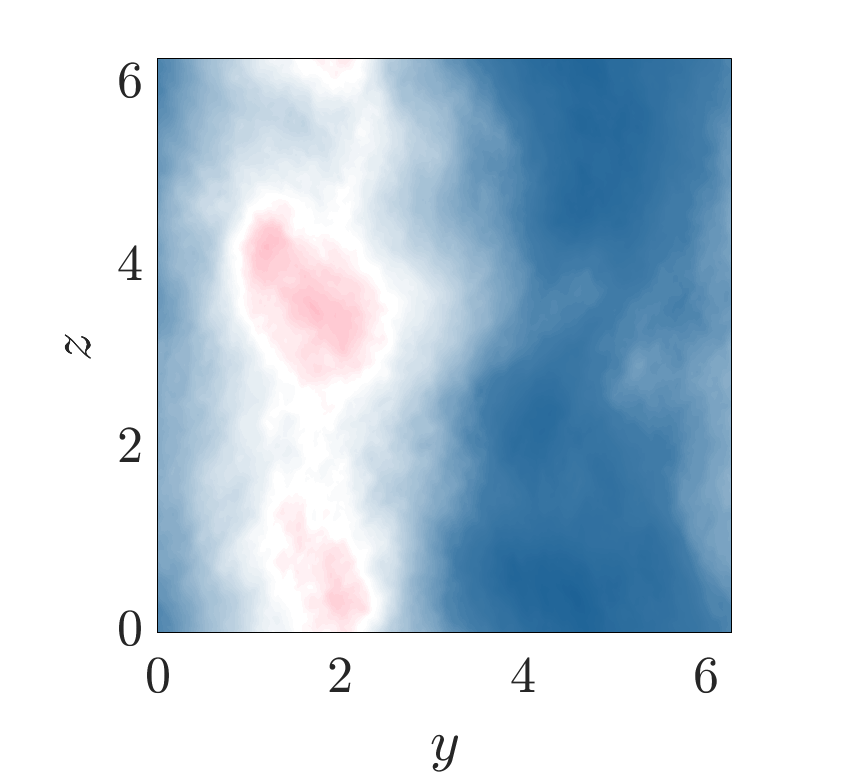}
\includegraphics[width=0.32\textwidth]{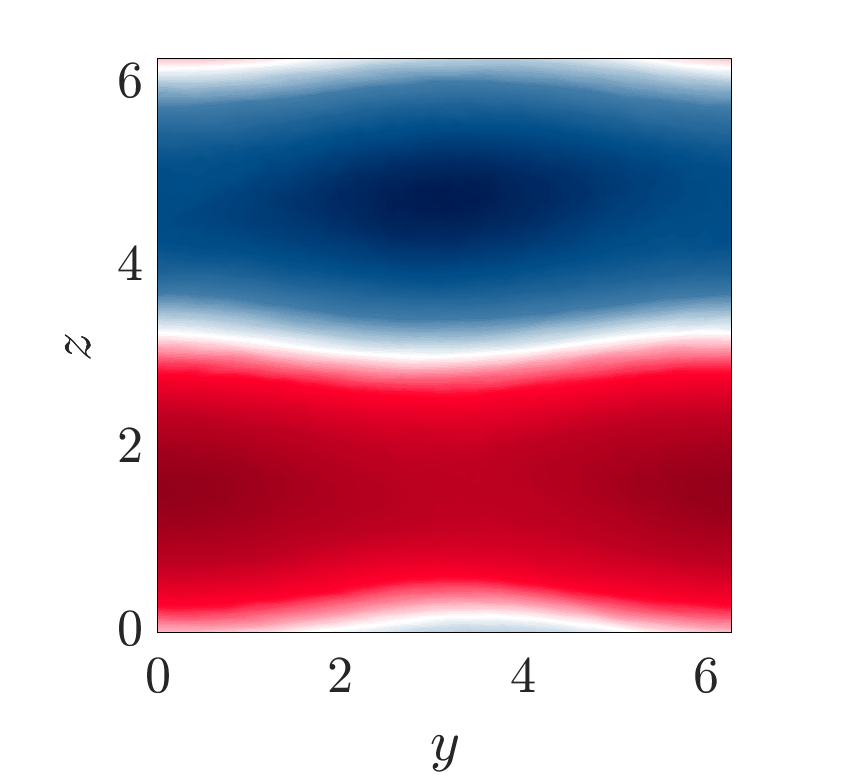}
\includegraphics[width=0.32\textwidth]{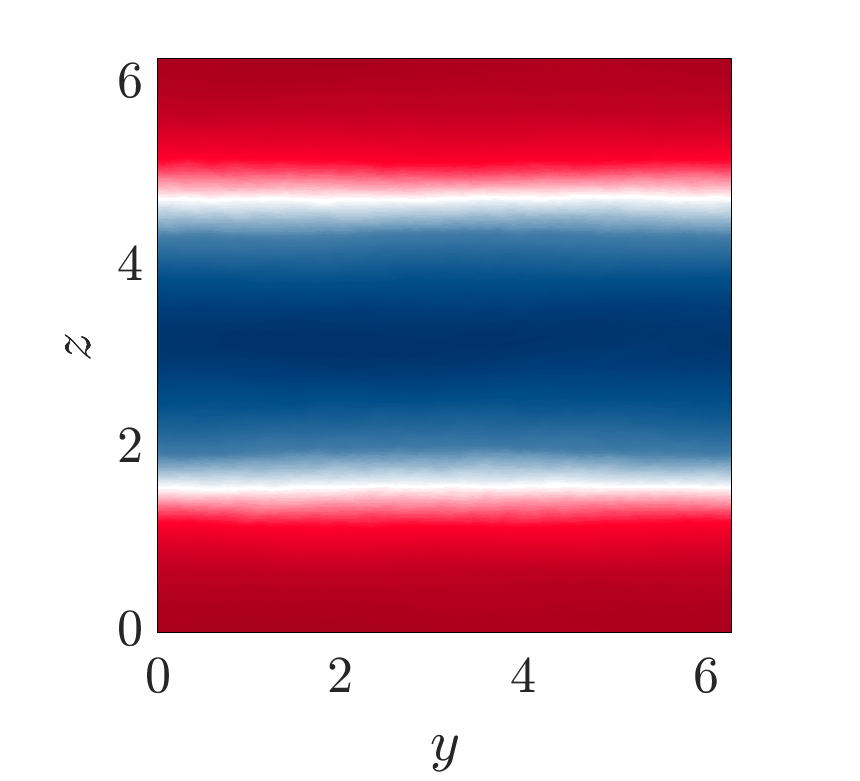}
\includegraphics[width=0.32\textwidth]{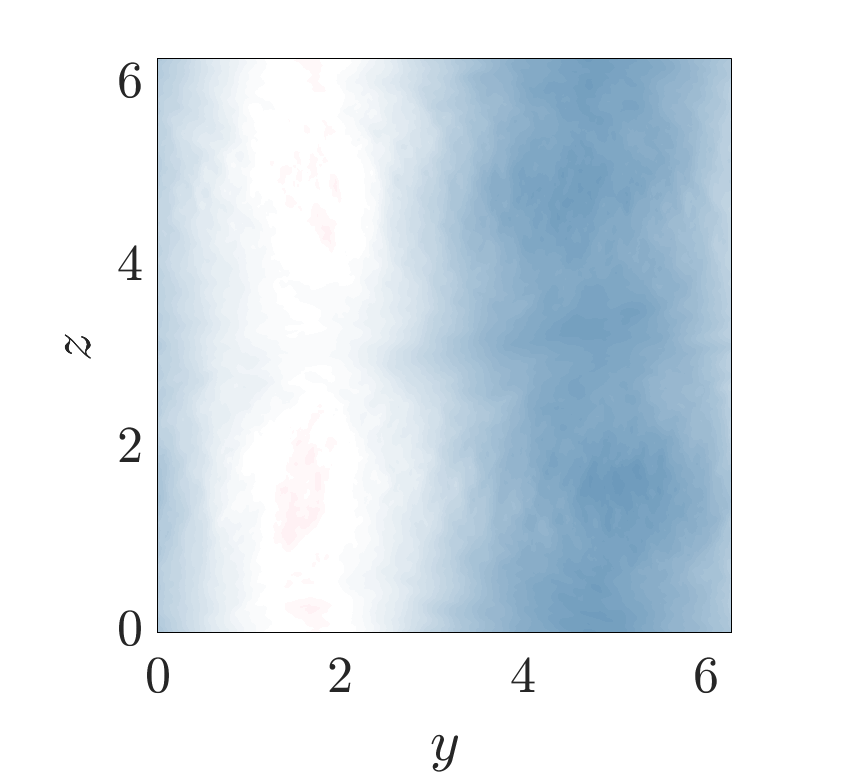}
\caption{Mean flow velocity components on the $x=L/2$ plane. Left: $U$. Middle: $V$. Right: $W$. Top: single phase. Middle: $D/\eta=32$ and $M=0.3$. Bottom: $D/\eta=32$ and $M=0.6$.}
\label{fig:Mvel}
\end{figure}

Figure \ref{fig:Mvel} plots in the $x=L/2$ plane the three components of the mean-flow velocity field, i.e. (left) $U$, (middle) $V$, and (right) $W$, for (top) the single-phase case, and for the $D/\eta=32$ particulate cases with (middle) $M=0.3$ and (bottom) $M=0.6$.
In the single-phase case the mean flow resembles the laminar ABC profile, i.e. $(U,V,W) \approx (A \sin(z) + C \cos(y), B \sin(x) + A \cos(z), C \sin(y) + B \cos(x))$ with $A=B=C=1$, similarly to what observed for the Kolmogorov flow by other authors \citep{borue-orszag-1996}; note for example the cellular pattern in the map of $U$. In the particle-laden case with $M=0.3$ (regime A), the mean flow retains a similar pattern, in agreement with the isotropic and marginal mean-flow modulation observed in figure \ref{fig:energyM}. In regime B, instead, the structure of the mean flow changes; for $M=0.6$ figure \ref{fig:Mvel} shows that the $U$ and $V$ velocity components are strongly enhanced, while the $W$ component is attenuated. Moreover, the mean flow does not show the ABC cellular pattern anymore (see the $U$ velocity component): the dependence on $x$ and $y$ is almost lost, while the sinusoidal dependence on $z$ is retained, i.e. $(U,V,W) \approx (\sin(z),\cos(z),0)$ \citep[see][for further details]{chiarini-etal-2023}.

\begin{figure}
\centering
\includegraphics[width=0.49\textwidth]{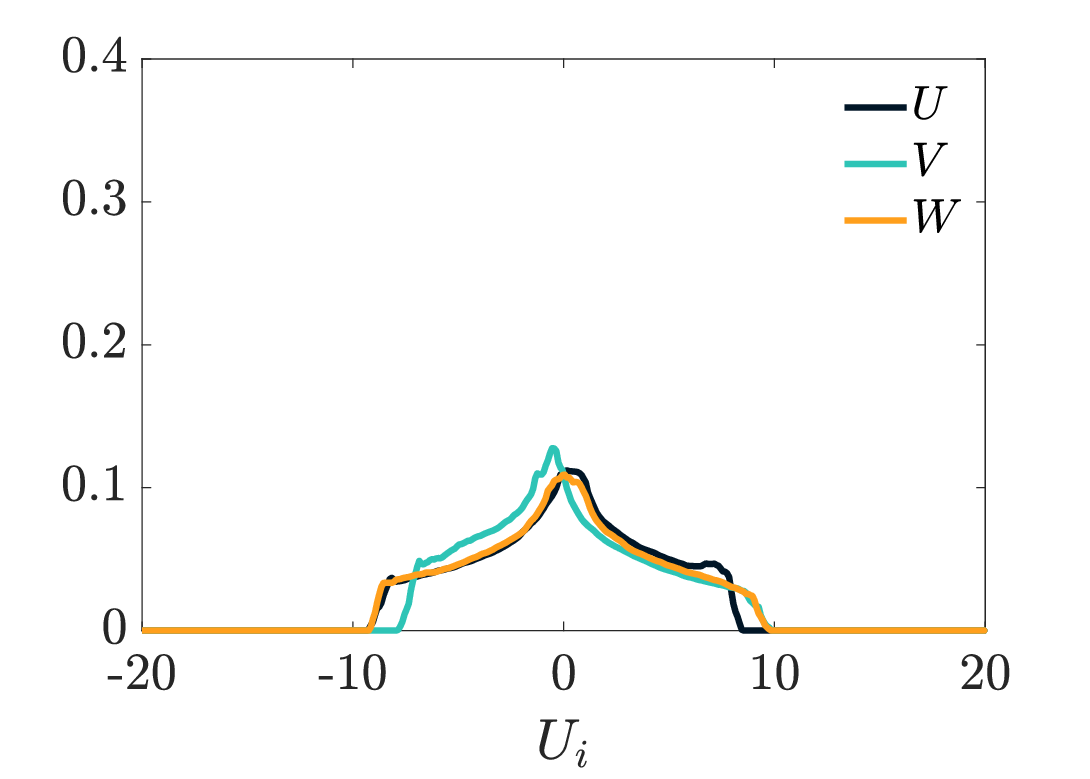}
\includegraphics[width=0.49\textwidth]{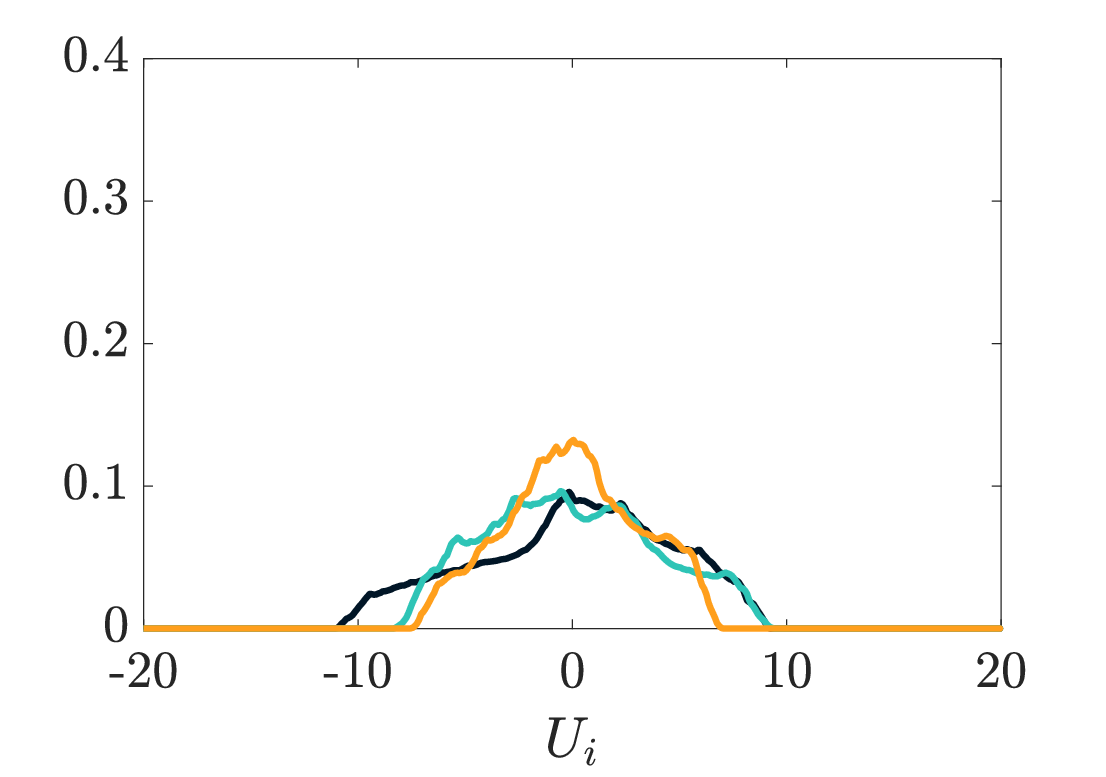}
\includegraphics[width=0.49\textwidth]{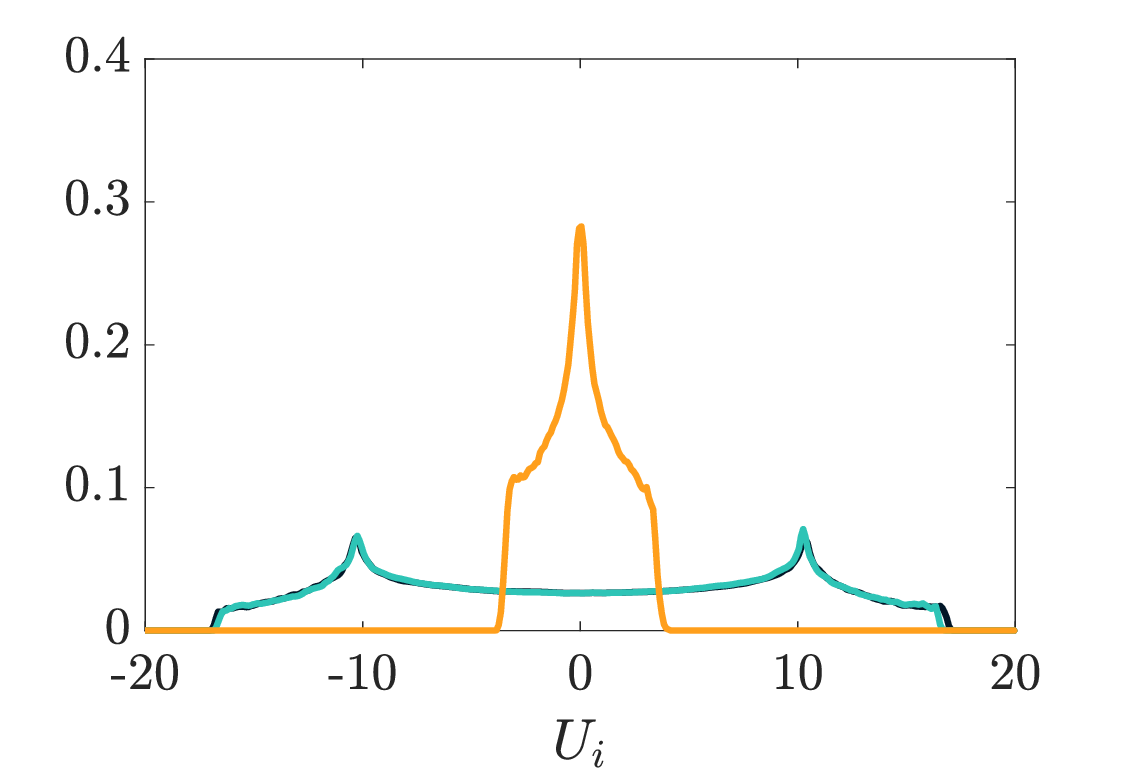}
\includegraphics[width=0.49\textwidth]{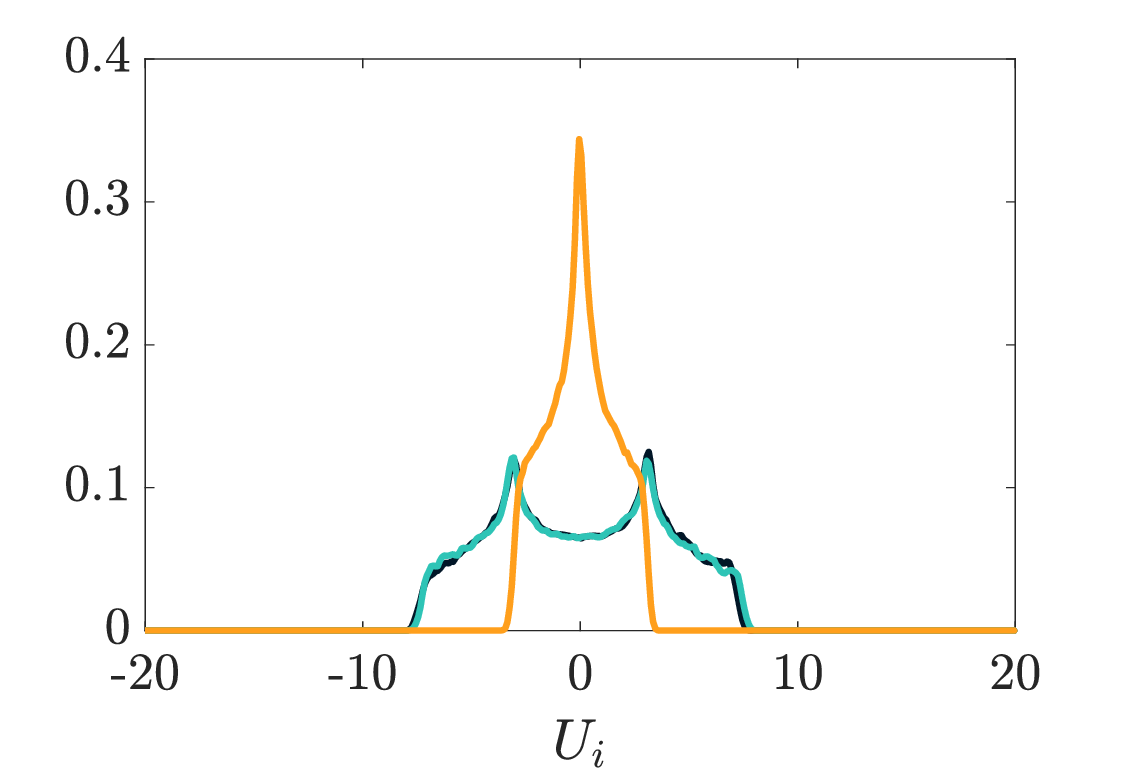}
\includegraphics[width=1.0\textwidth]{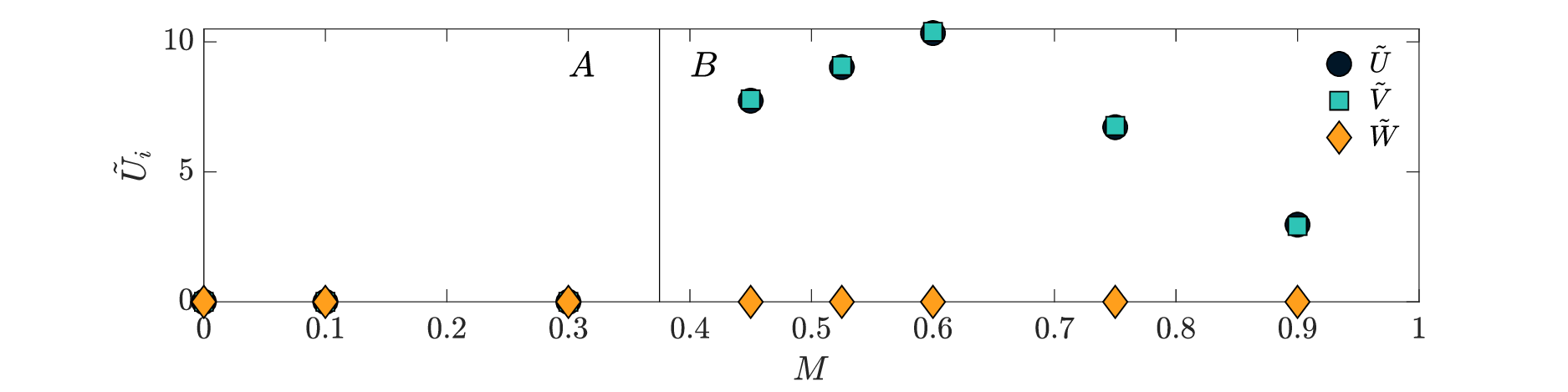}
\caption{Time-average flow velocity $\bm{U}(\bm{x})$ for $D/\eta=32$. Histogram of the spatial distribution of the three components of $\bm{U}(\bm{x})$ for the single phase (top left), and the $D/\eta=32$ particulate cases with $M=0.3$ (top right), $M=0.6$ (centre left) and $M=0.9$ (centre right). Black is for $U$, green for $V$ and orange for $W$. The bottom panel shows the dependence of the modes of the three velocity components ($\hat{U}$, $\hat{W}$ and $\hat{W}$) on $M$.}
\label{fig:velf_pdf}
\end{figure}

For a quantitative description of the mean-flow modulation, figure \ref{fig:velf_pdf} shows the histogram of the spatial distribution of the three velocity components $U(\bm{x})$, $V(\bm{x})$ and $W(\bm{x})$ for the $D/\eta=32$ particle-laden cases. Recall that $\bm{U}(\bm{x}) \equiv \overline{ \bm{u}(\bm{x},t) }$ is the space-dependent fluid velocity field averaged in time.
In regime A ($M < 0.45$) the modulation of $\bm{U}(\bm{x})$ is isotropic and the distribution of the three velocity components almost overlap, showing an unimodal distribution centred in $\hat{U} = \hat{V} = \hat{W} = 0$. In regime B ($M \ge 0.45$), the $W$ component retains a similar unimodal distribution, but the distribution becomes narrower in agreement with the progressive attenuation of its spatial fluctuations. In agreement with the spatial distribution shown in figure \ref{fig:Mvel}, instead, the $U$ and $V$ velocity components show a symmetric bimodal distribution. The modes $\pm \hat{U} = \pm \hat{V}$ change with $M$, and can be used as an estimate of the mean-flow anisotropy; see the bottom panel of figure \ref{fig:velf_pdf}.

\subsection{Turbulence modulation}

\subsubsection{Energy spectra}
\label{sec:spectra}

\begin{figure}
\centering
\includegraphics[trim={0 33 0 0},clip,width=1\textwidth]{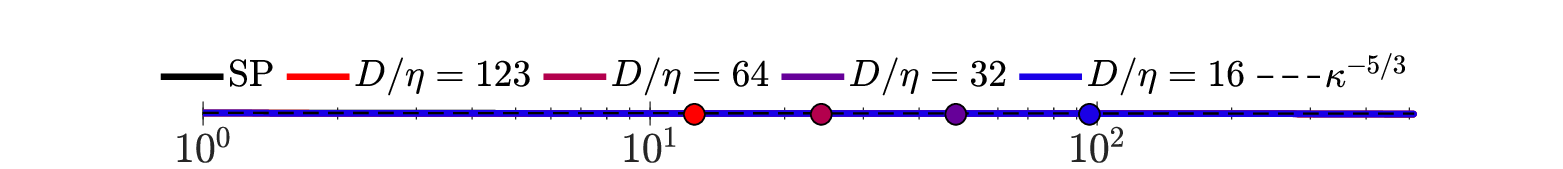}
\includegraphics[width=0.9\textwidth]{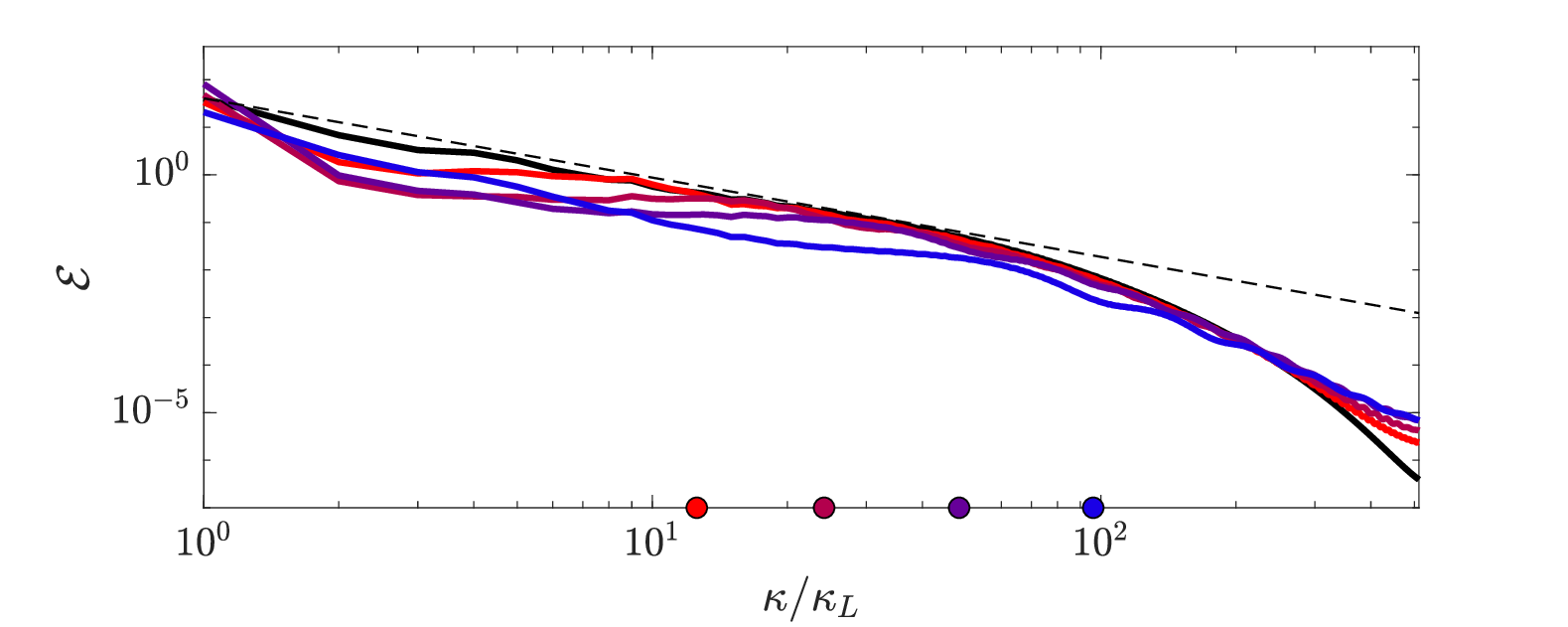}
\includegraphics[width=0.9\textwidth]{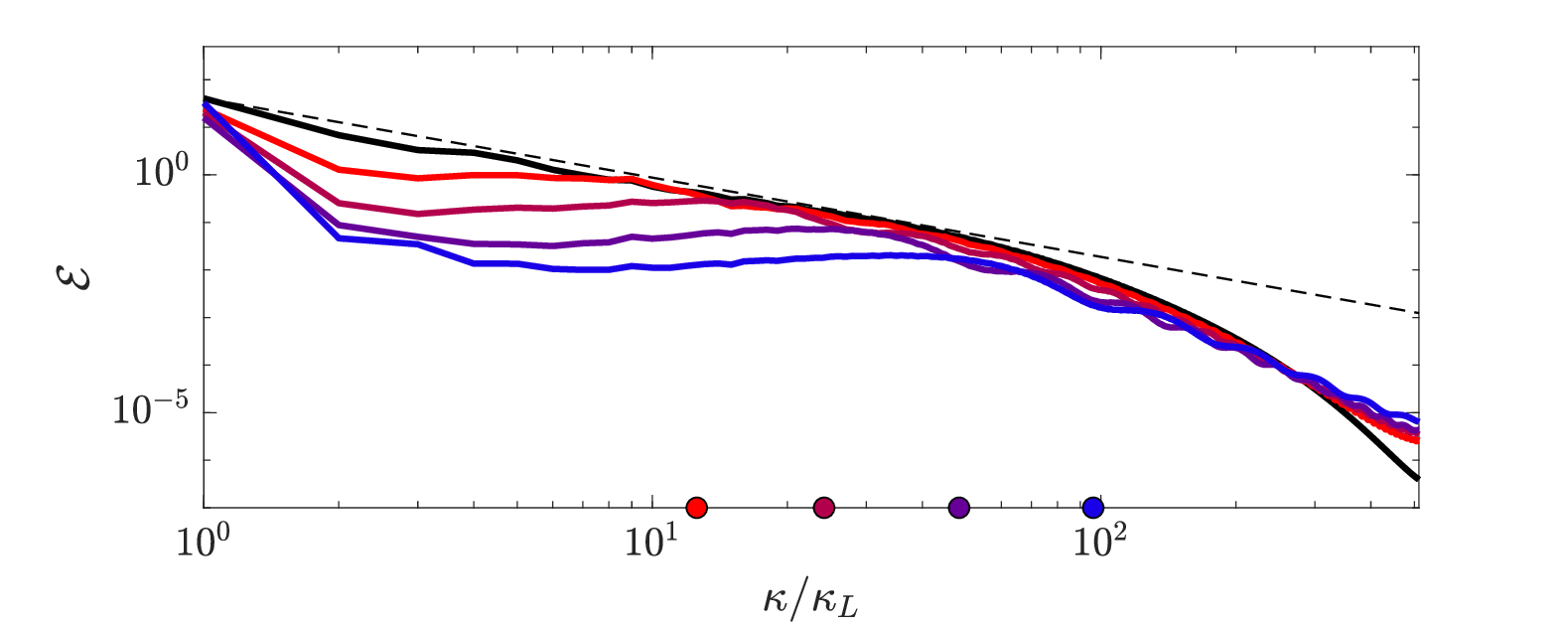}
\caption{Dependence of the energy spectrum on the particles size $D/\eta$ for (top) $M=0.6$ and (bottom) $M=0.9$. The circles identify the particle diameter wavenumbers $\kappa_d = 2 \pi/D$. Here and hereinafter $\kappa_L = 2 \pi /L$.}
\label{fig:spec_D}
\end{figure}
\begin{figure}
\centering
\includegraphics[trim={0 15 0 0},clip,width=1\textwidth]{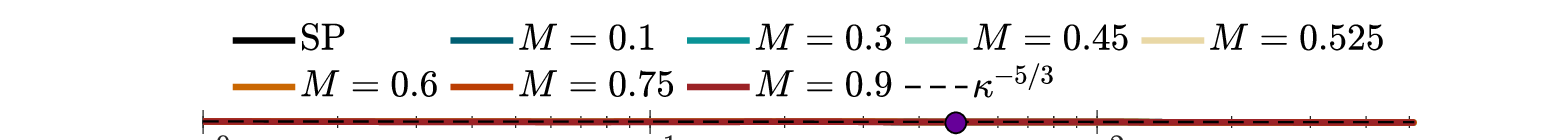}
\includegraphics[width=0.9\textwidth]{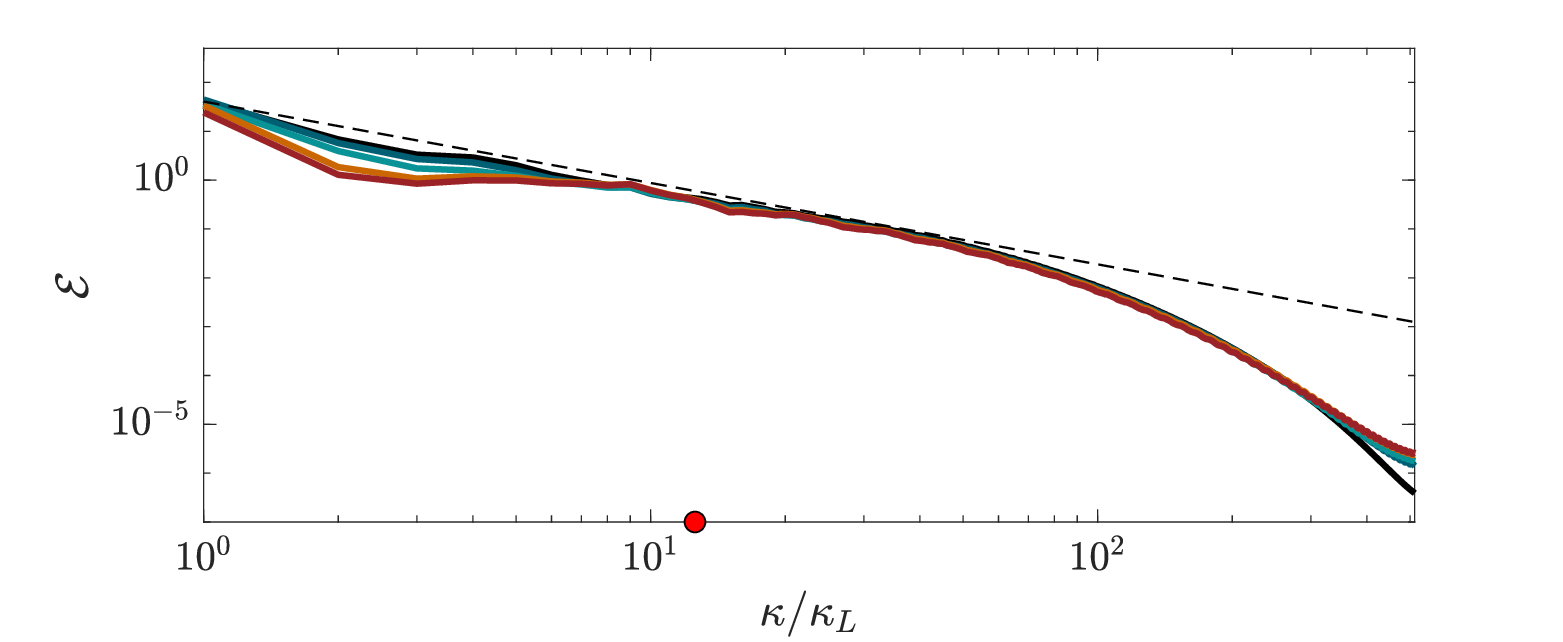}
\includegraphics[width=0.9\textwidth]{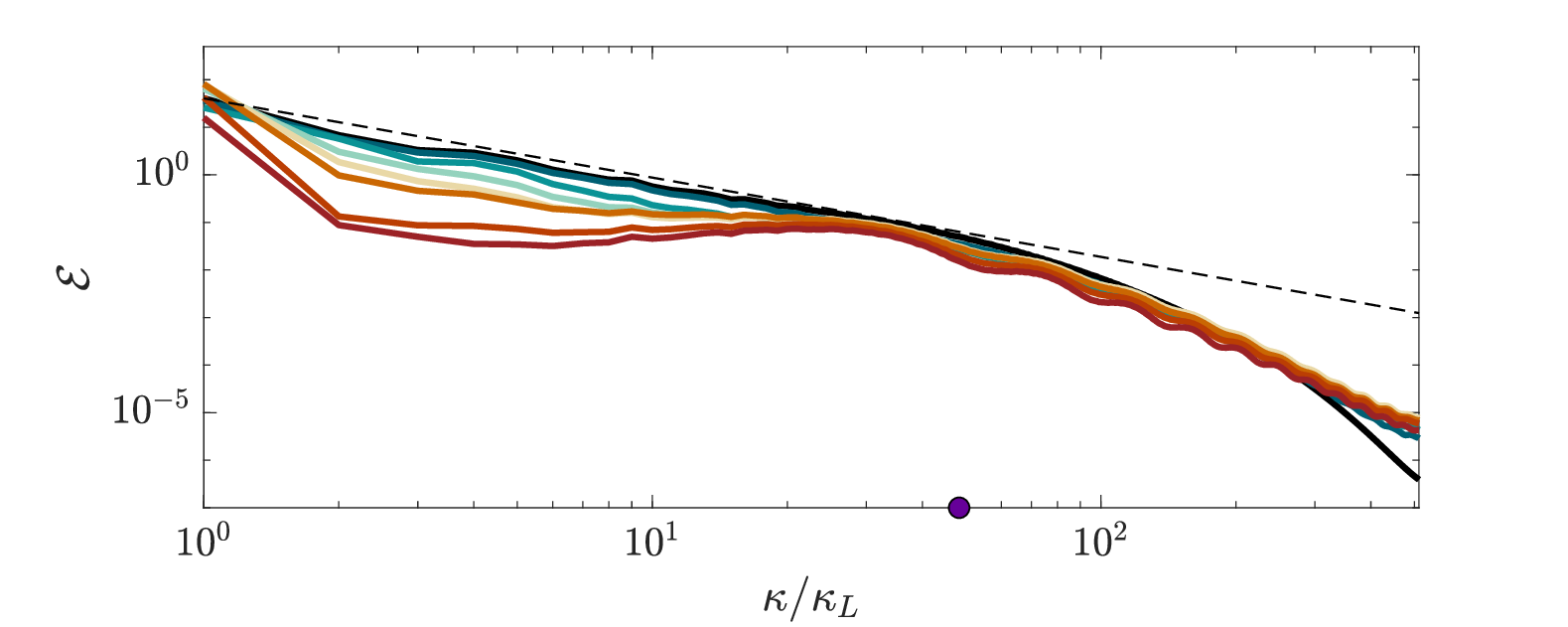}
\caption{Dependence of the energy spectrum on the mass fraction for  (top) $D/\eta=123$ and (bottom) $D/\eta=32$. The circles identify the particle diameter wavenumbers $\kappa_d = 2 \pi/D$.}
\label{fig:spec_M}
\end{figure}

To shed light on the mechanism with which the particles modulate the flow at smaller scales, we investigate the influence of the solid phase on the scale-by-scale energy content of the carrier flow, i.e. the energy spectrum.

It is worth recalling that, although for some $M$ and $D$ the solid phase modulates the largest scales ($\kappa/\kappa_L=1$, where $\kappa$ is the wave number and $\kappa_L = 2 \pi/L$) towards an anisotropic state (regime B), smaller scales with $\kappa/\kappa_L>1$ are isotropic and homogeneous for all cases (see figure \ref{fig:energyM}). The flow modulation for $\kappa/\kappa_L>1$, indeed, does not depend on how energy is injected in the system, and on how particles modify the structure of the largest scales. This is clearly shown in appendix \ref{sec:Pope}, where we present results from additional simulations carried out using the forcing introduced by \cite{eswaran-pope-1988} that, unlike the ABC forcing, does not generate a coherent and inhomogeneous shear at the largest scales $\kappa/\kappa_L=1$.

The presence of the solid phase modifies the energy spectrum of the carrier flow in a way that largely depends on the size and density of the particles. As first observed by \cite{tenCate-etal-2004}, particles drain energy from the fluid phase at scales larger than $D$, and inject it back in the fluid at smaller scales by means of their wake. This results into an energy depletion at low wavenumbers, i.e. large scales, and energy enhancement at large wavenumbers, i.e. small scales. Figures \ref{fig:spec_D} and \ref{fig:spec_M} consider separately the effect of the particle size and density on the energy spectrum. Note in passing that the single phase energy spectrum (black line) shows that the Reynolds number considered in this work leads to a proper separation of scales, with an inertial range of scales that extends to almost two decades of wavenumbers. Also, the oscillations at large wavenumbers appear as we compute the spectra using the fluid velocity field at all the mesh points of the computational domain, including those that are inside the particles \citep{lucci-etal-2010}. However, when dealing with structure functions in the real space (see section \ref{sec:inter}), we have verified that the results do not change when the points within the particles are neglected.

We start investigating the dependence of the energy spectrum on the particle size (figure \ref{fig:spec_D}). For large particles, the energy depletion is limited at the largest scales $\kappa < \kappa_{p,1} \lessapprox \kappa_d = 2 \pi/D$, the energy spectrum recovers the $\kappa^{-5/3}$ decaying predicted by the Kolmogorov theory for intermediate $\kappa $, and energy is enhanced at the smallest scales $\kappa > \kappa_{p,2}$. For example, for $D/\eta=123$ the energy depletion is observed for $ \kappa/\kappa_L \le \kappa_{p,1}/\kappa_L \approx 5$, while energy enhancement occurs for $\kappa/\kappa_L \ge \kappa_{p,2}/\kappa_L \approx 230$. Smaller particles amplify the overall mechanism. They drain energy from a wider range of scales, and, therefore, inject back a larger amount of energy at a wider range of small scales. For smaller particles, indeed, $\kappa_{p,1}$ increases, while $\kappa_{p,2}$ decreases. For $M=0.6$, for example, we measure that $\kappa_{p,1}/\kappa_L \approx 49, 25, 17$ and $8$ and $\kappa_{p,2}/\kappa_L \approx 189, 199, 219$ and $230$ for $D/\eta=16, 32, 64$ and $123$; $\kappa_{p,1}$ well correlates with the particle diameter wavenumber, being $\kappa_d/\kappa_L \approx 96, 45, 25$ and $12.5$ respectively. Interestingly, for $D/\eta \le 32$ the $\kappa^{-5/3}$ decay is not observed, indicating that in non-dilute suspensions of small particles the inertial energy cascade is substantially modified (see section \ref{sec:energy-budget}). 

We now consider the effect of the mass fraction on the energy spectrum (figure \ref{fig:spec_M}). When fixing the size of the particles, the cutoff wavenumbers $\kappa_{p,1}$ and $\kappa_{p,2}$ almost do not vary with the mass fraction: the range of scales where particles drain or release energy does not change.
However, heavier particles interact more effectively with the fluid phase, leading to a stronger large-scale energy depletion --- that explains the larger attenuation of the fluctuating energy discussed in section \ref{sec:integ} ---, and to a larger energy enhancement at the smallest scales \citep{yeo-etal-2010}. Heavier particles, indeed, result into an increase of the fluid-particle system inertia \citep{balachandar-eaton-2010}. 

Note that, for $D/\eta=32$ and $0.45 \le M \le 0.6$, the energy content at $\kappa/\kappa_L = 1$ is larger compared to the single-phase case, in agreement with the enhancement of the mean flow (largest scales) previously discussed in section \ref{sec:integ}.

\subsubsection{Scale-by-scale energy transfer}
\label{sec:energy-budget}

\begin{figure}
\centering
\includegraphics[trim={0 10 0 0},clip,width=0.65\textwidth]{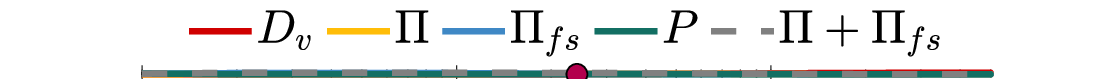}
\includegraphics[width=0.49\textwidth]{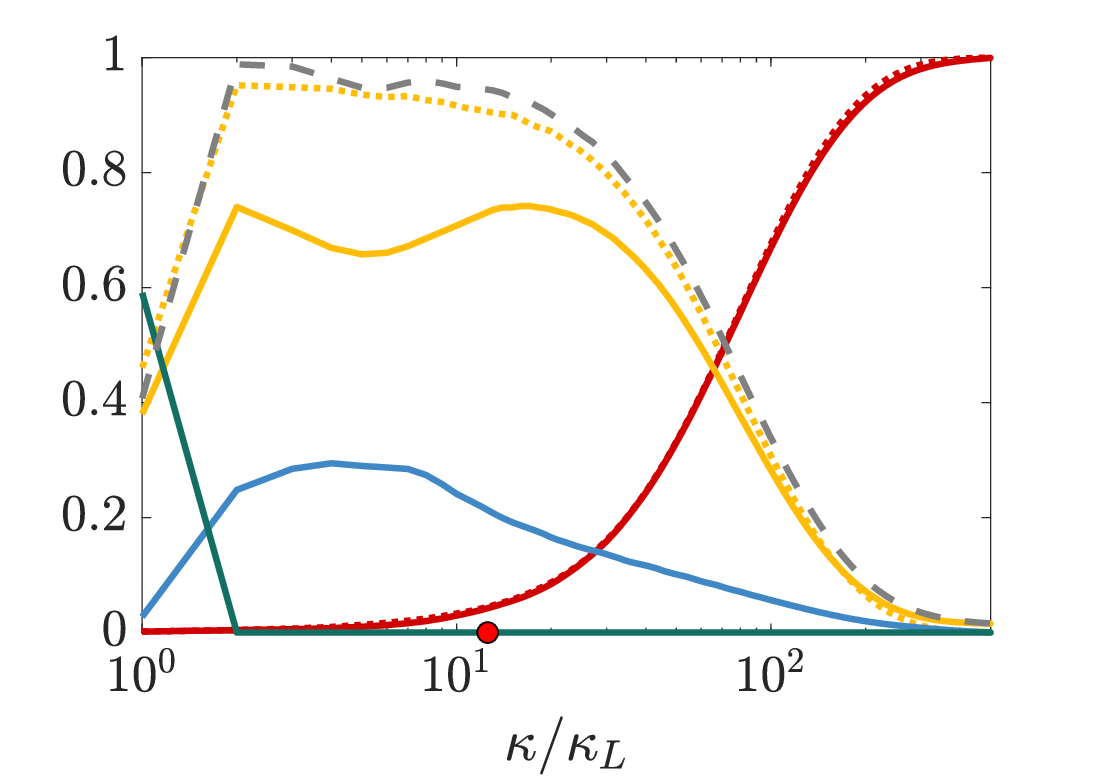}
\includegraphics[width=0.49\textwidth]{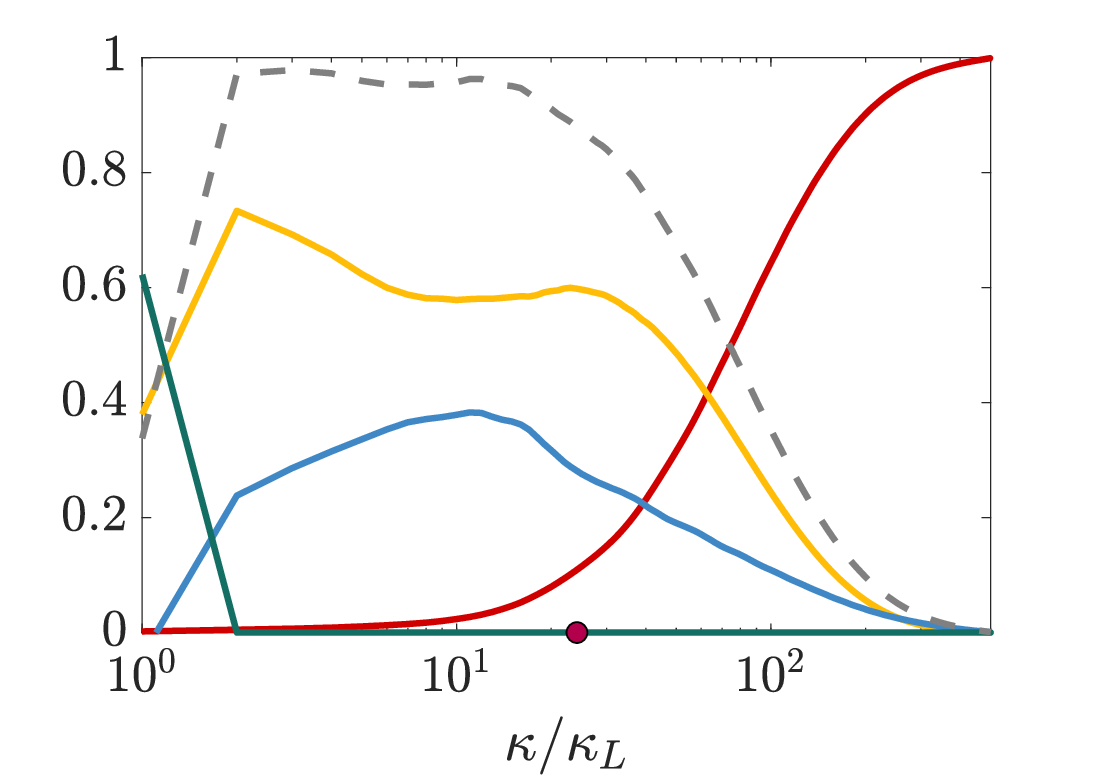}
\includegraphics[width=0.49\textwidth]{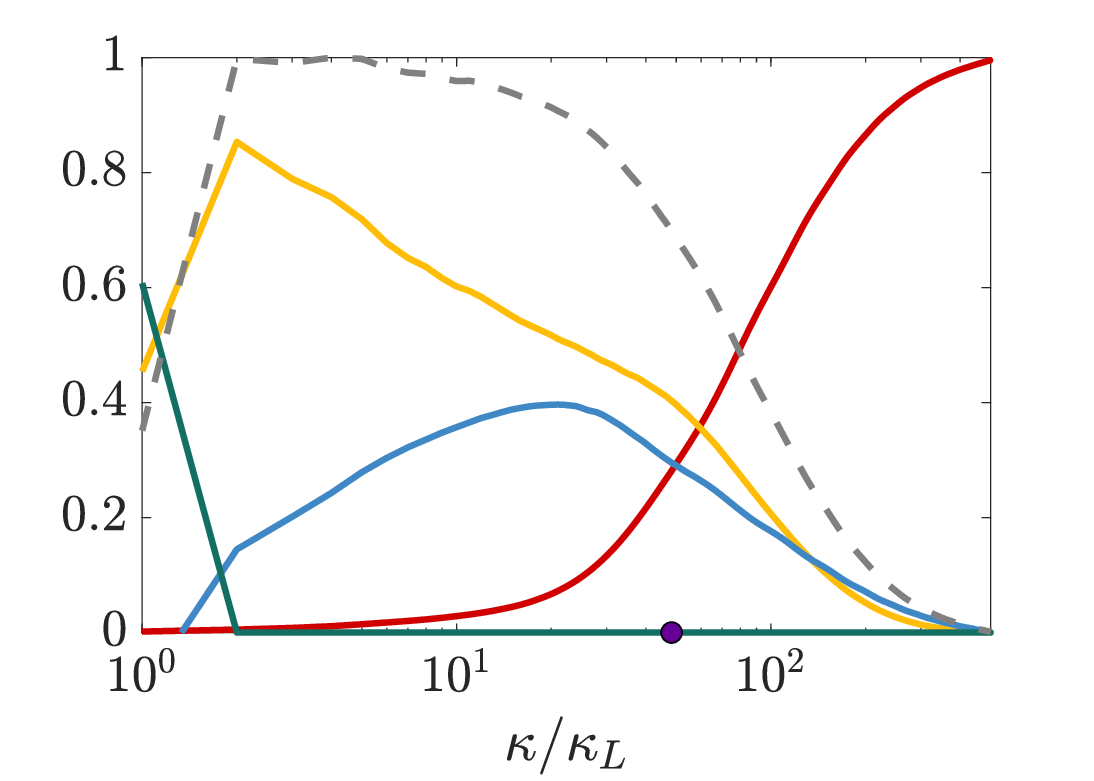}
\includegraphics[width=0.49\textwidth]{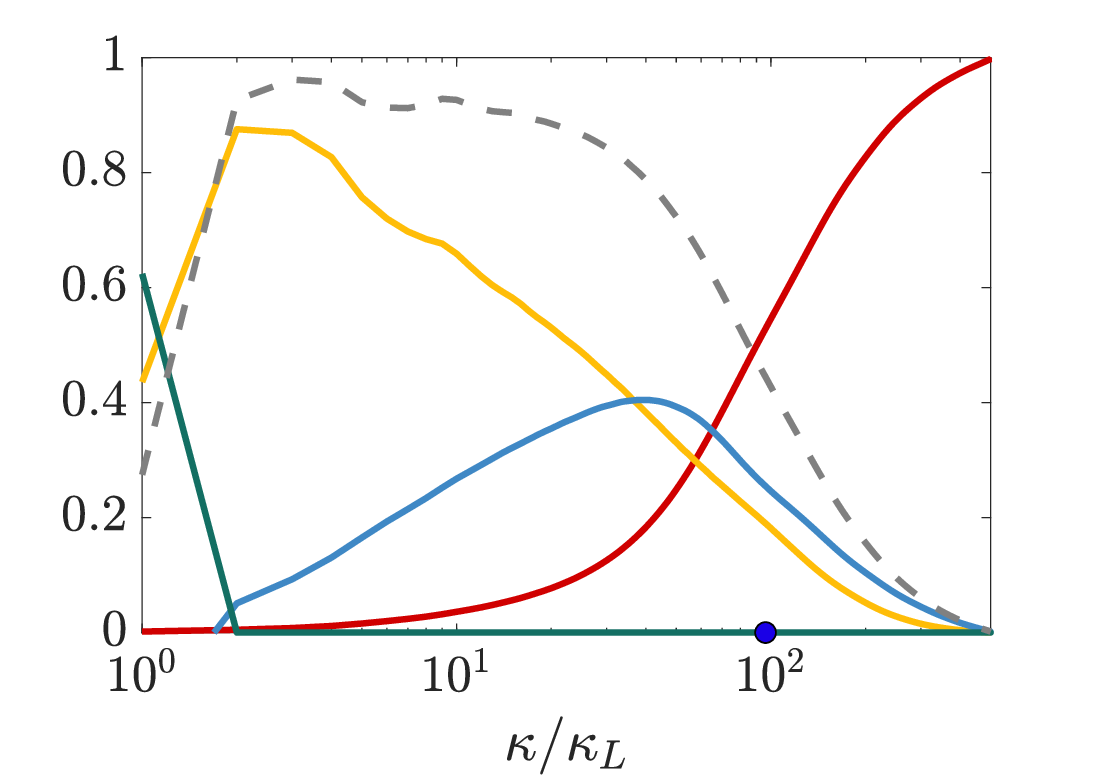}
\caption{Scale-by-scale energy transfer balance for $M=0.3$ and (top left) $D/\eta=123$, (top right) $D/\eta=64$, (bottom left) $D/\eta=32$ and (bottom right) $D/\eta=16$ . The circles identify the particle diameter wavenumbers. The dotted lines in the top left panel are the results for the single-phase case, to be used as reference.}
\label{fig:bud_gamma5}
\end{figure}
\begin{figure}
\centering
\includegraphics[trim={0 10 0 0},clip,width=0.65\textwidth]{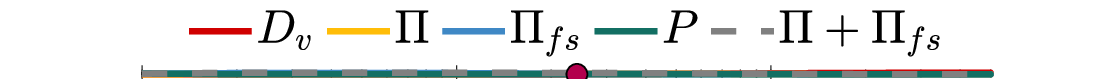}
\includegraphics[width=0.49\textwidth]{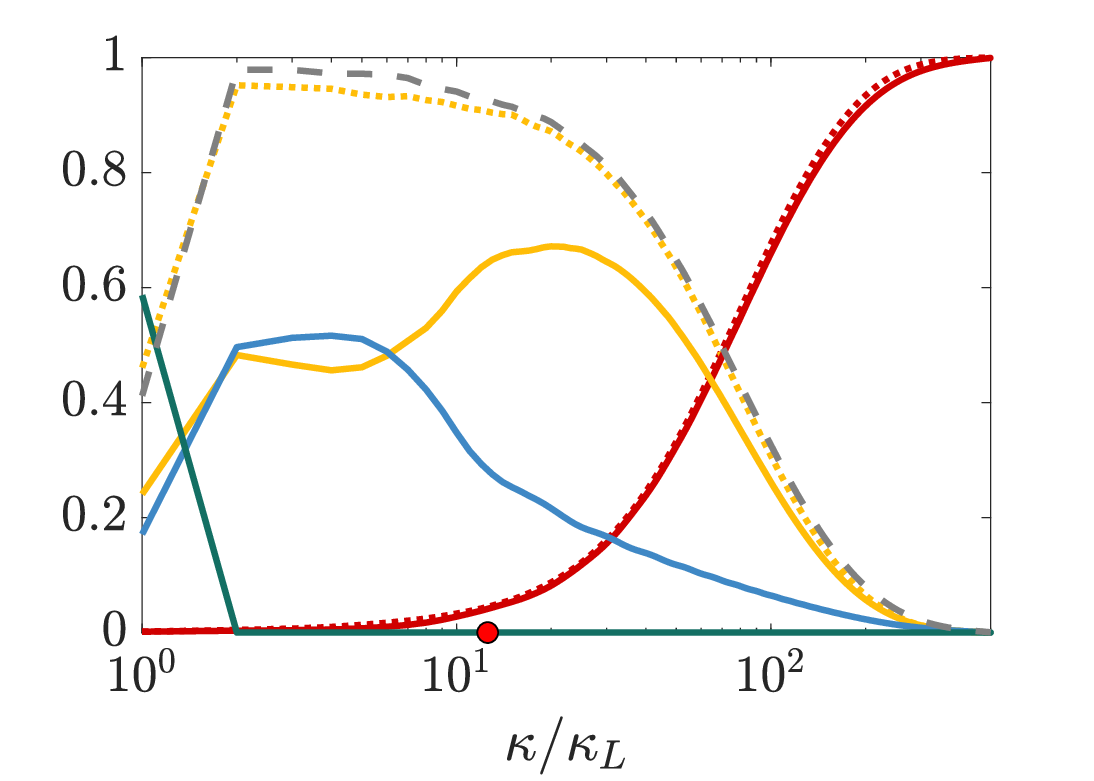}
\includegraphics[width=0.49\textwidth]{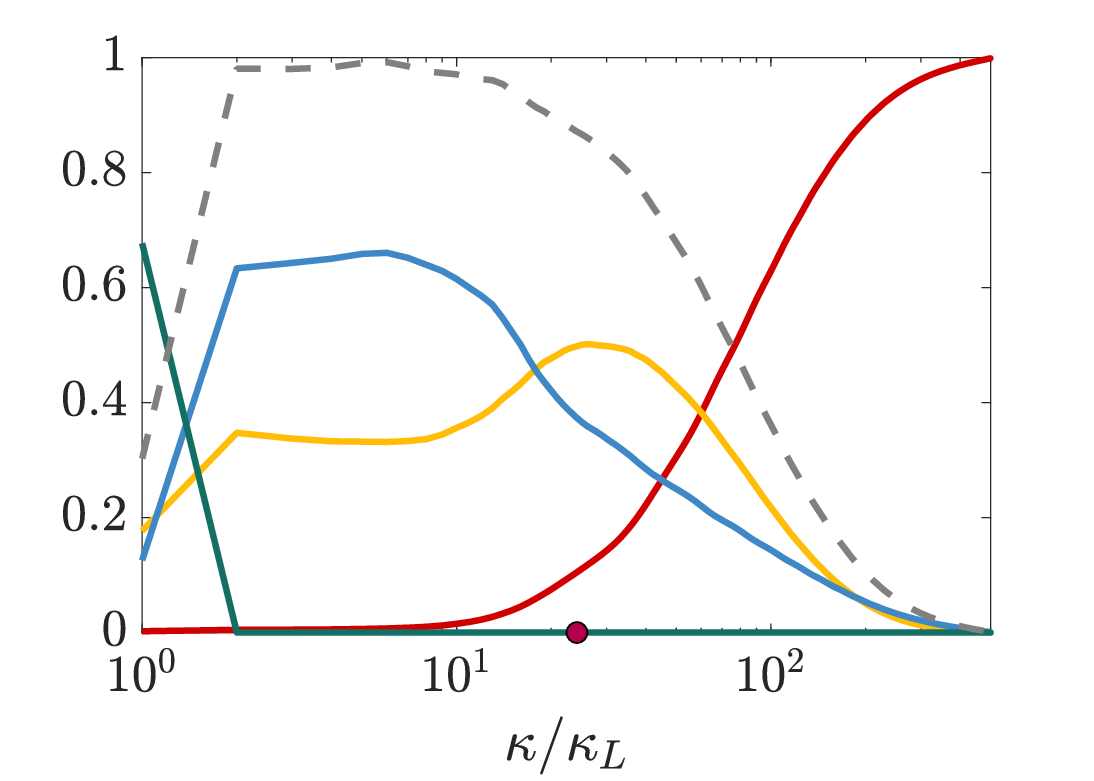}
\includegraphics[width=0.49\textwidth]{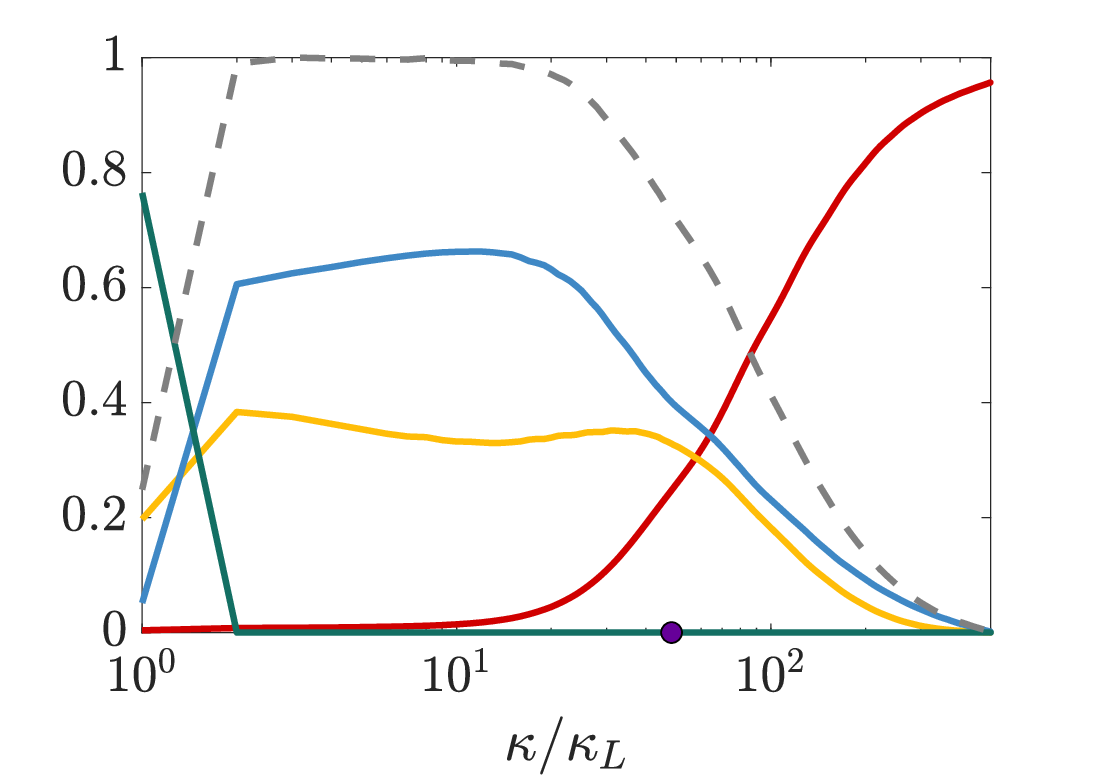}
\includegraphics[width=0.49\textwidth]{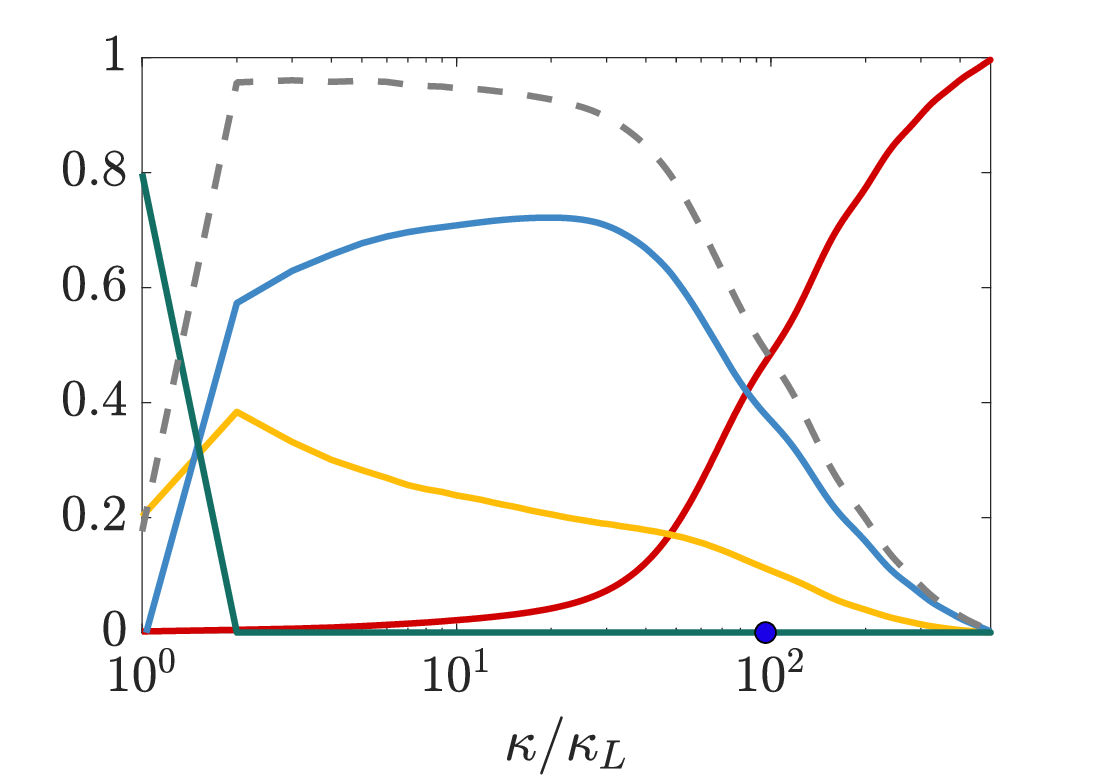}
\caption{Same as figure \ref{fig:bud_gamma5}, but for $M=0.6$.}
\label{fig:bud_gamma17}
\end{figure}
\begin{figure}
\centering
\includegraphics[trim={0 10 0 0},clip,width=0.65\textwidth]{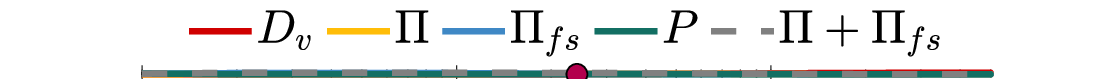}
\includegraphics[width=0.49\textwidth]{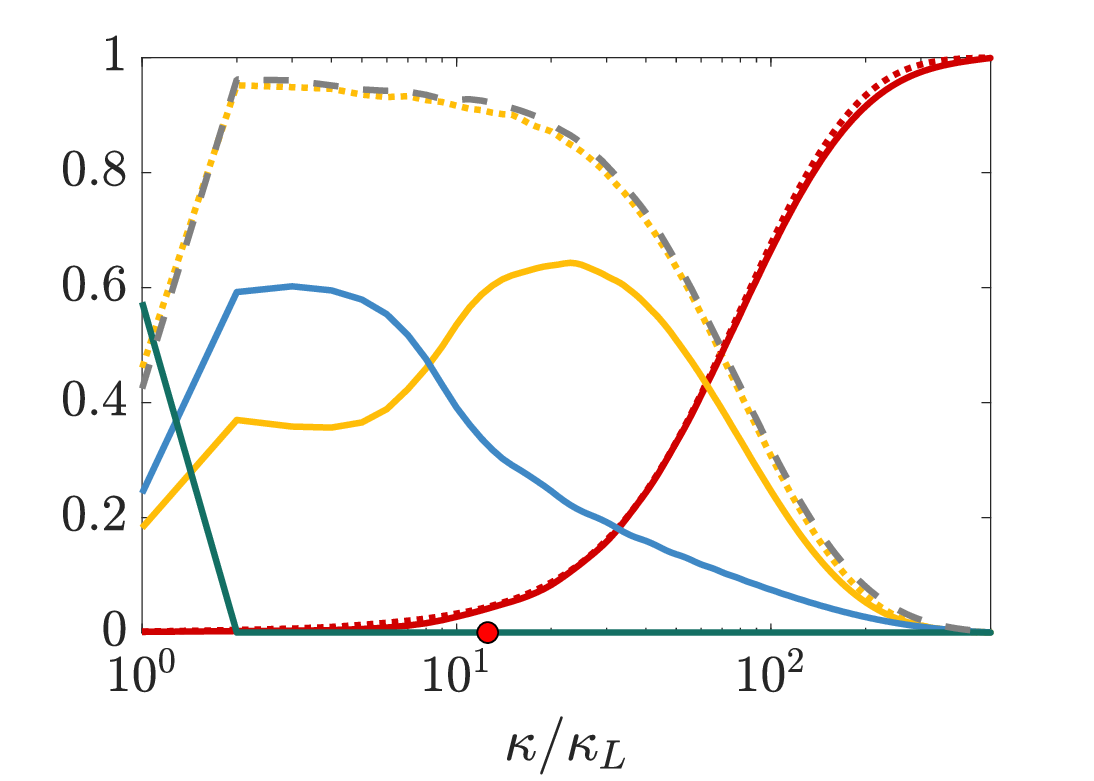}
\includegraphics[width=0.49\textwidth]{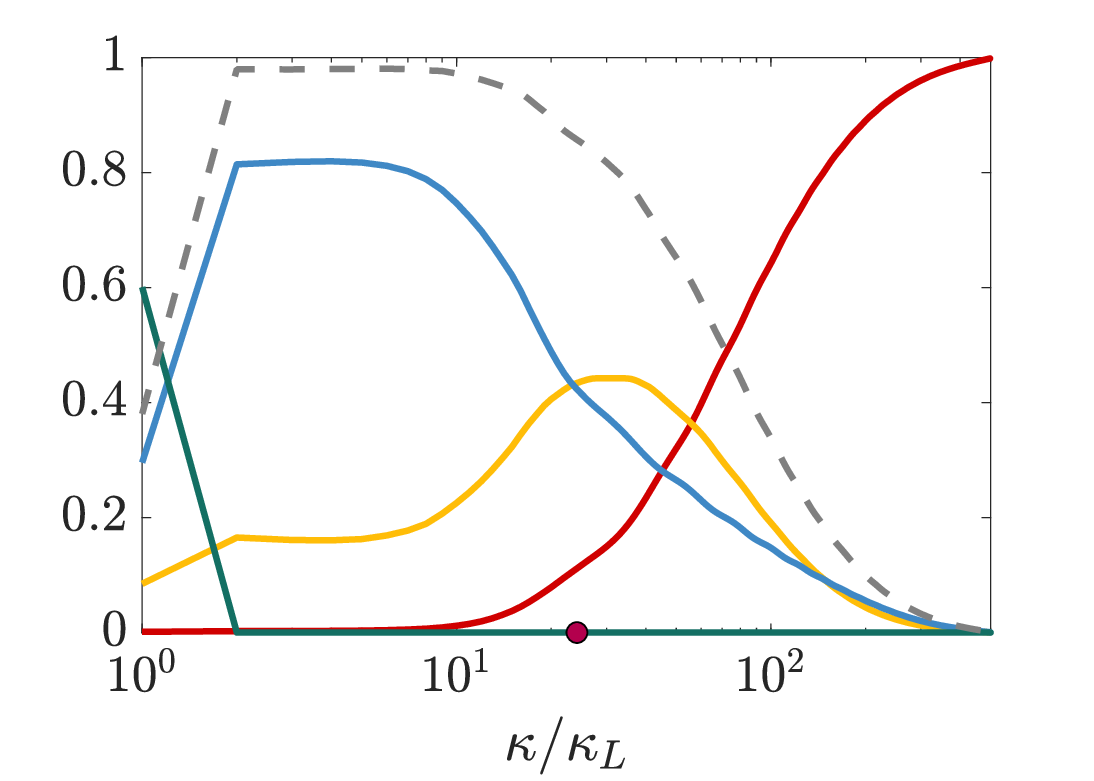}
\includegraphics[width=0.49\textwidth]{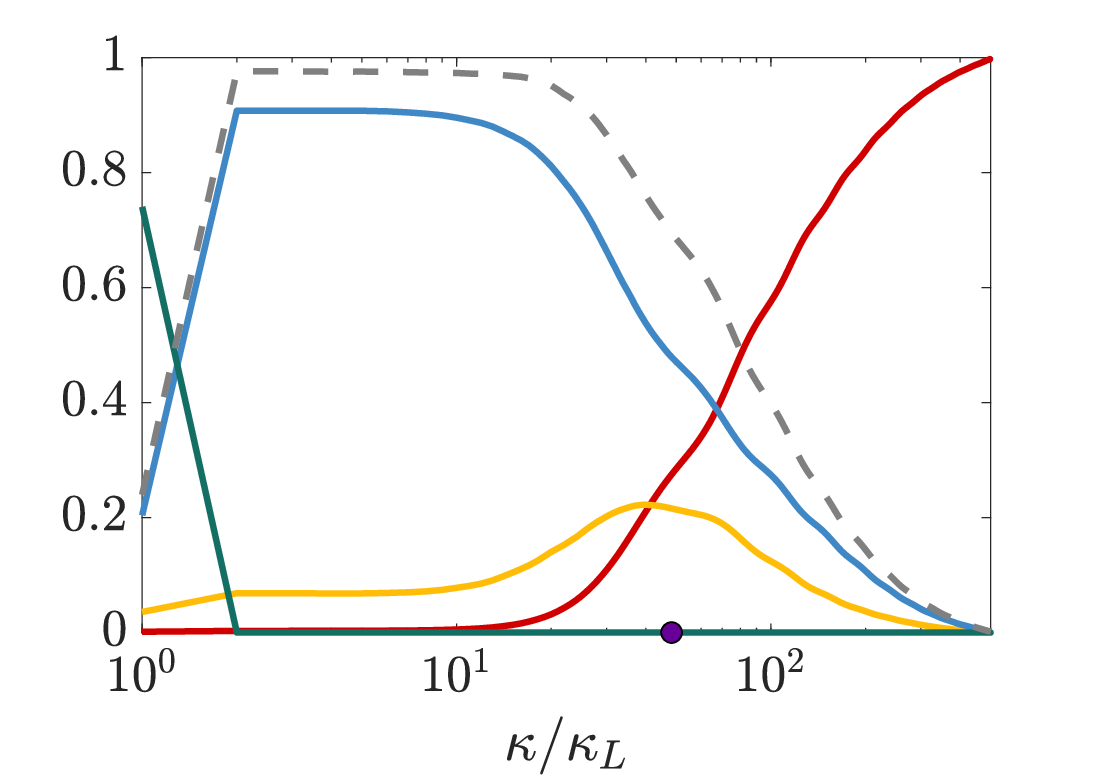}
\includegraphics[width=0.49\textwidth]{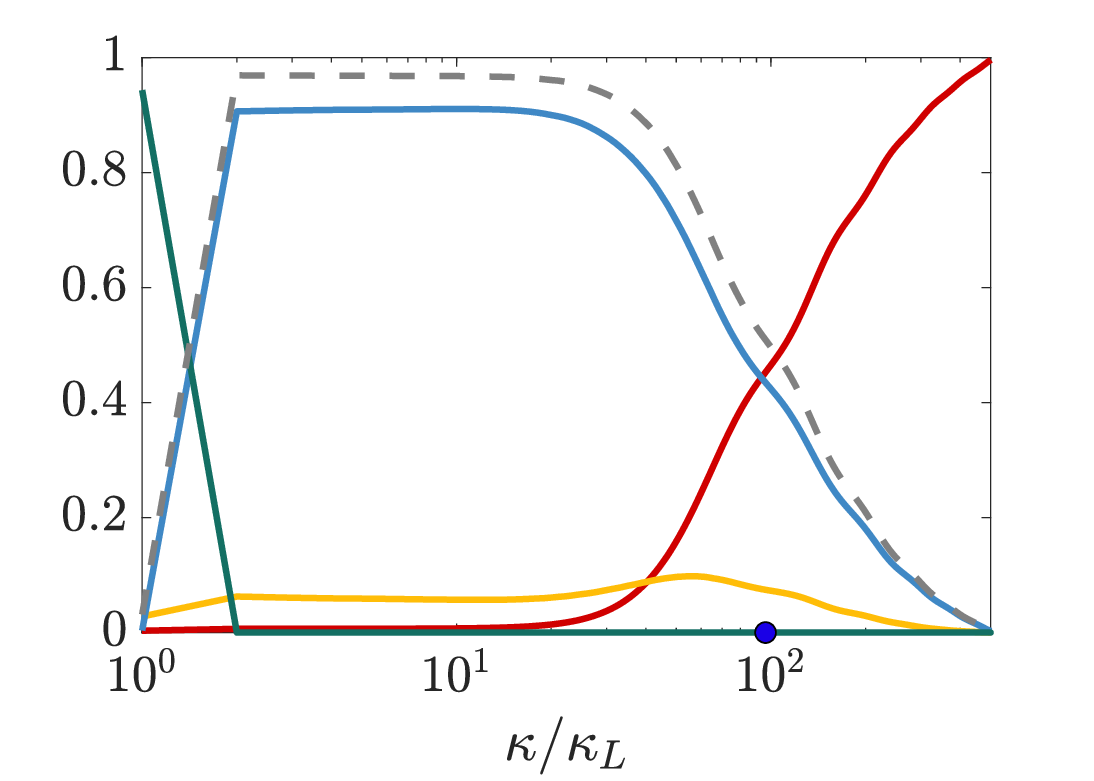}
\caption{Same as figure \ref{fig:bud_gamma5}, but for $M=0.9$.}
\label{fig:bud_gamma100}
\end{figure}

For a more detailed insight of the influence of the dispersed phase on the energy distribution mechanism, we look at the scale-by-scale energy transfer balance. It is obtained after some manipulations of the Fourier-transformed form of the Navier--Stokes equations \ref{eq:Navier-Stokes}. The energy balance can be compactly written as
\begin{equation}
  P(\kappa) + \Pi(\kappa) + \Pi_{fs}(\kappa) + D_v(\kappa) = \epsilon,
  \label{eq:budget}
\end{equation}
where $P(\kappa)$ is the scale-by-scale turbulent energy production due to the external forcing, $\Pi(\kappa)$ and $\Pi_{fs}(\kappa)$ are the scale-by-scale energy fluxes associated with the non linear convective term and with the fluid-solid coupling term, and $D_v(\kappa)$ is the scale-by-scale viscous dissipation. They are defined as
\begin{align*}
  P(\kappa) &= \int_\kappa^\infty \frac{1}{2} \left( \hat{\bm{f}} \cdot \hat{\bm{u}}^* + \hat{\bm{f}}^* \cdot \hat{\bm{u}} \right) \text{d}k, \\
  \Pi(\kappa) &= \int_\kappa^\infty -\frac{1}{2} \left( \hat{\bm{G}} \cdot \hat{\bm{u}}^* + \hat{\bm{G}}^* \cdot \hat{\bm{u}} \right) \text{d}k, \\
  \Pi_{fs}(\kappa) &= \int_\kappa^\infty \frac{1}{2} \left( \hat{\bm{f}}^{\leftrightarrow p} \cdot \hat{\bm{u}}^* + \hat{\bm{f}}^{\leftrightarrow p,*} \cdot \hat{\bm{u}} \right) \text{d}k, \\
  D_v(\kappa) &= \int_0^\kappa \left( -2 \nu k^2 \mathcal{E} \right) \text{d} k.
\end{align*}
Here $\hat{\cdot}$ denotes the Fourier transform operator, and the superscript $\cdot^*$ denotes complex conjugate. $\hat{\bm{G}}$ is the Fourier transform of the non linear term $\bm{\nabla} \cdot (\bm{u} \bm{u})$. For a detailed derivation of the budget equation we refer the reader to \cite{pope-2000}. The production term and the fluxes are obtained integrating from $\kappa$ to $\infty$, while the dissipation term is integrated from $0$ to $\kappa$, to obtain a positive quantity that matches $\epsilon$. $\Pi(\kappa)$ and $\Pi_{fs}(\kappa)$ do not produce nor dissipate energy at any scale, but redistribute it among scales by means of the classical energy cascade and the fluid-solid interaction. As such, when integrating from $\kappa = 0$ to $\kappa = \infty$, both $\Pi(0)$ and $\Pi_{fs}(0)$ have to be null, meaning that, in a statistical sense, the amount of energy the fluxes drain from the flow at certain scales has to match the amount of energy they release back to the flow at other scales ($\Pi_{fs}$ has to be null for $\kappa = \infty$, as in the present case the particle-particle interaction is subdominant).
Figures \ref{fig:bud_gamma5}, \ref{fig:bud_gamma17} and \ref{fig:bud_gamma100} show the dependence of the scale-by-scale energy budget on the particle size for $M=0.3$, $M=0.6$ and $M=0.9$.

In the single-phase case, the energy transfer mechanism is qualitatively and quantitatively well described by the Kolmogorov theory \citep{kolmogorov-1941}. Energy is injected at the largest scales of the flow by the external forcing $P(\kappa)$, and is transferred by means of the classical direct energy cascade --- identified in equation \ref{eq:budget} by the non linear convection term $\Pi(\kappa)$ --- to the smallest scales, where it is dissipated by viscosity $D_v(\kappa)$; see the dotted lines in the top left panels of figures \ref{fig:bud_gamma5}, \ref{fig:bud_gamma17} and \ref{fig:bud_gamma100}. The presence of the solid phase introduces an additional energy flux, associated with the fluid-solid interaction $\Pi_{fs}(\kappa)$, and the relevance of $\Pi(\kappa)$ and $\Pi_{fs}(\kappa)$ at the different scales changes with $D$ and $M$, in agreement with the different modulation of the energy spectrum.

For light particles, $M \le 0.3$, $\Pi_{fs}$ is small for all $\kappa$ and the fluid-solid interaction only marginally influences the overall scale energy transfer (see figure \ref{fig:bud_gamma5}). For large particles with $D/\eta \ge 32$, indeed, the fluid-solid coupling term is subdominant at all scales. For small particles with $D/\eta=16$, instead, $\Pi_{fs}$ slightly overcomes $\Pi$ for $\kappa/\kappa_L \gtrapprox 34$.

When heavier particles are considered ($M \ge 0.6$), the relevance of $\Pi_{fs}$ increases and the scale energy transfer mechanism due to the fluid-solid interaction progressively gains importance for all particle sizes (see figures \ref{fig:bud_gamma17} and \ref{fig:bud_gamma100}).
For large and heavy particles with $D/\eta \ge 64$ and $M \ge 0.6$, the fluid-solid coupling contribution $\Pi_{fs}(\kappa)$ dominates at small wavenumbers $\kappa \le \kappa_p $, while the non linear term $\Pi(\kappa)$ dominates at larger wavenumbers $\kappa>\kappa_p$, with the cross-over $\kappa_p$ increasing with the mass fraction and the particle wavenumber. For $M=0.6$ and $M=0.9$, we measure $\kappa_p/\kappa_L \approx 18$ and $24$ for $D/\eta=64$, and $\kappa_p/\kappa_L \approx 6$ and $8$ for $D/\eta=123$. Both the $\Pi(\kappa)$ and $\Pi_{fs}(\kappa)$ fluxes present a plateau, meaning that at the corresponding wavenumbers (in average) they do not drain nor release energy from the flow, but simply transfer it from larger to smaller scales. The scenario is the following: particles drain energy from the flow at scales larger than $D$, where $-\text{d}\Pi_{fs}/\text{d}\kappa<0$, and released it by means of their wake at smaller scales with $\kappa \gtrapprox \kappa_p \approx \kappa_d$, where $-\text{d}\Pi_{fs}/\text{d}\kappa>0$. The non linear energy transfer $\Pi(\kappa)$, instead, drives the energy transfer at smaller scales $\kappa > \kappa_p$, where the classical energy cascade is recovered; see in figures \ref{fig:spec_D} and \ref{fig:spec_M} that here the spectrum follows the Kolmogorov $\kappa^{-5/3}$ decay. $\Pi(\kappa)$ drains energy from the flow ($-\text{d}\Pi/\text{d}\kappa<0$) for $\kappa \lessapprox \kappa_d$, and transfers it to the smallest and dissipative scales ($-\text{d}\Pi/\text{d}\kappa>0$). 

For smaller particles with $D/\eta \le 32$ and $M \ge 0.6$ the relevance of the fluid-solid coupling term increases. In fact, the plateau of $\Pi_{fs}$ progressively increases, and the range of scales dominated by the fluid-particle interaction widens. For small and heavy particles ($D/\eta \le 32$ and $M>0.6$), the fluid-solid coupling term $\Pi_{fs}$ dominates at all scales, while the non linear term $\Pi$ is largely attenuated. In this case the classical energy cascade is almost annihilated, and the solid phase generates a direct link between the largest and energetic scales of the flow and the smallest and dissipative ones. The energy injected at the largest scales is drained by the particles, and by means of their wake it is directly transferred to the smallest scales where it is dissipated. This is consistent with the modification of the energy spectrum observed in figure \ref{fig:spec_D}.

A last comment regards the influence of the size and density of the particles on the dissipative range of scales. We denote with $\kappa_{diff}$ the wavenumber above which the dissipative term $D_v(\kappa)$ dominates. Overall, our data show that $\kappa_{diff}$ is only marginally influenced by $D$ and $M$. In fact, for large particles with $D/\eta \ge 32$, $\kappa_{diff}/\kappa_L \approx 60$ for all $D$ and $M$. For small particles with $D/\eta=16$, instead, $\kappa_{diff}$ increases with $M$ up to $\kappa_{diff}/\kappa_L \approx 94$: for heavier particles the dissipative range of scales shrinks, consistently with the enlargement of the plateau of $\Pi_{fs}$.

\subsubsection{Structure functions and intermittency}
\label{sec:inter}

\begin{figure}
\centering
\includegraphics[trim={0 14 0 0},clip,width=1\textwidth]{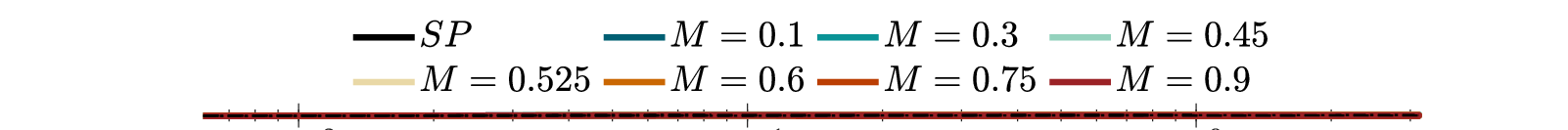}
\includegraphics[width=0.9\textwidth]{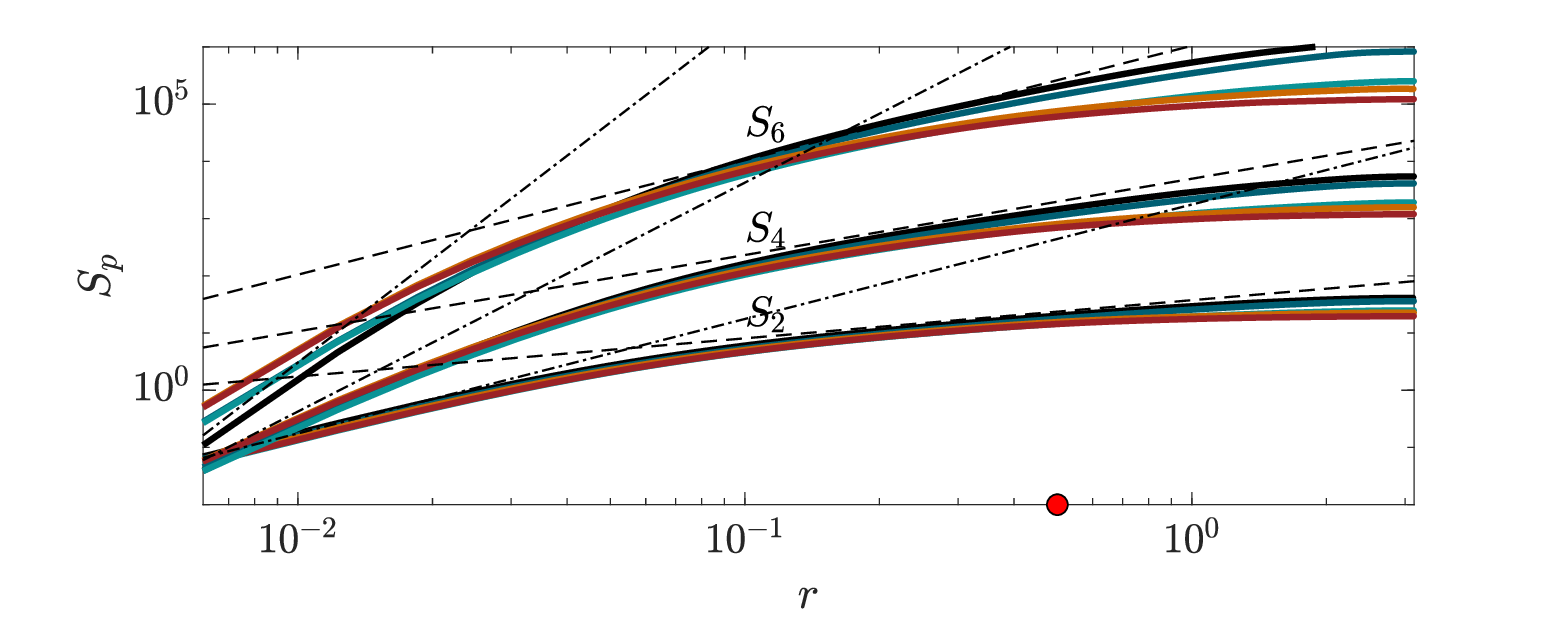}
\includegraphics[width=0.9\textwidth]{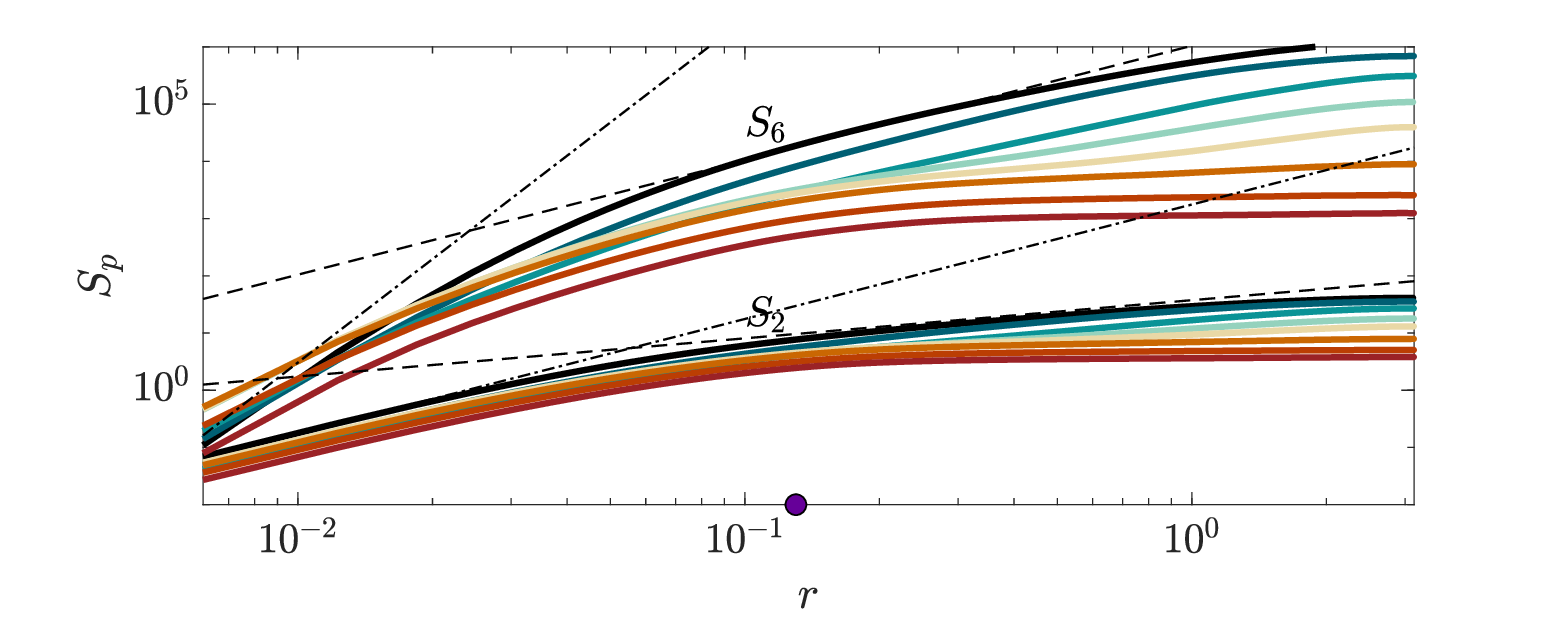}
\caption{Dependence of the $p-$order structure function $\aver{\delta u^p}$ on the mass fraction for (top) $D/\eta=123$ and (bottom) $D/\eta=32$. For $D/\eta=32$ $S_4$ is not plotted for sake of clarity. The dashed lines represent $r^{p/3}$, while the dash-dotted lines $r^{p}$. The circles identify the particle diameter $D$.}
\label{fig:Sp}
\end{figure}
To extend the analysis done in the spectral domain, we compute the longitudinal structure functions, defined as $S_p(r) = \overline{\aver{ \left( \delta u(r) \right)^p }}$, where $p$ is the order of the structure function and $\delta u(r) = u(\bm{x}+r)-u(\bm{x})$ is the velocity increment across a length scale $r$; see figure \ref{fig:Sp}. 

For the single-phase case, $S_p(r) \sim r^p$ at small scales and $S_p(r) \sim r^{p/3}$ in the inertial range, as predicted by the Kolmogorov theory \citep{kolmogorov-dissipation-1941}, with some deviations due to the flow intermittency \citep{frisch-1995,pope-2000}.
With the dispersed phase, the structure functions deviate from the single phase behaviour and the deviation progressively increases as the mass fraction increases and the particles size decreases, accordingly with the modulation of the energy spectrum. The even order structure functions flatten for $r \ge r_d \approx D$, progressively approaching a $r^0$ power-law, denoting that velocity fluctuations with scale larger than the particle size decorrelate and lose their coherency.

\begin{figure}
\centering
\includegraphics[trim={0 133 0 0},clip,width=1\textwidth]{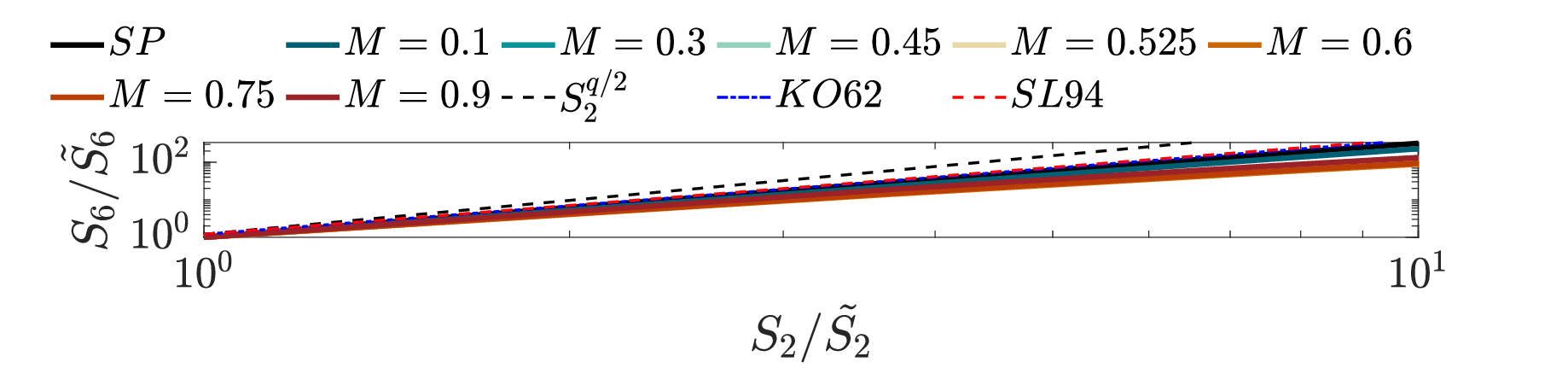}
\includegraphics[width=0.49\textwidth]{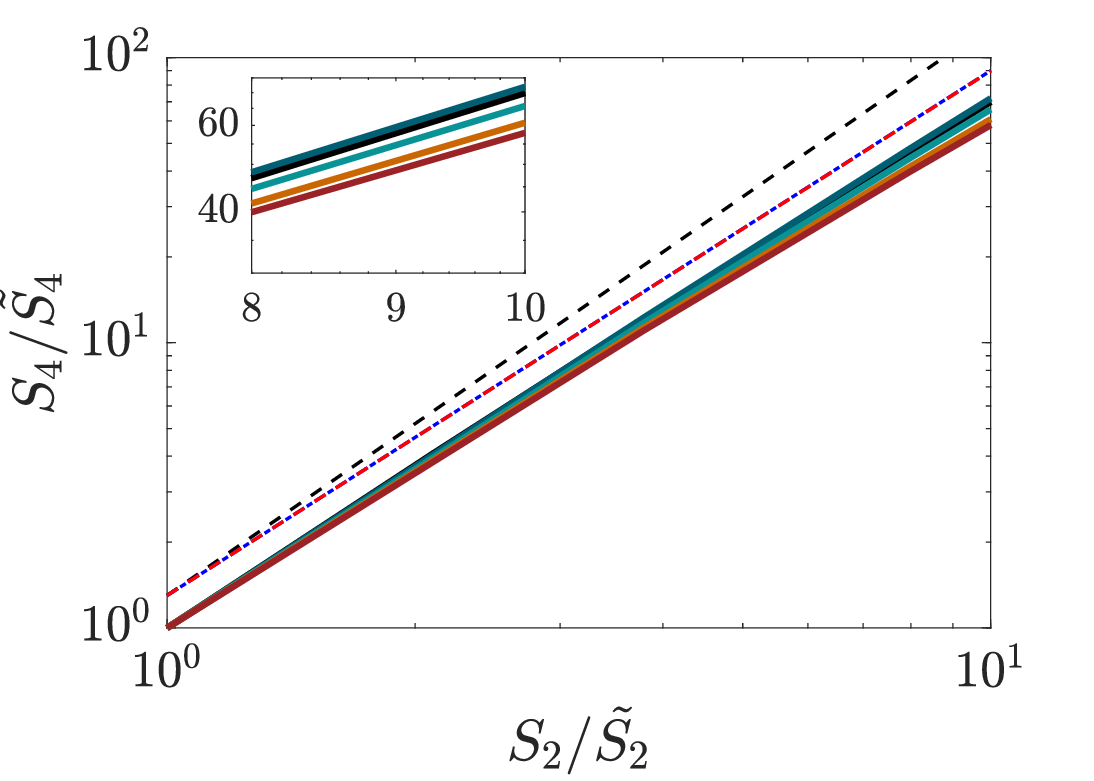}
\includegraphics[width=0.49\textwidth]{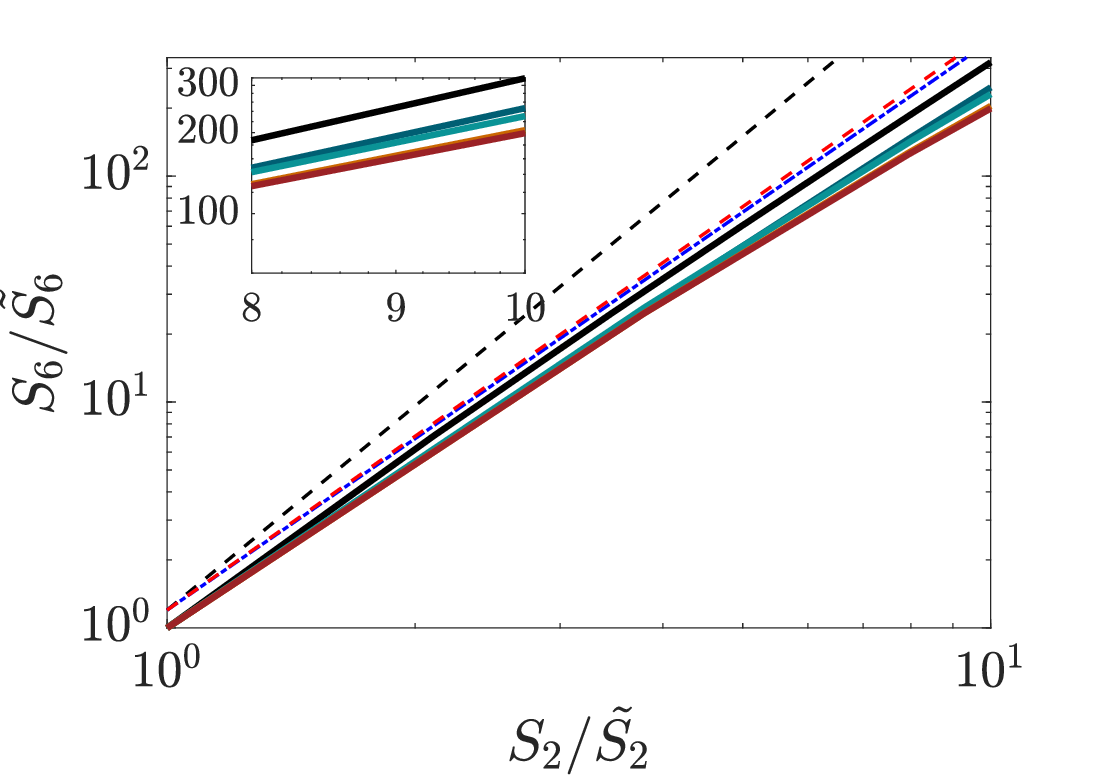}
\includegraphics[width=0.49\textwidth]{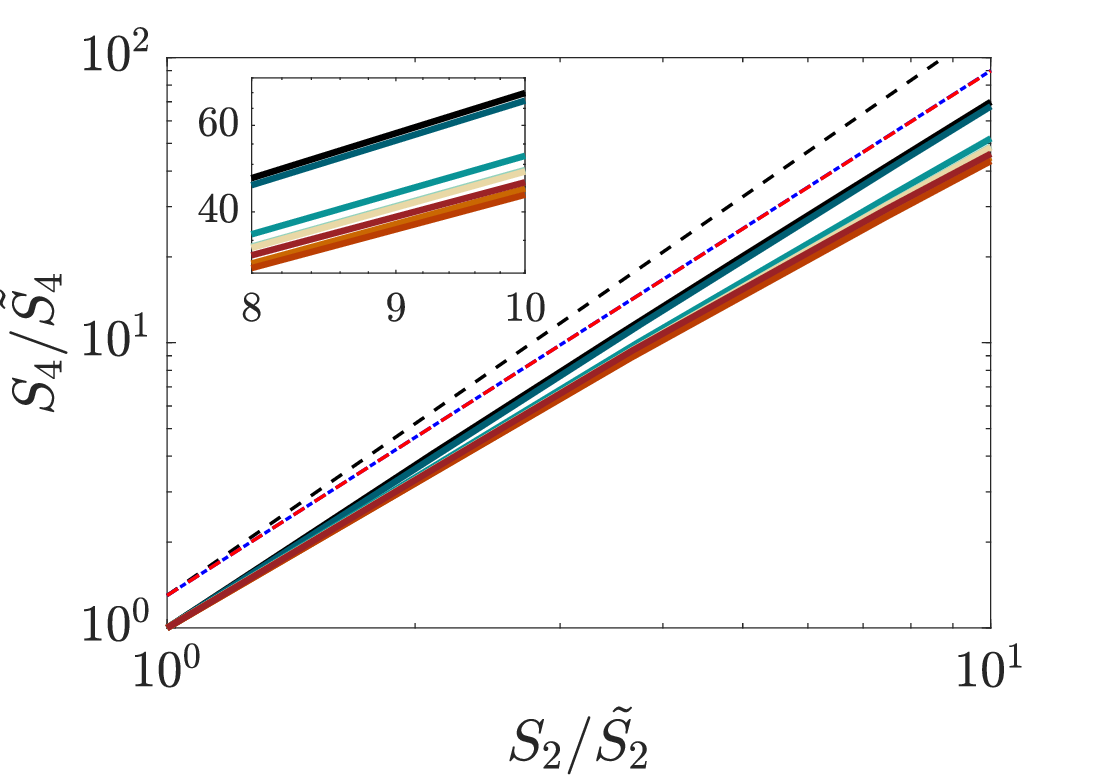}
\includegraphics[width=0.49\textwidth]{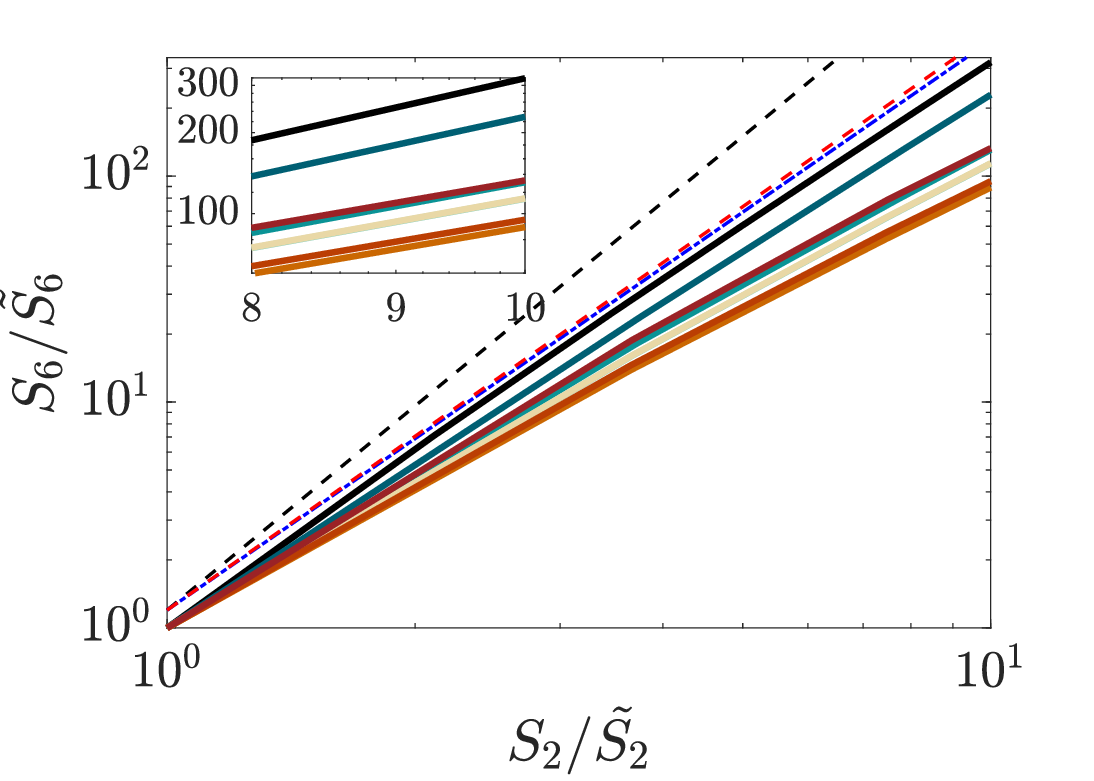}
\caption{Extended similarity for (top) $D/\eta=123$ and (bottom) $D/\eta=32$. Left: $S_4$ against $S_2$. Right: $S_6$ against $S_2$. The $\tilde{\cdot}$ is for $\tilde{S}_q=S_q(2\Delta x)$.}
\label{fig:ESS_M}
\end{figure}

\begin{figure}
\centering
\includegraphics[trim={0 180 0 0},clip,width=0.75\textwidth]{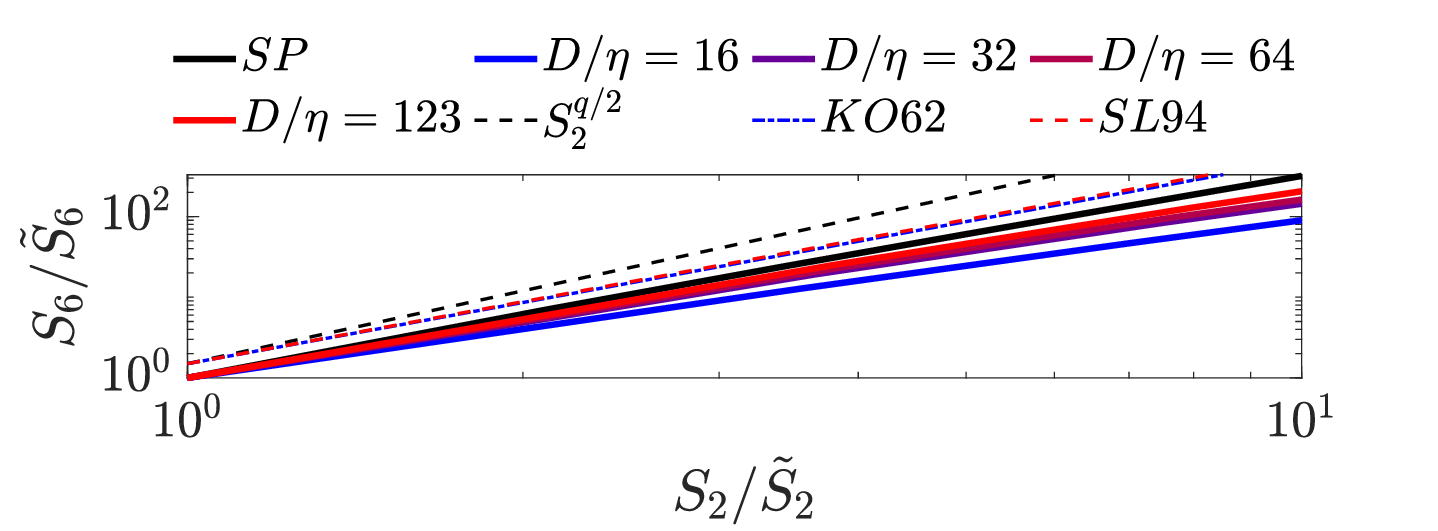}
\includegraphics[width=0.49\textwidth]{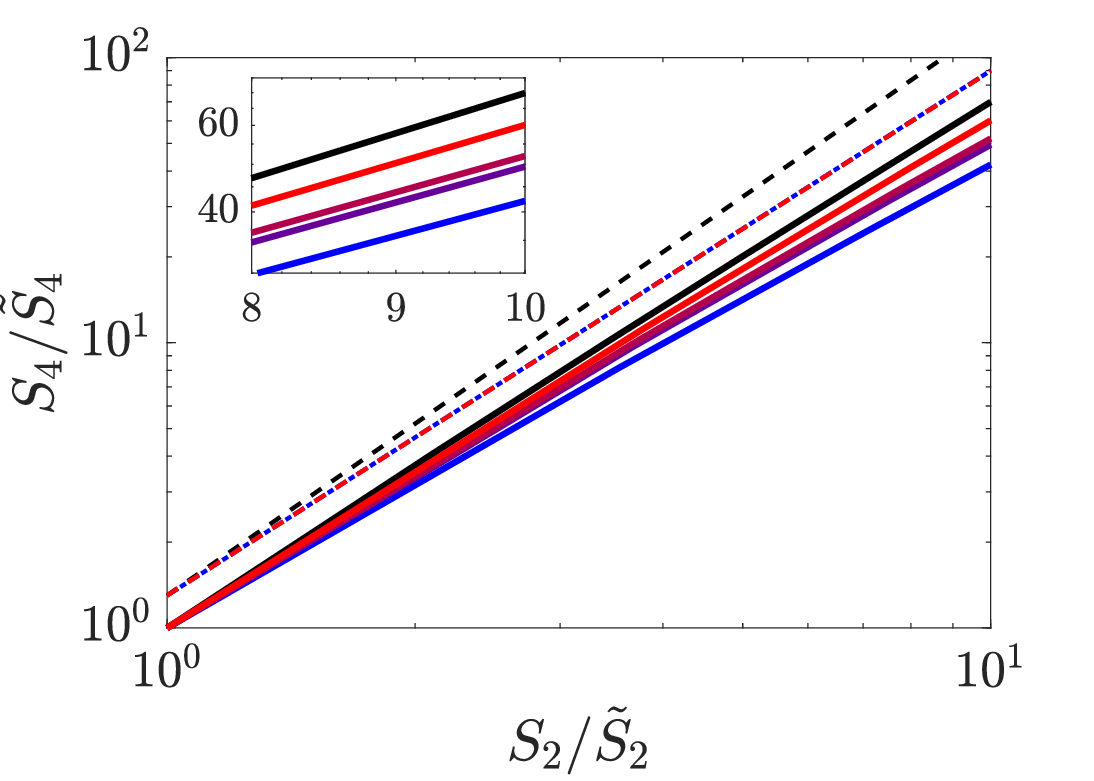}
\includegraphics[width=0.49\textwidth]{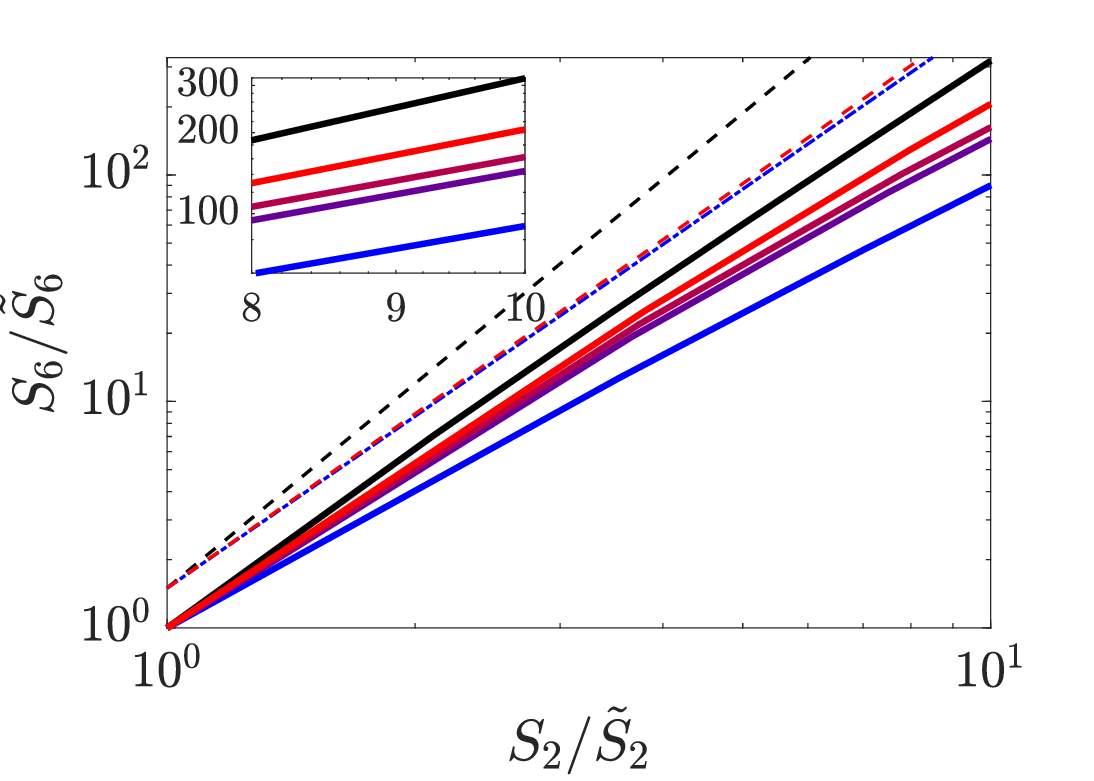}
\caption{Extended similarity for $M=0.9$. Left: $S_4$ against $S_2$. Right: $S_6$ against $S_2$. The $\tilde{\cdot}$ is for $\tilde{S}_q=S_q(2\Delta x)$.}
\label{fig:ESS_D}
\end{figure}

For the single-phase, the deviation of the high-order moment $S_p(r)$ from the Kolmogorov theory is commonly associated with the intermittency of the flow \citep{frisch-1995,pope-2000}, i.e. extreme events which are localised in space and time that break the Kolmogorov similarity hypothesis. In different words, as stated by \cite{biferale-2003}, the flow intermittency implies that the probability distribution function of the velocity fluctuations deviates from a Gaussian distribution and that this deviation increases by decreasing the scale. These extreme events correspond to tails in the velocity increment distributions and make vast contribution to the high-order moments. The larger deviation from the Kolmogorov behaviour observed for the particle-laden cases, suggests that the dispersed phase influences the flow intermittency as well. To estimate this, we use the extended self-similarity form \citep{benzi-1993}, and plot one structure function against the other. In figures \ref{fig:ESS_M} and \ref{fig:ESS_D}, we plot $S_q$ against $S_2$ for $q=4$ and $6$, and investigate the effect of $M$ and $D$ on the flow intermittency. 
In the limit case where extreme events do not occur, the $S_q \sim S_2^{q/2}$ power-law holds; the deviation from this behaviour is a measure of the flow intermittency. Before investigating the influence of the dispersed phase, it is worth noting that, as expected, $S_4$ and $S_6$ deviate from the theoretical prediction already in the single-phase case. However, at least for these values of $q$, the corrections proposed by \cite{kolmogorov-1962} and \cite{she-leveque-1994} provide a good prediction of the data. 

Moving to the effect of the dispersed phase, figure \ref{fig:ESS_M} shows that an increase of the mass fraction generally leads to a larger deviation from the $S_q \sim S_2^{q/2}$ scaling. For $D/\eta=123$ the deviation from the single-phase is monotonous: heavier particles lead to a stronger flow intermittency. For smaller $D/\eta$, instead, the deviation is not monotonic. The deviation from the single phase increases when the mass fraction is increased up to $M=0.6$, but decreases for further heavier particles. This happens because the increase of the intermittency is due to the no-slip and no-penetration boundary conditions at the surface of the particles, which lead to strong velocity gradients events and widen the tails of the velocity increment distribution. The level of intermittency is, therefore, expected to increase with the relative velocity between the fluid and the solid phase $|\Delta \bm{u}|$. As shown in the following, however, $|\Delta \bm{u}|$ does not show a monotonic dependence on $M$, but for the largest particles (see figure \ref{fig:deltau}). For $D/\eta=123$, it increases monotonically with $M$, while for $D/\eta=32$ it increases for mass fractions up to $M=0.6$, and decreases for larger $M$, in agreement with the trends shown in figure \ref{fig:ESS_M}.

Figure \ref{fig:ESS_D} shows the dependence of the flow intermittency on the particle size using $M=0.9$ as an example. When the particle size decreases the deviation from the $S_q \sim S_2^{q/2}$ scaling progressively increases, and the flow becomes more intermittent. Indeed, when the particle size decreases the number of particles increases (recall that $\Phi_V$ is constant), together with the total surface area of the solid phase. Therefore, the number of events associated with the particles boundary conditions increases, and the tails of the velocity increments distribution become more relevant. 


\section{Near-particle flow}
\label{sec:near-par}

In this section we focus on the influence of a single particle on the surrounding flow, since the characterisation of the fluid modulation around isolated (and non isolated) particles plays a relevant role for the development of accurate particle-fluid interaction models. The classical point-particle models, indeed, are based on the simplifying assumption of separation between all turbulence scales and the particle size, the absence of hydrodynamic interactions, and the existence of a simple parametrisation of the fluid-particle interaction that does not change with the particle size and density \citep{brandt-coletti-2022}. These models properly perform for very dilute suspensions of small particles ($\Phi_V \lessapprox 10^{-4}$ and $D/\eta \lessapprox 1$), but are unable to describe the complex fluid-solid interaction of non-dilute suspensions of finite-size particles \citep{hwang-eaton-2006}. In these conditions, particles modify the surrounding flow in a way that changes with their size ($D/\eta$) and density ($\rho_p/\rho_f$), and with the volume fraction of the suspension \citep{burton-etal-2005,tanaka-eaton-2010,botto-prosperetti-2012}. For small and light particles, for example, the particle-fluid relative velocity is weak, and the flow does not separate from their surface. In this case, the effect of the particles on the surrounding flow is limited in space in the vicinity of the surface of the particles, and large part of the energy drained by the particles from the fluid is dissipated within the boundary layer that develops over their surface. For larger and heavier particles, instead, the fluid-particle relative velocity increases, and the flow separates from the surface of the particles with vortices being possibly shed downstream in the wake. In this case, part of the energy drained by the particles is injected back in their wake, and is then dissipated farther away from the particle. In case of non-dilute suspensions, this increased kinetic energy can also interact with downstream particles, affecting their wake and modulating their shedding. All these effects are not captured by the classical point-particle models. Also, the flow around a particle moving in a turbulent flow shows a more complex dynamics than the flow past a sphere in free stream \citep[see for example][]{johnson-patel-1999,constantinescu-squires-2004,yun-etal-2006,rodriguez-etal-2011}. Indeed, different mechanisms are expected to be at play, depending on the inertia of the particles and on the intensity of the flow fluctuations \citep{mittal-2000,zeng-etal-2009}.

We investigate the near-particle flow modulation by means of a conditional average of the flow. For each particle, we define a local Cartesian reference system $(\xi,\eta,\zeta)$, which is, at each time step, centred with the particle and has the $\xi$ axis aligned with the relative fluid-particle velocity $\Delta \bm{u}$. We use spherical coordinates $(r,\phi,\theta)$, where $\xi= r \cos(\phi) \cos(\theta)$, $\eta = r \cos(\phi) \sin(\theta)$ and $\zeta = r \sin(\phi)$  ($r \in [0, \infty)$, $\theta \in [0, 2\pi)$, $\phi \in [-\pi/2, \pi/2]$), and define the conditional average of the generic quantity $a$, as
\begin{equation}
  \aver{a}_{cp}( r, \theta, \phi) = \lim_{T \rightarrow \infty} \frac{1}{T} \int_t^{t+T} \left( \frac{1}{N_p} \sum_{ip=1}^{N} a_{ip}( r, \phi,\theta,\tau) \right) \text{d} \tau.
\end{equation} 
Radial profiles away from the particle surface are then obtained by averaging over the surface of spherical shells centred with the particle. For the generic quantity $a$, we define the radial profile $\aver{a}_{cp,r}$ as
\begin{equation}
  \aver{a}_{cp,r}(r) = \frac{1}{ 4 \pi r^2} \int_0^{2 \pi} \int_{-\pi/2}^{\pi/2}  \aver{a}_{cp}(r,\theta,\phi) r^2 \sin(\phi) \text{d}\phi \text{d} \theta.
\end{equation}

In this section we present the results in terms of $\rho_p/\rho_f$, instead of $M$. Note however, that the two quantities are interchangeable as $\Phi_V$ is maintained constant.

\subsection{Relative velocity and particle Reynolds number}
\label{sec:Rep}

\begin{figure}
\centering
\includegraphics[width=0.7\textwidth]{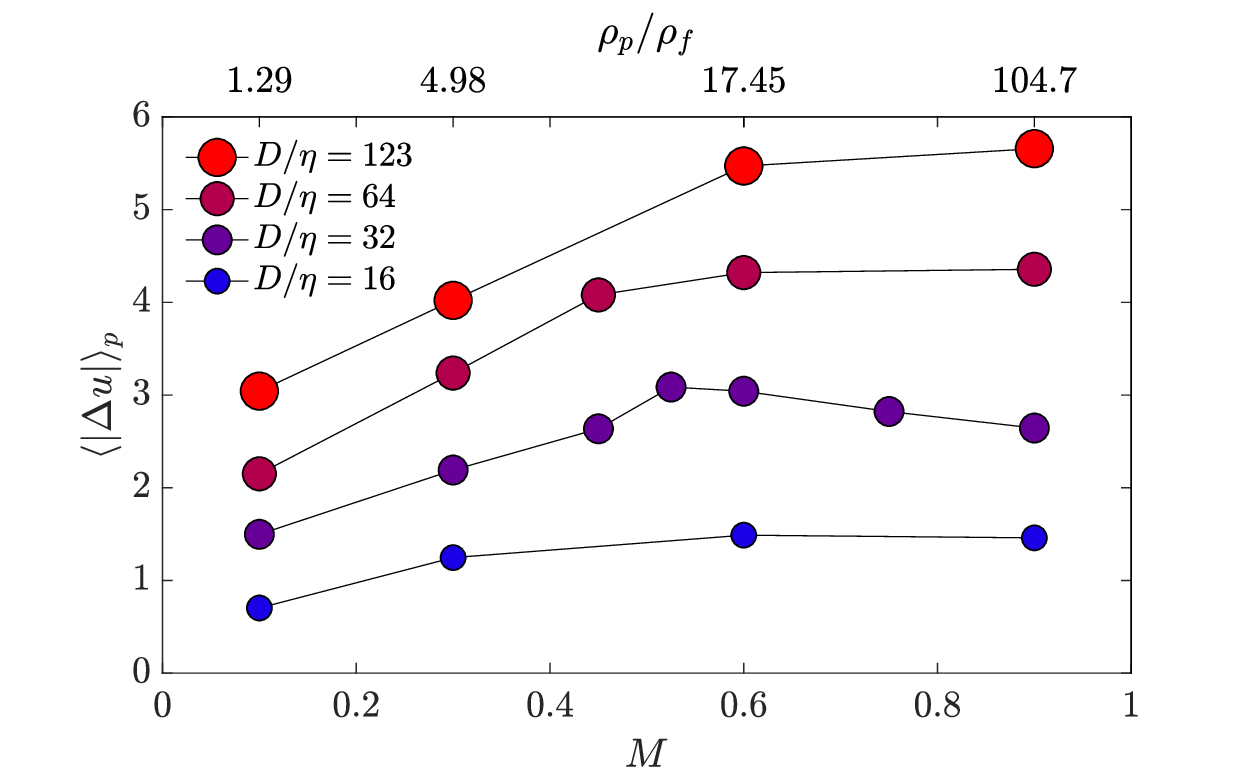}
\caption{Fluid-particle average velocity difference as a function of the mass fraction (or particle density) and of the particle size.}
\label{fig:deltau}
\end{figure}
The modulation of the flow around the particles depends on the relative motion between the fluid and the solid phase, which, indeed, is a key quantity of interest for modelling the motion of the particles. We define the velocity difference between the $i_{th}$ particle and the fluid as $\Delta \bm{u}^i = \bm{u}_p^i - \bm{u}_f^i$, where $\bm{u}_p^i$ is the velocity of the $i_{th}$ particle, and $\bm{u}_f^i$ is the fluid velocity seen by the particle. Different approaches have been proposed to estimate $\bm{u}_f^i$, see for example \cite{bagchi-balachandar-2003,lucci-etal-2010, kidanemariam-etal-2013, uhlmann-doychev-2014,cisse-etal-2015,oka-goto-2022}. We follow the works by \cite{kidanemariam-etal-2013,uhlmann-chouippe-2017,oka-goto-2022}, and evaluate $\bm{u}_f^i$ as the average of the fluid velocity within a shell centred with the particle at $\bm{x}_p^i$, and delimited by $\bm{x}_p^i + R$ and $\bm{x}_p^i + R_{sh}$. The external $R_{sh}$ radius should be large enough to avoid the average fluid velocity to equal the particle velocity as the no-slip boundary is approached, and small enough to avoid considering portions of the fluid that are not relevant for the particle motion. Here we chose $R_{sh} = 3 R$ as in \cite{uhlmann-chouippe-2017}, but we have verified that the results qualitatively do not change when considering $ 2 R \le R_{sh} \le 3 R$.

Figure \ref{fig:deltau} shows the dependence of $\aver{|\Delta \bm{u}|}_p$ on $\rho_p/\rho_f$ and $D$; here $\aver{\cdot}_p$ indicates average over particles and in time. For all densities, the fluid-solid relative velocity increases with the particle size, in agreement with the increase of the particle inertia, and the fact that particles become less able to follow the fluid. Generally, $\aver{|\Delta \bm{u}|}_p$ increases also with the density of the particle when fixing $D$. Note, however, that for $D/\eta \le 64$, $\aver{|\Delta \bm{u}|}_p$ decreases for $\rho_p/\rho_f \gtrapprox 17$ ($M \gtrapprox 0.6$), consistently with the results shown in section \ref{sec:inter}.

\begin{figure}
\centering
\includegraphics[trim={0 10 0 0},clip,width=0.8\textwidth]{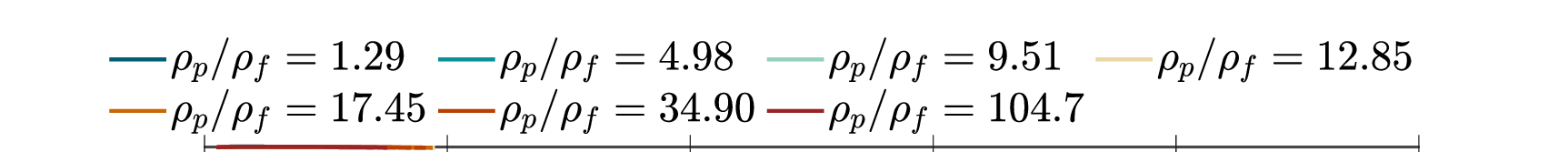}
\includegraphics[width=0.49\textwidth]{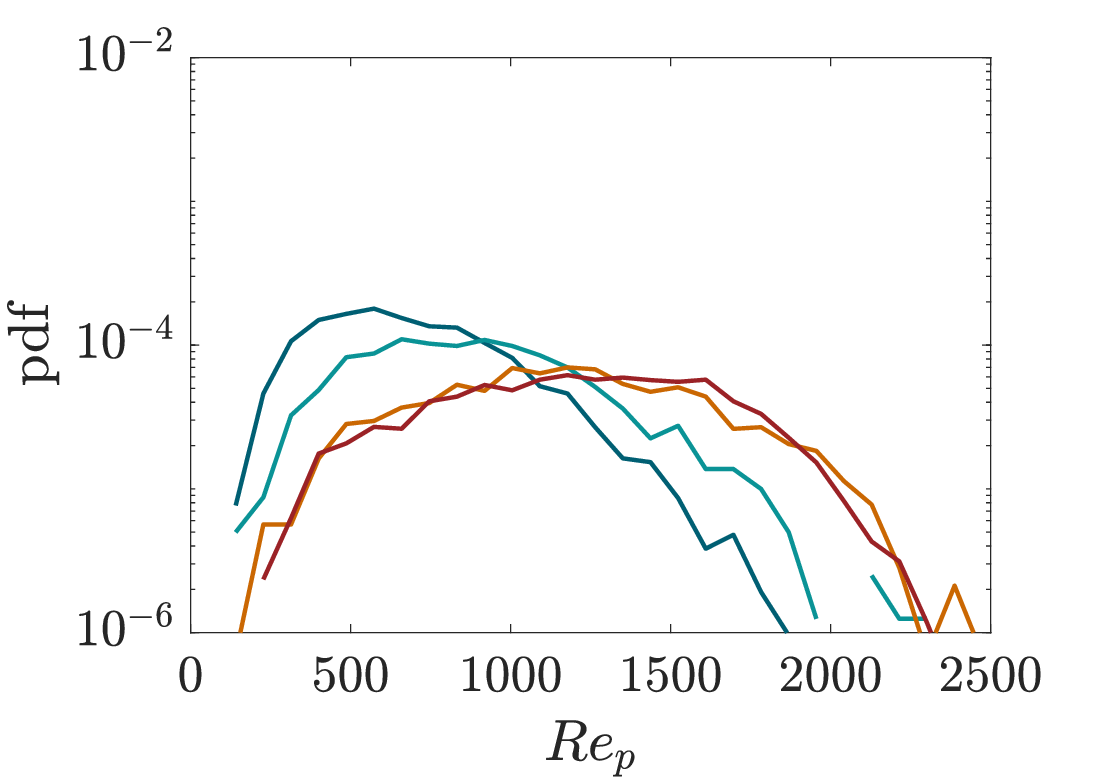}
\includegraphics[width=0.49\textwidth]{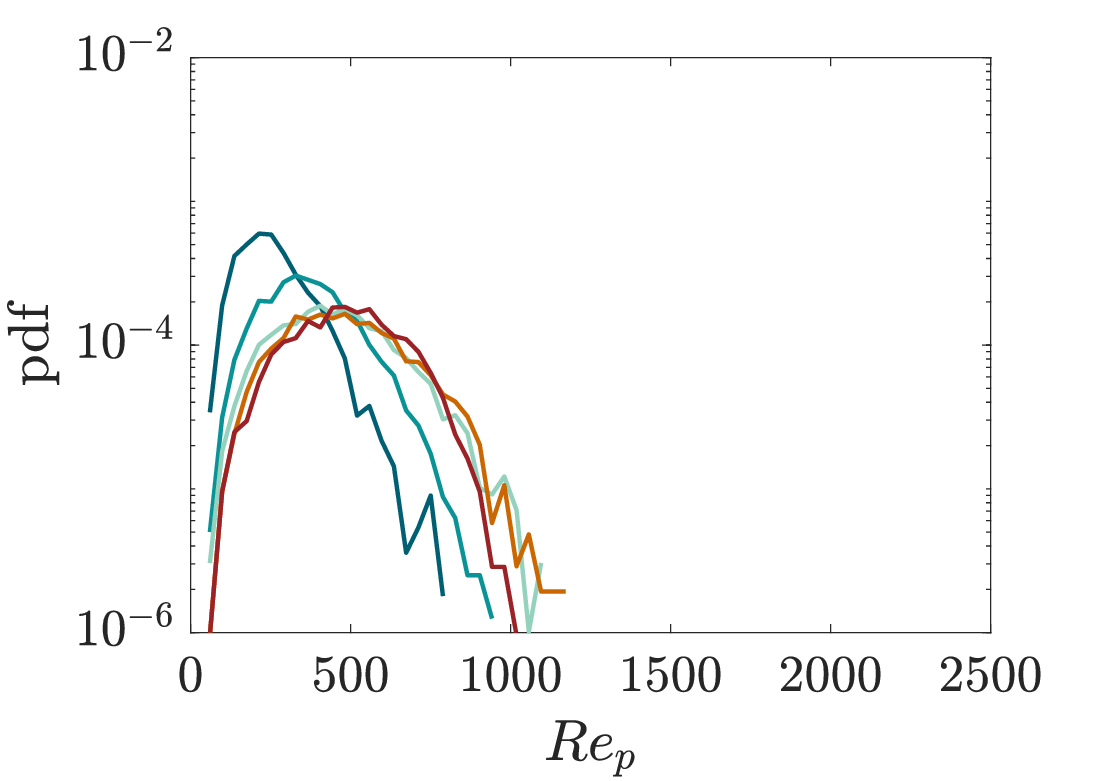}
\includegraphics[width=0.49\textwidth]{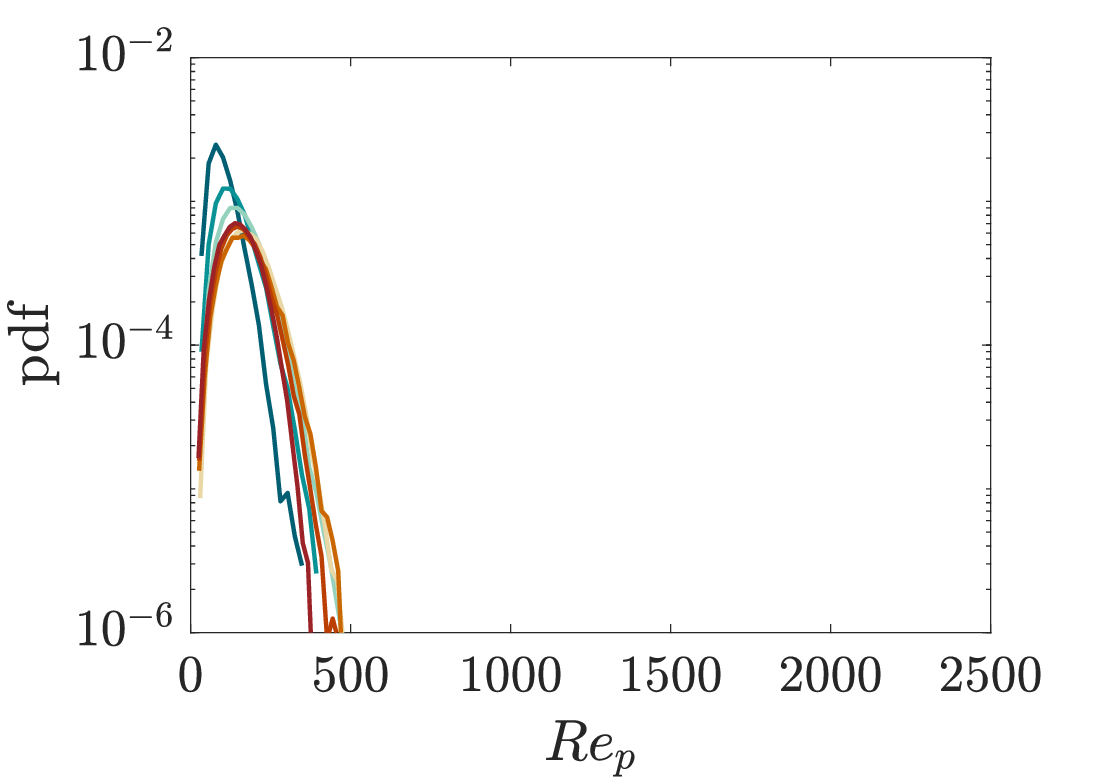}
\includegraphics[width=0.49\textwidth]{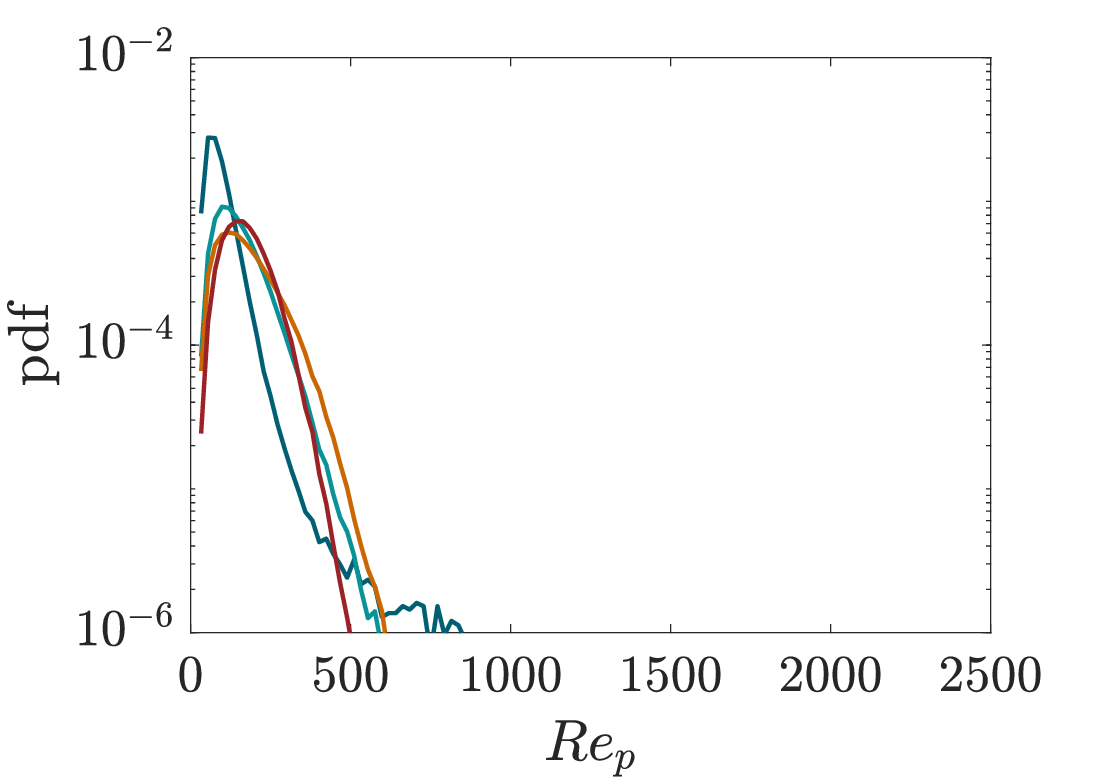}
\caption{Probability density function of the particle Reynolds number $Re_p = \aver{|\Delta \bm{u}|}_p D / \nu$ for $1.29 \le \rho_p/\rho_f \le 104.7$. Top left: $D/\eta=123$. Top right: $D/\eta=64$. Bottom left: $D/\eta=32$. Bottom right:$D/\eta=16$.}
\label{fig:Rep_pdf}
\end{figure}
We define the particle Reynolds number as $Re_p = \aver{|\Delta \bm{u}|}_p D / \nu$. Figure \ref{fig:Rep_pdf} shows the probability density functions of $Re_p$. Similarly to what shown by \cite{uhlmann-chouippe-2017}, for small $\rho_p/\rho_f$ the probability density function of $Re_p$ is skewed since the components of $\aver{|\Delta \bm{u}|}_p$ have large tails. When $\rho_p/\rho_f$ increases, however, the probability density function becomes less skewed for all particle sizes. Heavier particles, indeed, are less able to follow the flow and thus the fluid-particle relative velocity increases, and the relevance of the tails of $\aver{|\Delta \bm{u}|}_p$ decreases. Overall, the range of $Re_p$ largely varies with the particle size and density, suggesting that in the considered range of parameters, the flow regime changes. Note, moreover, that for all cases considered, large values of $Re_p$, that are sufficient for the formation of vortical structures in the vicinity of the particles, are instantaneously attained (see the following discussion). This suggests that a quasi-steady linear drag assumption, commonly adopted in point-particle models, is not sufficient for modelling suspensions at the present conditions.

\subsection{Radial profiles}

We start investigating the average radial profiles of the kinetic energy and dissipation rate. The aim is to quantify the attenuation of the energy fluctuations, and the enhancement of the dissipation rate at the particle surface for different values of $D$ and $\rho_p/\rho_f$.

\begin{figure}
\centering
\includegraphics[trim={0 10 0 0},clip,width=0.8\textwidth]{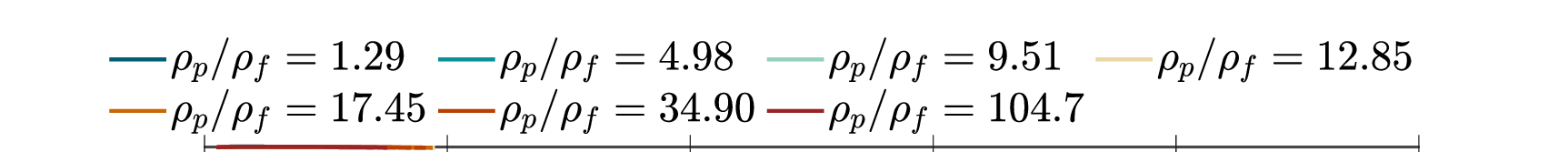}
\includegraphics[width=0.49\textwidth]{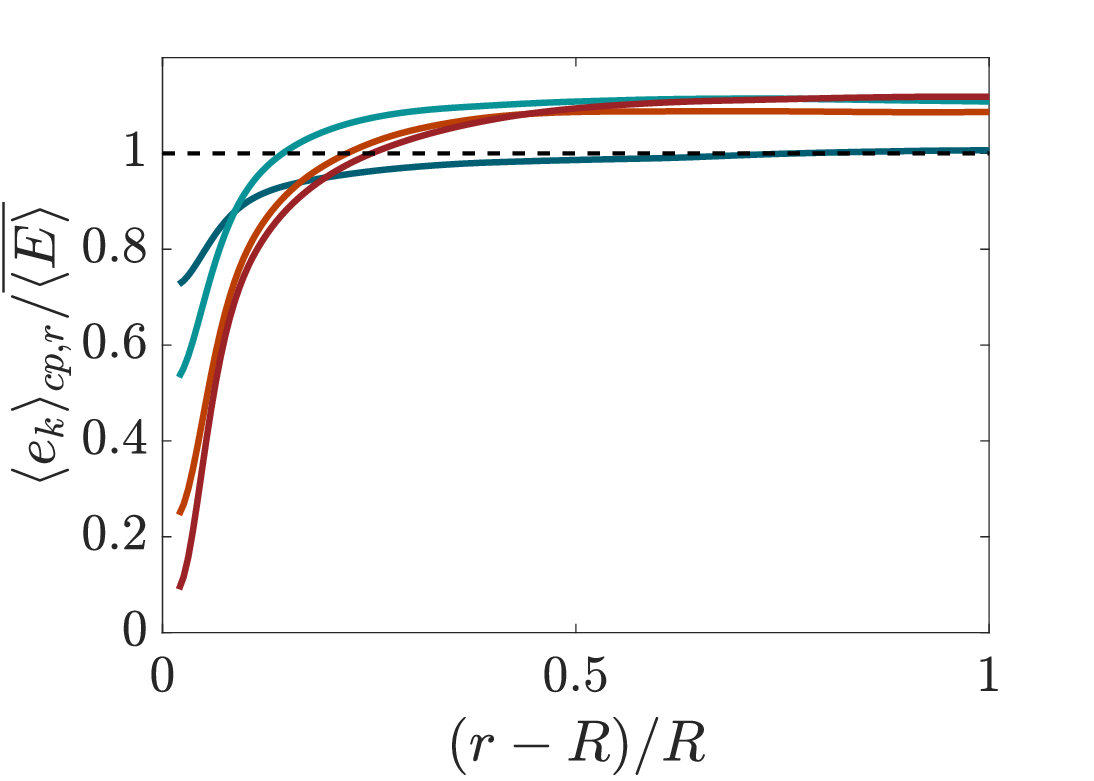}
\includegraphics[width=0.49\textwidth]{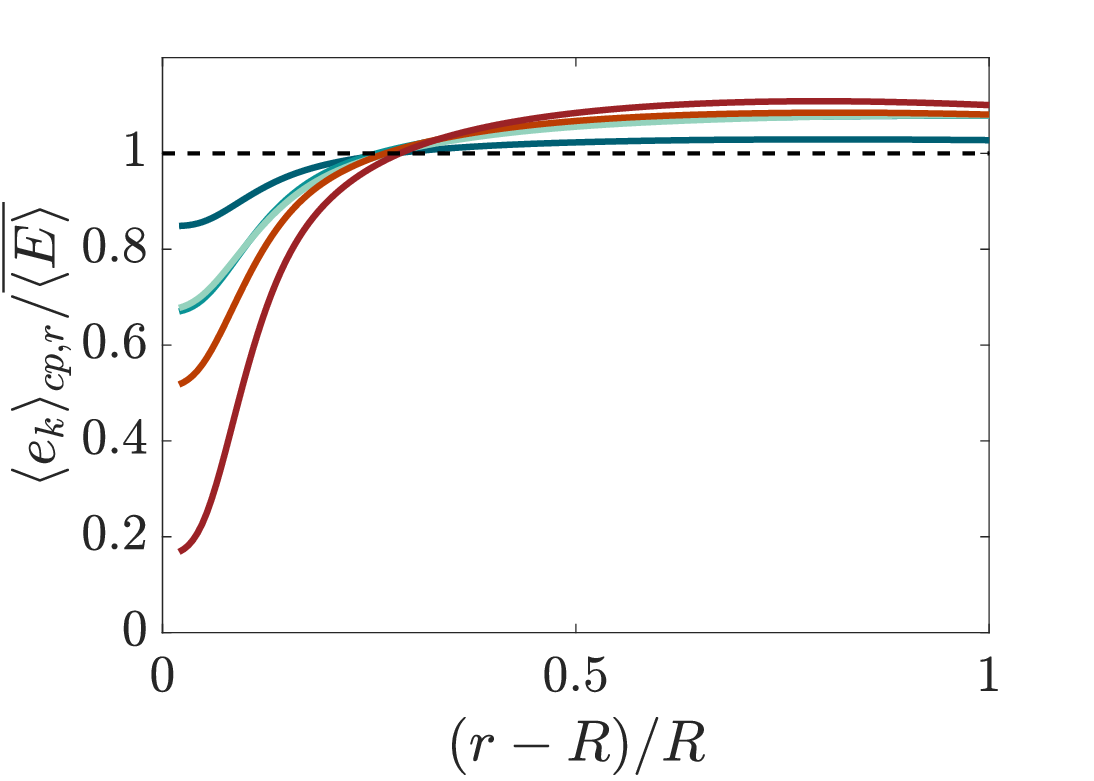}
\includegraphics[width=0.49\textwidth]{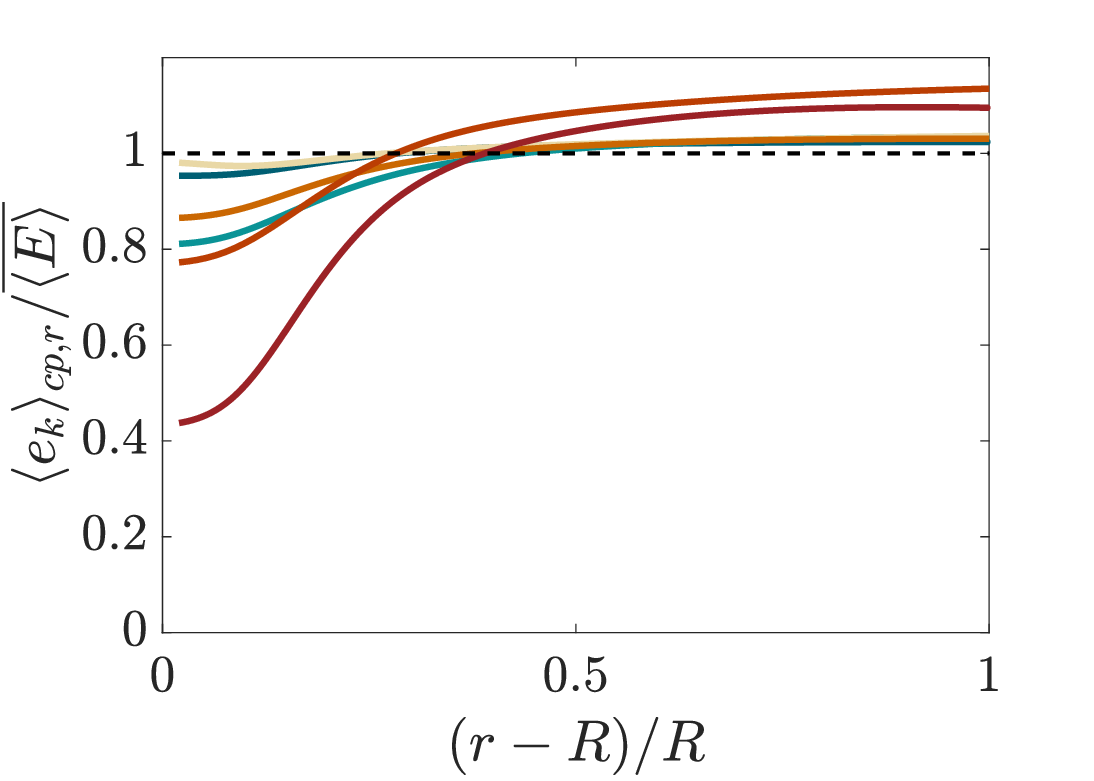}
\includegraphics[width=0.49\textwidth]{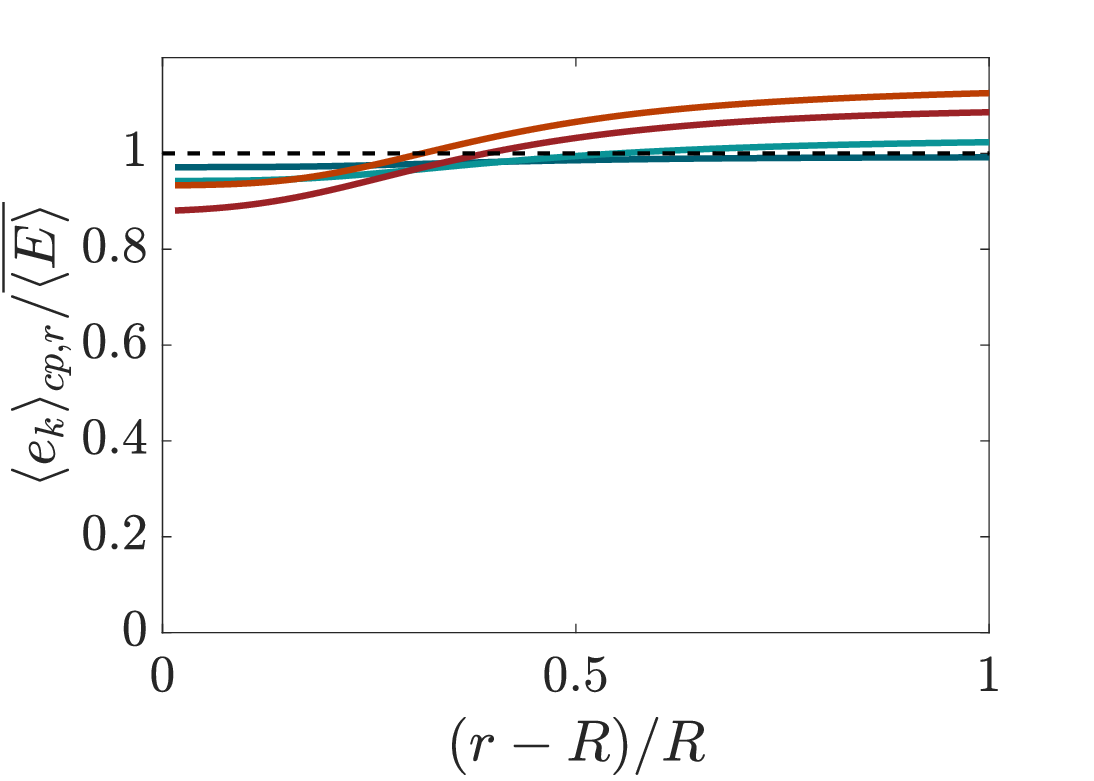}
\caption{Energy distribution of the flow energy as a function of the distance from the particle surface. Top left: $D/\eta=123$. Top right: $D/\eta=64$. Bottom left: $D/\eta=32$. Bottom right: $D/\eta=16$.}
\label{fig:parLocEner}
\end{figure}

Figure \ref{fig:parLocEner} plots the average radial flow energy distribution away from the particle surface $\aver{e_k}_{cp,r}$.
The turbulent kinetic energy is strongly attenuated in the neighbourhood of the particles. The region of strong energy reduction extends out to less than $0.5$ the particle radius, with a size that widens as $D$ decreases, being approximately $0.2R$, $0.25R$, $0.4R$ and $0.5R$ for $D/\eta =123$, $D/\eta=64$, $D/\eta=32$ and $D/\eta=16$; see the regions where $\aver{e_k}_{cp,r}/\overline{\aver{E}}<1$, which can be regarded as the flow region mostly influenced by the particles. This local attenuation of the kinetic energy is due to the particle-fluid density difference, and scales with the average relative fluid-particle velocity $\aver{|\Delta \bm{u}|}_p$. Particles with larger inertia respond only to the turbulent eddies with large time- and length-scales, and act as obstacles for fluctuations with smaller scales, attenuating therefore more effectively the fluid fluctuations. The local attenuation compared to the box average value, indeed, increases with $\rho_p/\rho_f$ and $D$, up to approximately $80\%$ for $\rho_p/\rho_f = 104.7$ and $D/\eta=123$. Note, however, that for $D/\eta = 32$ the near-particle energy attenuation does not have a monotonic dependence on $\rho_p/\rho_f$ ($M$), as it decreases when the flow moves to the anisotropic and more energetic state (regime B) discussed in section \ref{sec:flow}. The energy attenuation increases with the density of the particles when $1.29 \le \rho_p/\rho_f \le 4.98$ ($0.1 \le M \le 0.3$) and $9.51 \le \rho_p/\rho_f \le 104.7$ ($0.45 \le M \le 0.9$), but it decreases when increasing $\rho_p/\rho_f$ from $\rho_p/\rho_f =4.98$ ($M=0.3$) to $\rho_p/\rho_f = 9.51$ ($M=0.45$), i.e. when moving from regime A to regime B. It is worth stressing that, while single particles with larger $D$ are more effective in reducing the turbulent fluctuations of the surrounding flow, when fixing the volume fraction of the suspension, smaller particles are more effective in reducing the global turbulent energy (see figure \ref{fig:energyM}), due to the increase of the number of particles.

\begin{figure}
\centering
\includegraphics[trim={0 10 0 0},clip,width=0.8\textwidth]{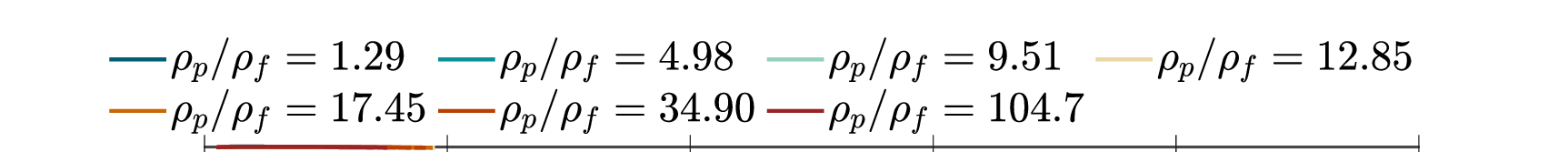}
\includegraphics[width=0.49\textwidth]{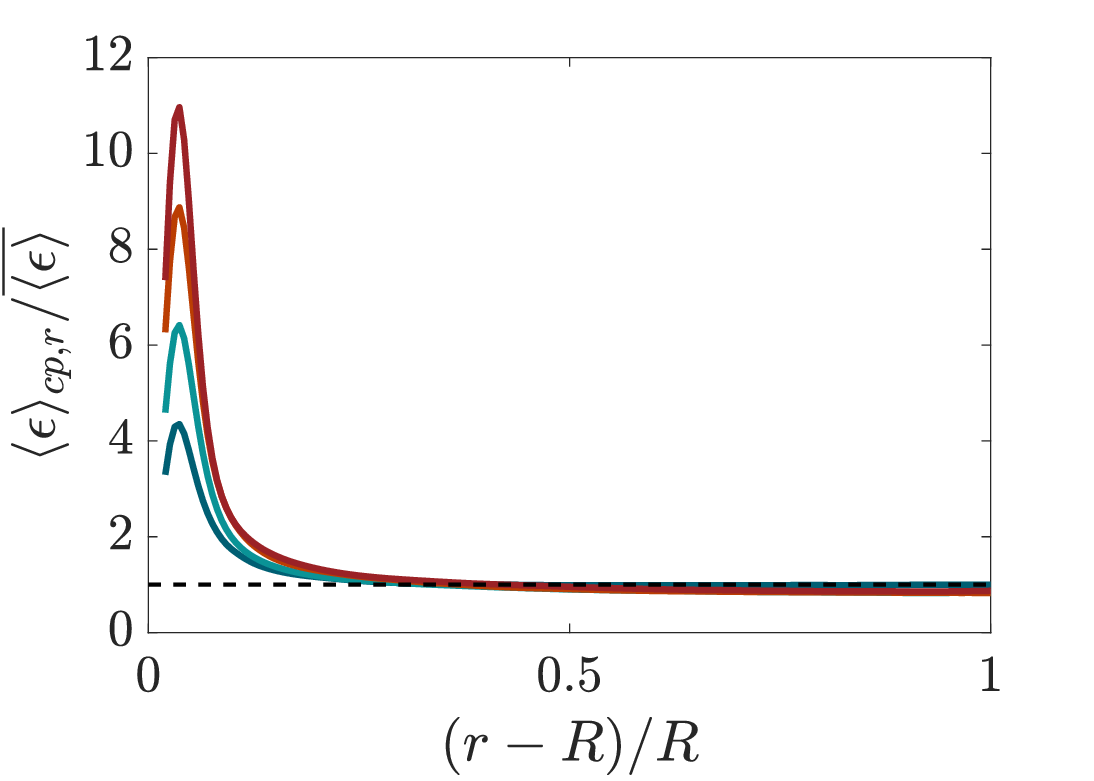}
\includegraphics[width=0.49\textwidth]{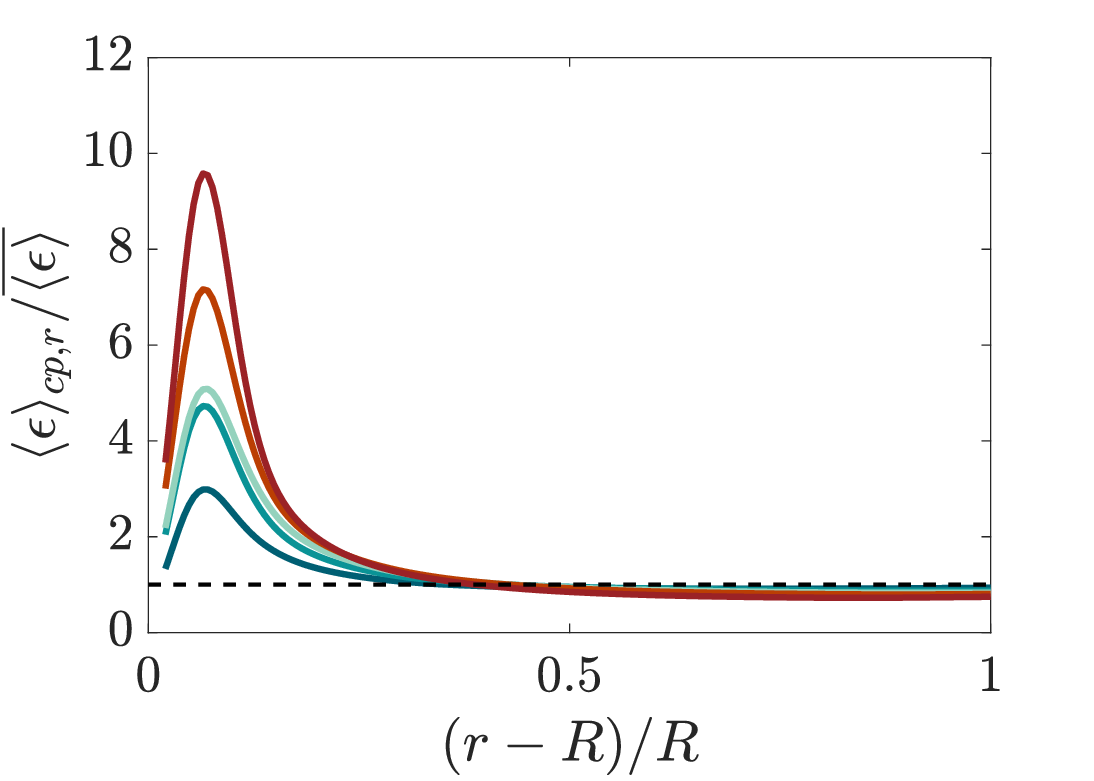}
\includegraphics[width=0.49\textwidth]{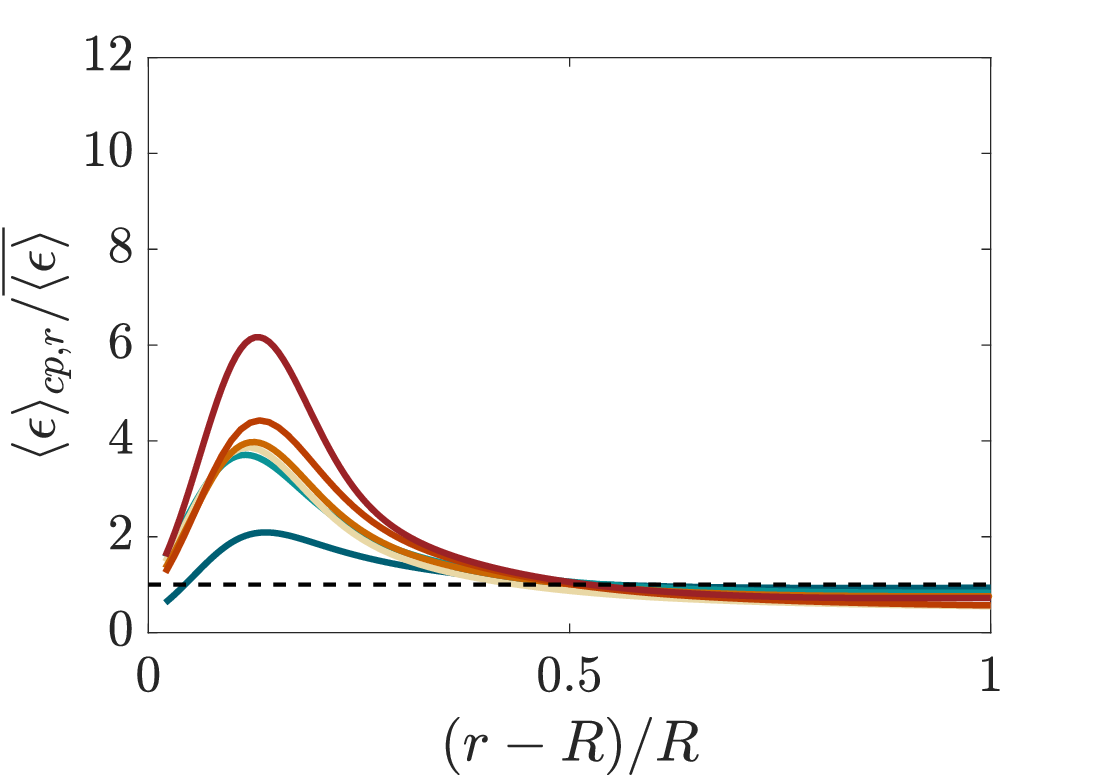}
\includegraphics[width=0.49\textwidth]{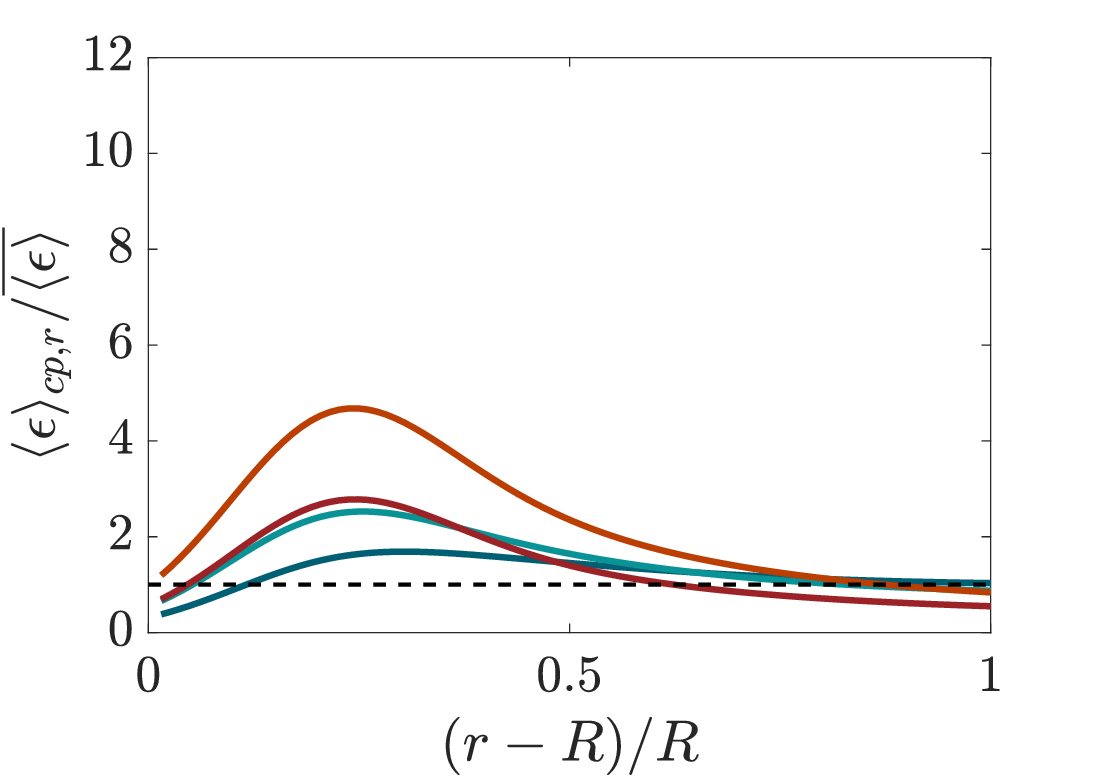}
\caption{Distribution of the dissipation rate as a function of the distance from the particle surface. Top left: $D/\eta=123$. Top right: $D/\eta=64$. Bottom left: $D/\eta=32$. Bottom right: $D/\eta=16$.}
\label{fig:parLocDiss}
\end{figure}

Figure \ref{fig:parLocDiss} shows the radial dissipation rate profile away from the particles surface.
Due to the large velocity gradients that develop within the boundary layer, a region of high dissipation rate arises in the neighbourhood of each particle. Part of the turbulent kinetic energy driving the relative motion between the fluid and the particle is dissipated here, giving rise to an unique cross-scale energy transfer from scales larger than $D$ to the particle-surface boundary scales. The intensity of the near-particle dissipation rate changes with the particle size and density, similarly to what observed by \cite{shen-etal-2022}. $\aver{\epsilon}_{cp,r}/\overline{\aver{\epsilon}}$ increases with $\rho_p/\rho_f$ and $D$ up to approximately $11\%$ for $D/\eta=123$ and $\rho_p/\rho_f = 104.7$. This is consistent with the increase of the the fluid-particle relative velocity with $\rho_p/\rho_f$ and $D$, that leads to more intense velocity gradients at the particle surface. The width of the $\aver{\epsilon}_{cp,r}/\overline{\aver{\epsilon}}>1$ region is a measure of the particle boundary layer thickness \citep{johnson-patel-1999}, and its dependence on $D$ and $\rho_p/\rho_f$ is relevant for modelling implications. The $\aver{\epsilon}_{cp,r}/\overline{\aver{\epsilon}}>1$ regions enlarges when the particle size and particle Reynolds number increase, while it only marginally changes with the density of the particles. This trend recalls what observed for the cutoff scale $2 \pi/\kappa_{p,2}$ in section \ref{sec:spectra}, suggesting a relation between these two quantities. Note, however, the wavenumber corresponding to the $\aver{\epsilon}_{cp,r}/\overline{\aver{\epsilon}}>1$ region decreases with the particle size, while $\kappa_{p,2}$ increases.

\subsection{Near-particle flow}

For a more detailed characterisation of the near-particle flow modulation, we now account for the direction of the relative particle-fluid velocity, and investigate the conditionally averaged flow field without averaging along $\theta$ and $\phi$. When choosing the local reference system of the $i_{th}$ particle, here we evaluate the fluid velocity seen by the particle $\bm{u}_f^{i}$ as the average fluid velocity within a shell centred with the particle and delimited by $R \le r \le R_{sh}$, where $R_{sh}$ corresponds to the distance from the particle centre at which $\aver{\epsilon}_{cp,r}/\overline{\aver{\epsilon}} = 1$ (see figure \ref{fig:parLocDiss}). We show the $D/\eta=32$ cases as an example; the results for the other particle sizes are similar and are omitted for the sake of conciseness.

\begin{figure}
\centering
\includegraphics[trim={0 240, 0, 0},clip,width=0.7\textwidth]{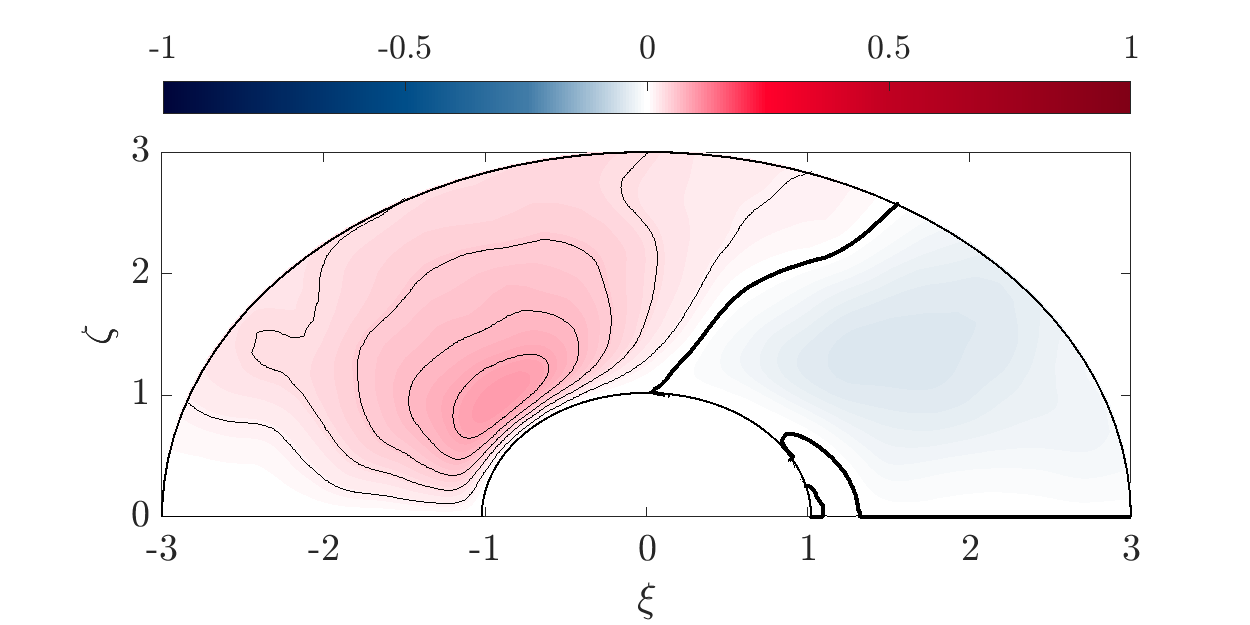} \\
\includegraphics[width=0.49\textwidth]{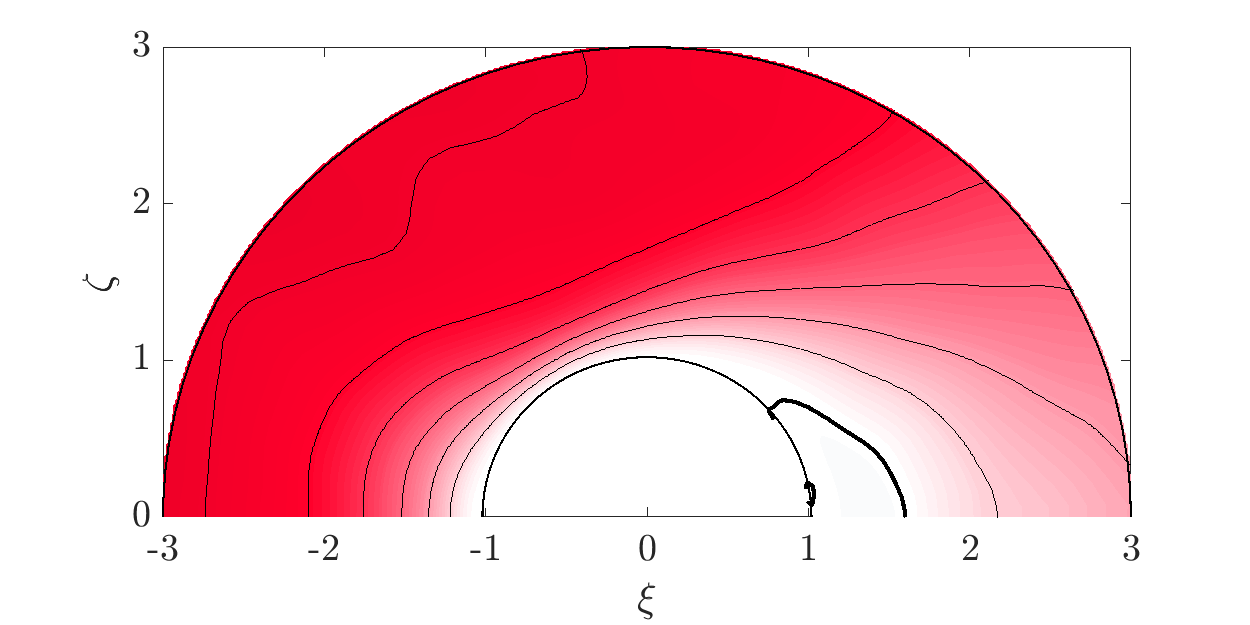}
\includegraphics[width=0.49\textwidth]{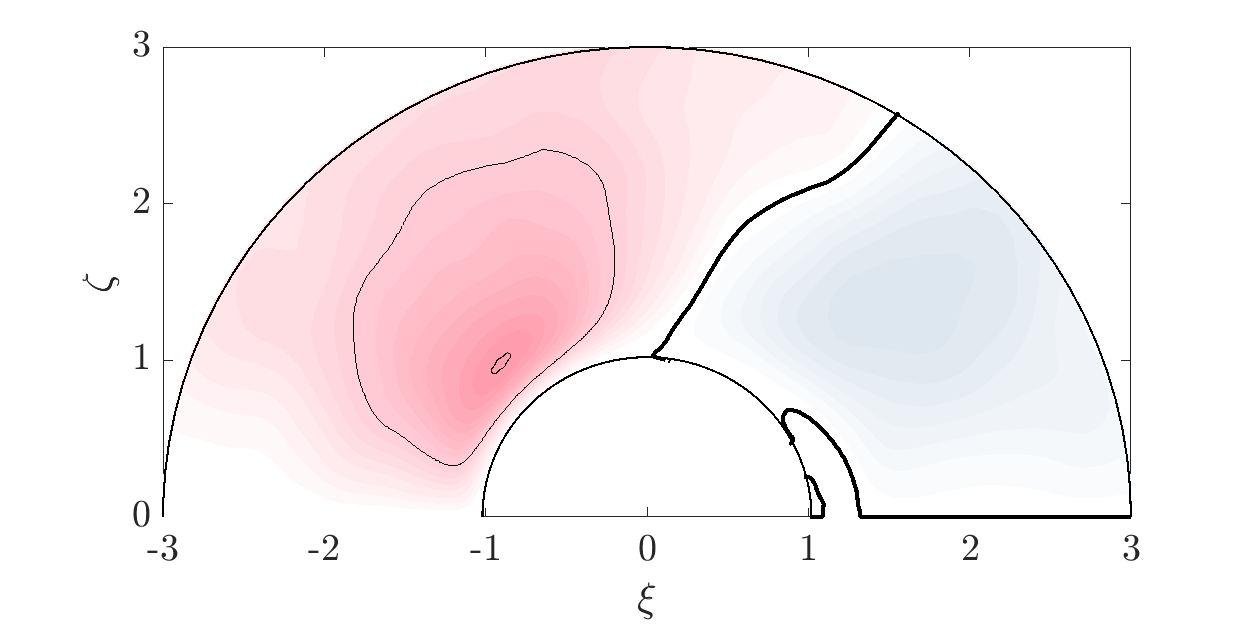}
\includegraphics[width=0.49\textwidth]{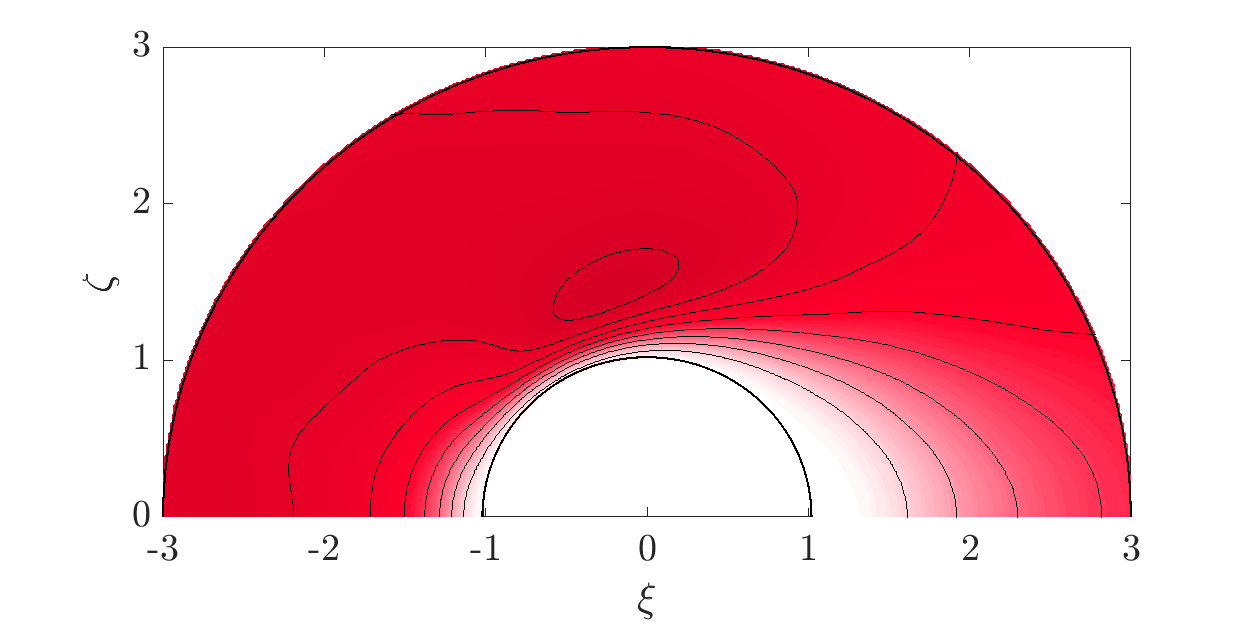}
\includegraphics[width=0.49\textwidth]{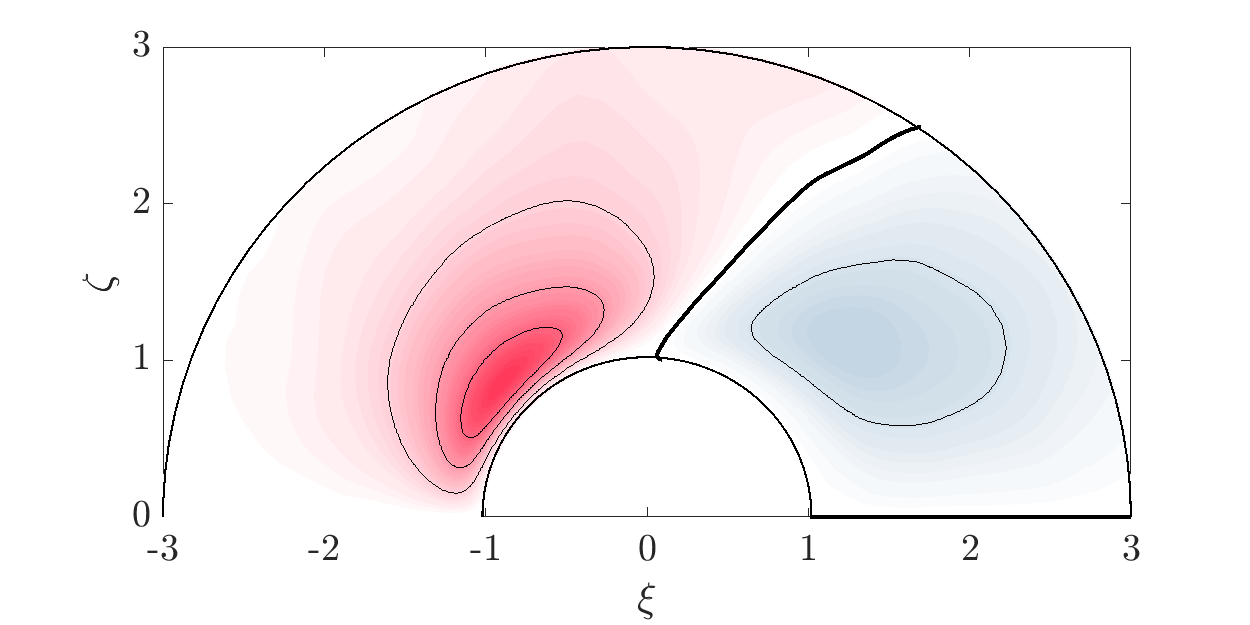}
\includegraphics[width=0.49\textwidth]{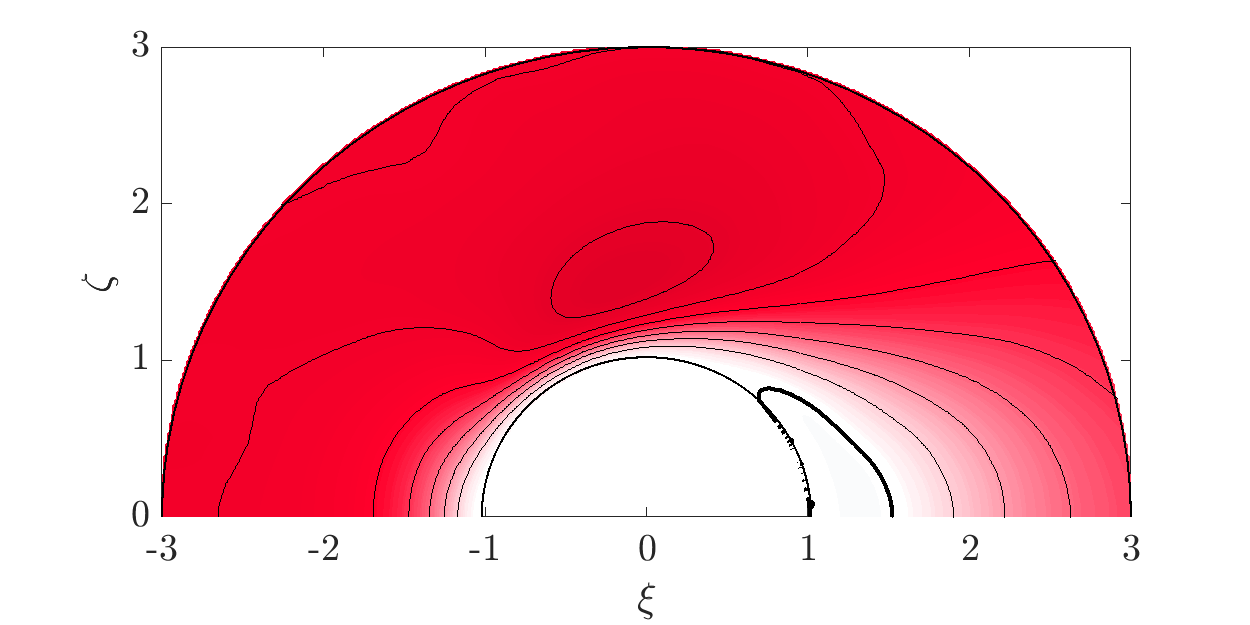}
\includegraphics[width=0.49\textwidth]{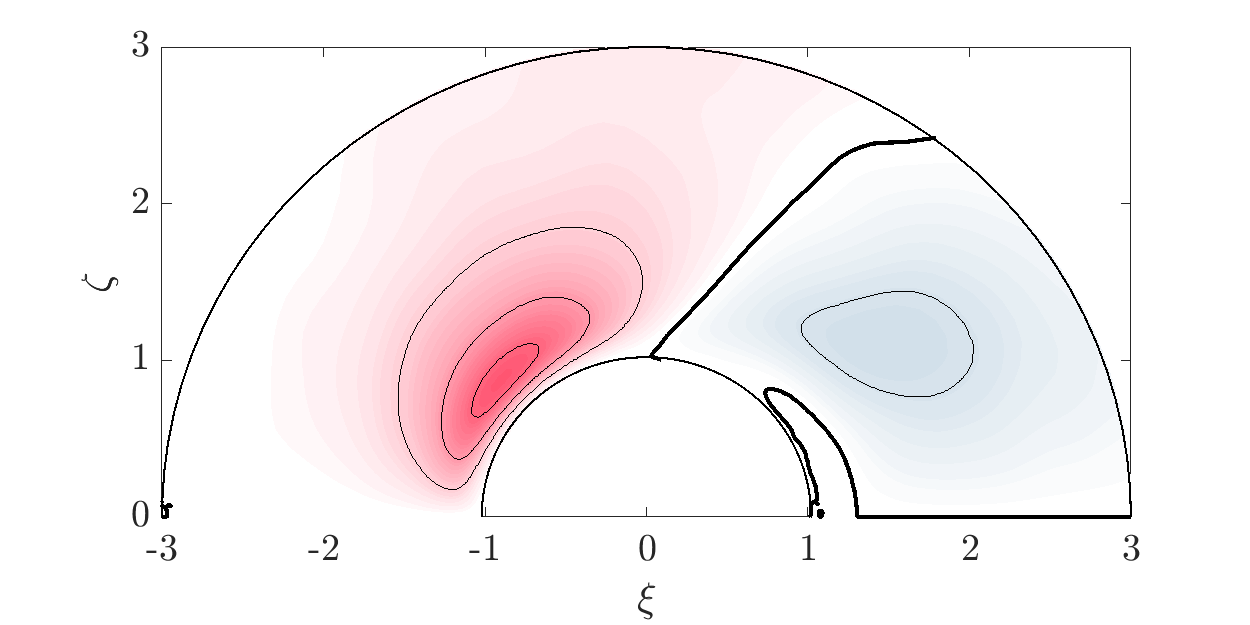}
\includegraphics[width=0.49\textwidth]{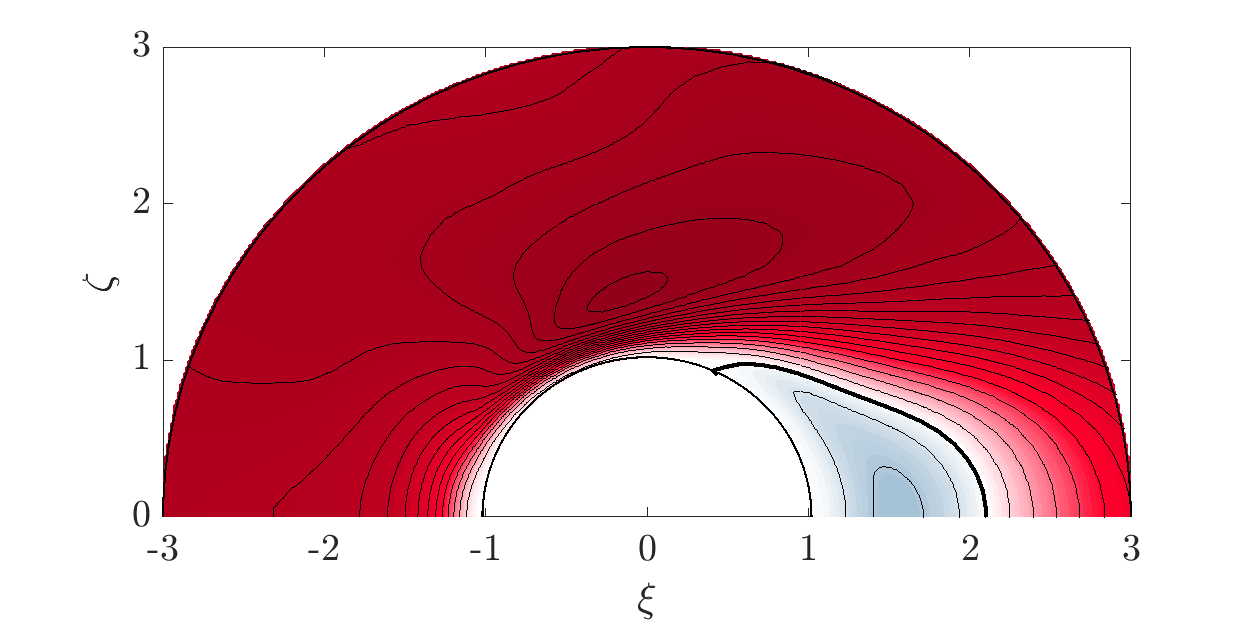}
\includegraphics[width=0.49\textwidth]{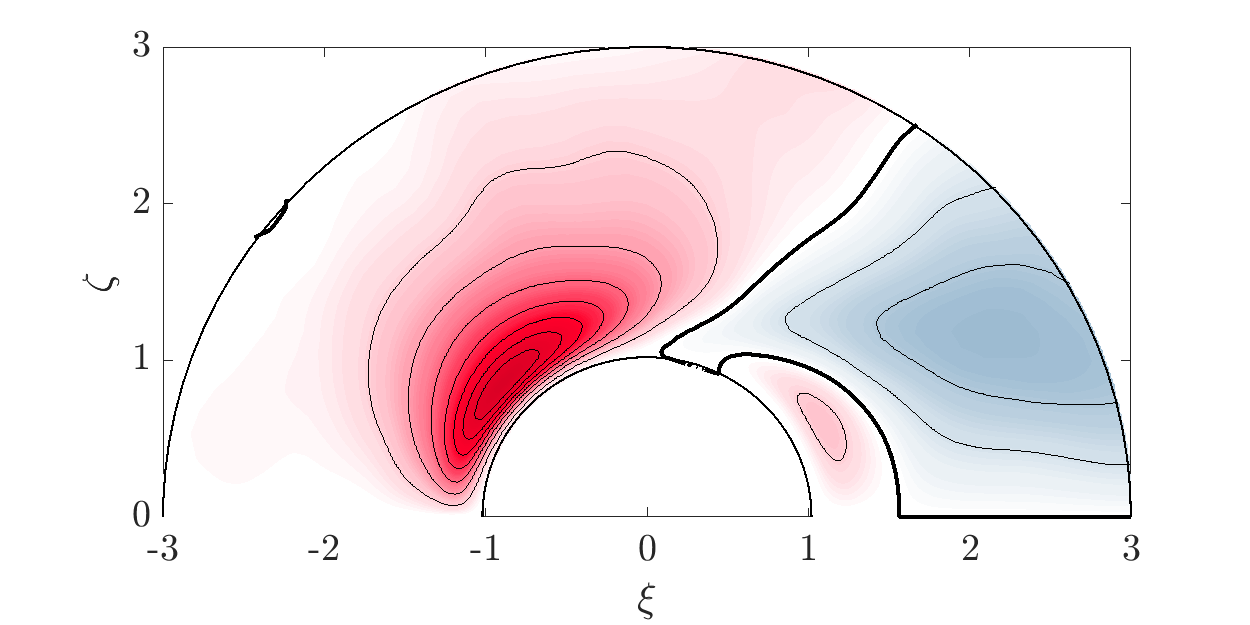}
\caption{Conditionally averaged velocity around a particle for $D/\eta=32$. Left: $\aver{u}_{cp,\ell}$. Right: $\aver{w}_{cp,\ell}$. From top to bottom: $\rho_p/\rho_f=1.29,4.98,17.45$ and $104.7$}
\label{fig:ParLocVel}
\end{figure}

\begin{figure}
\centering
\includegraphics[trim={0 240, 0, 0},clip,width=0.7\textwidth]{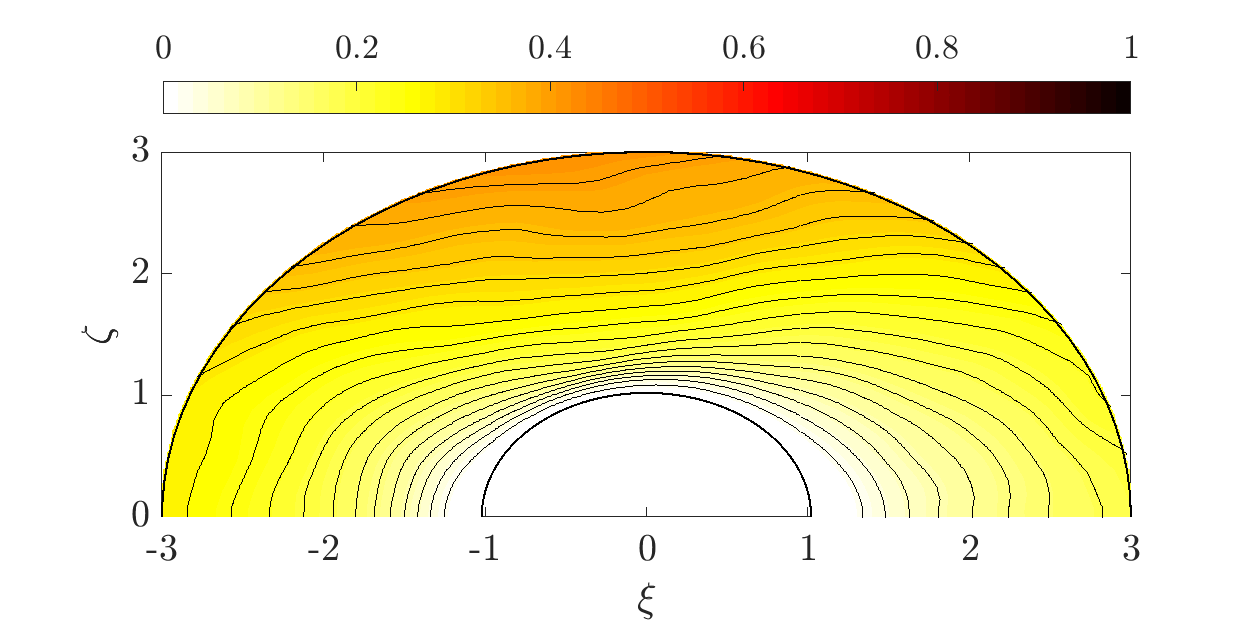} \\
\includegraphics[width=0.49\textwidth]{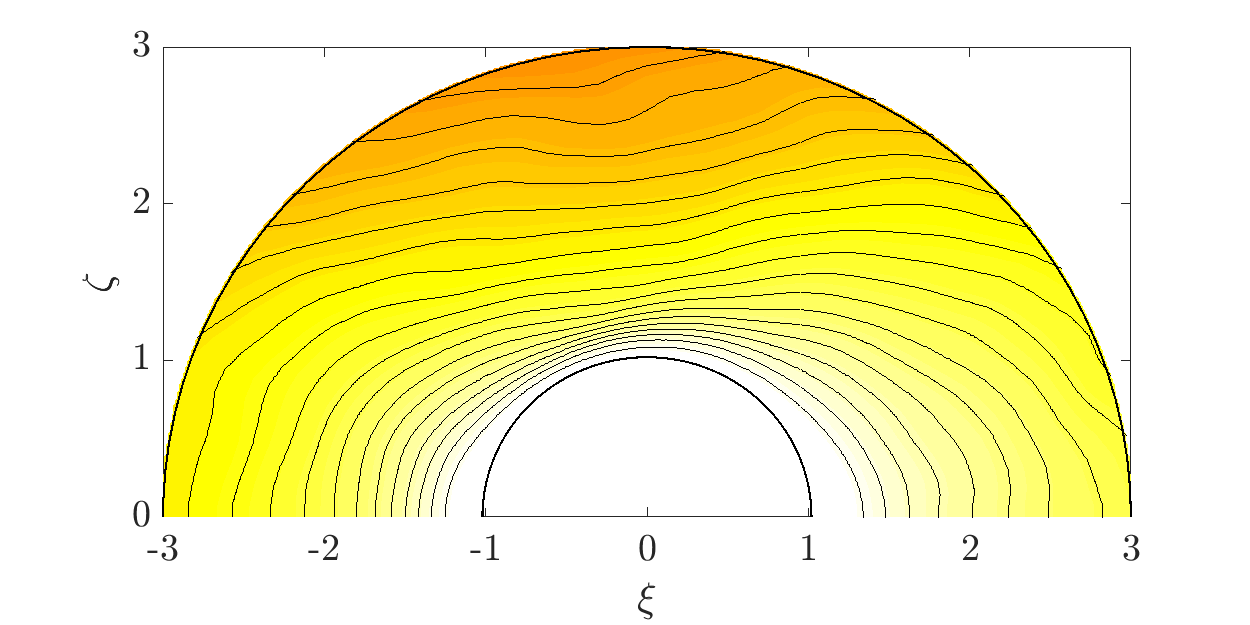}
\includegraphics[width=0.49\textwidth]{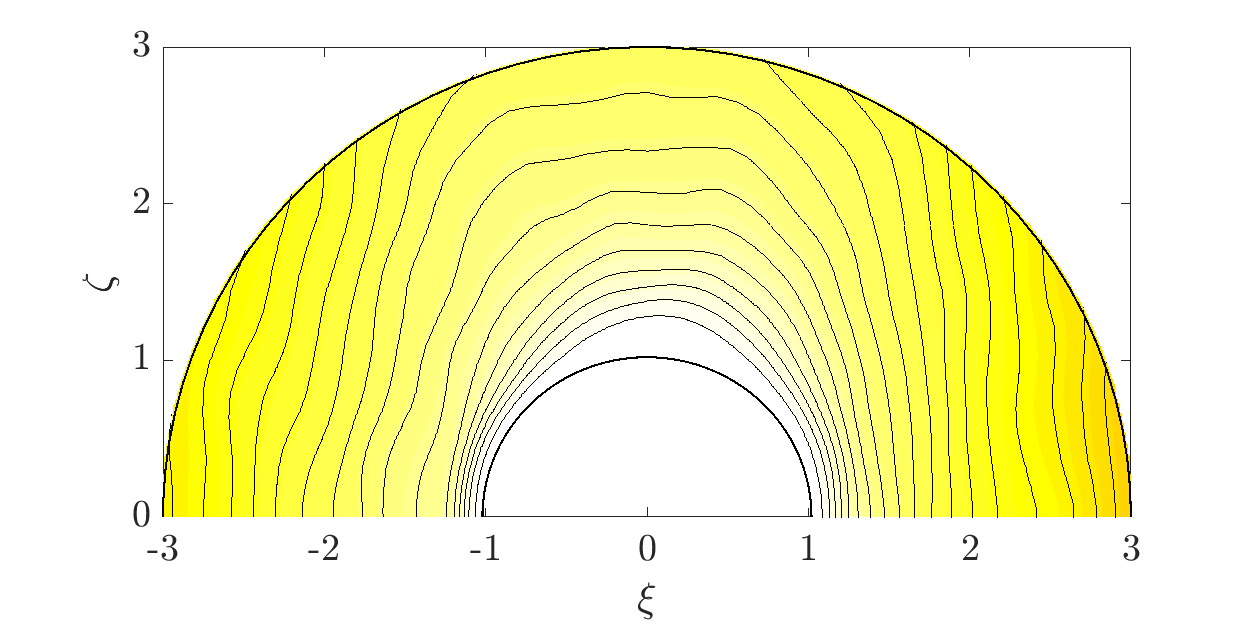}
\includegraphics[width=0.49\textwidth]{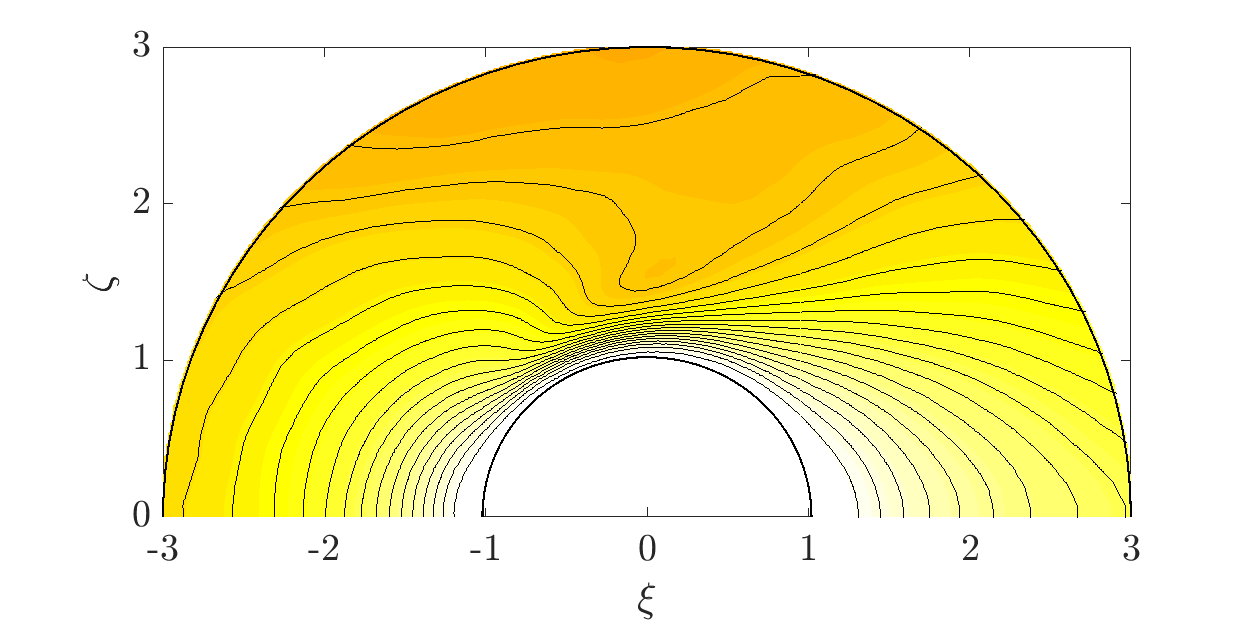}
\includegraphics[width=0.49\textwidth]{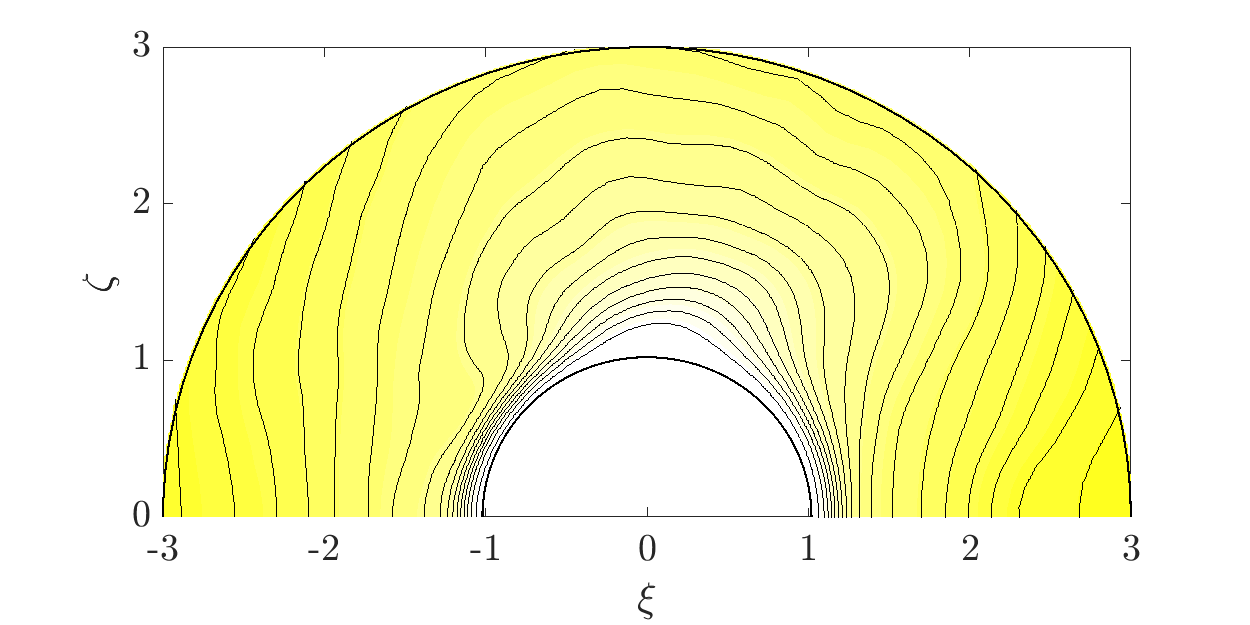}
\includegraphics[width=0.49\textwidth]{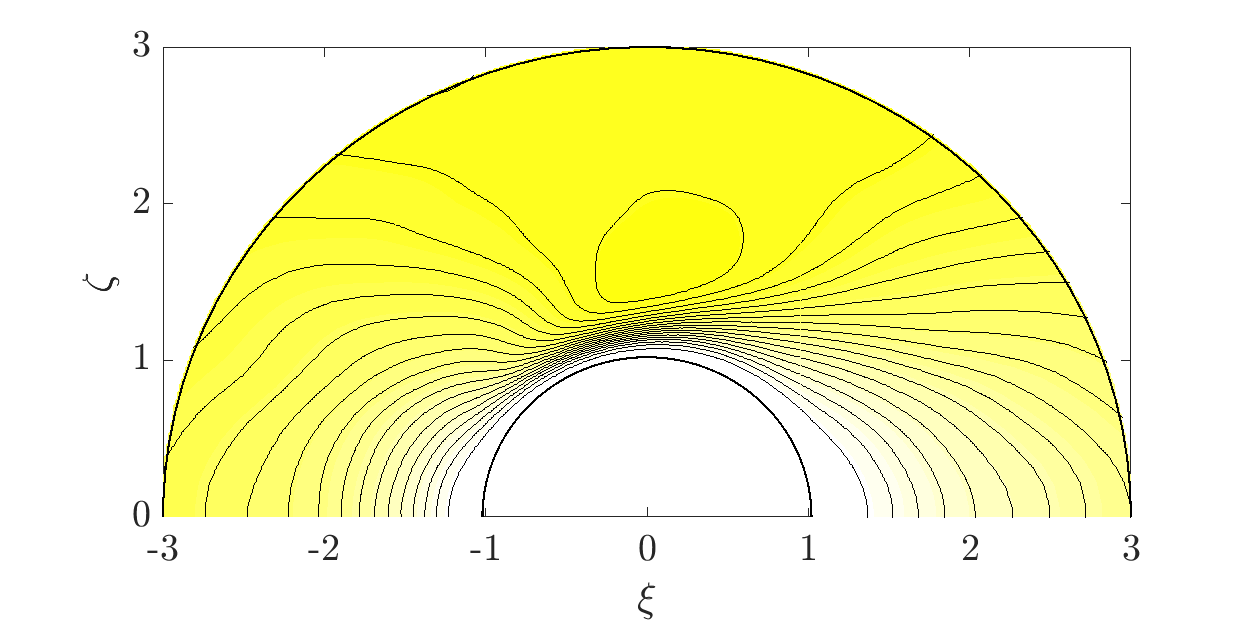}
\includegraphics[width=0.49\textwidth]{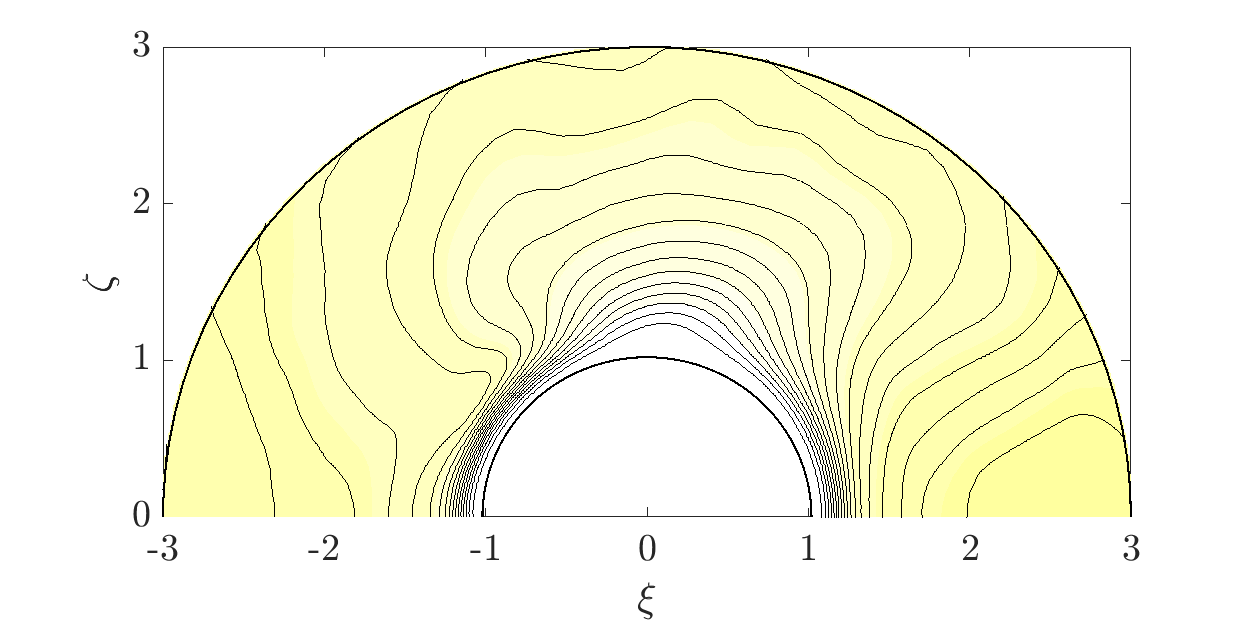}
\includegraphics[width=0.49\textwidth]{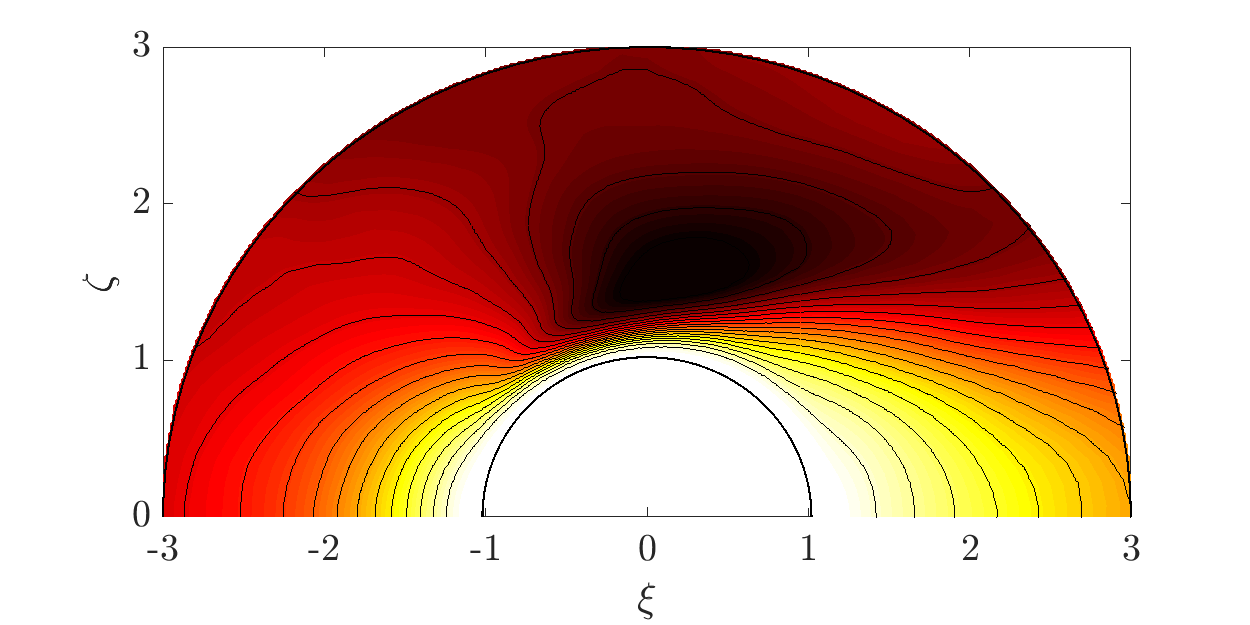}
\includegraphics[width=0.49\textwidth]{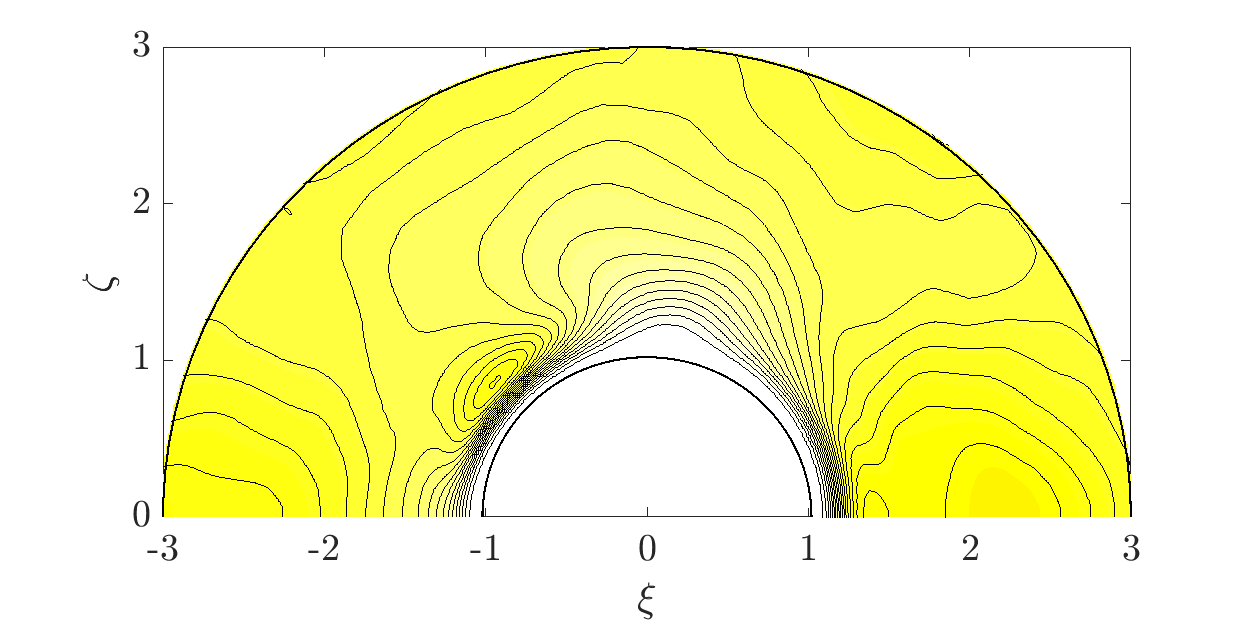}
\caption{Same as figure \ref{fig:ParLocVel}, but for the velocity variances $\aver{u'u'}_{cp,\ell}$ (left) and $\aver{w'w'}_{cp,\ell}$ (right).}
\label{fig:ParLocVarVel}
\end{figure}

Figure \ref{fig:ParLocVel} shows the average fluid velocity seen by the particle in the local reference system, i.e. $\aver{\bm{u}}_{cp,\ell} = \aver{\bm{u} - \bm{u}_f}_{cp}$; the $\cdot_\ell$ subscript refers to quantities evaluated in the (moving) local reference system of the particle. Due to the symmetries of the flow we consider only the $(\xi,\zeta)$ plane with $\eta=0$ (we show the maps for $\zeta \ge 0$, and use the $\zeta \le 0$ half plane to double the statistical sample). The left and right panels are for $\aver{u}_{cp,\ell}$ and $\aver{w}_{cp,\ell}$ respectively (here $u$ and $w$ denote the velocity components aligned with the $\xi$ and $\zeta$ axes) while from top to bottom the density of the particle is increased from $\rho_p/\rho_f = 1.29$ to $\rho_p/\rho_f = 104.7$. Figure \ref{fig:ParLocVarVel} shows instead the variance of the local velocity components, i.e. (left) $\aver{u'u'}_{cp,\ell}$ and (right) $\aver{w'w'}_{cp,\ell}$, to characterise the relative flow unsteadiness; here the apex refers to fluctuations around the conditional average value. 

When approaching the particle, the averaged flow progressively decelerates, having null velocity at the $(\xi,\zeta)=(-R,0)$ stagnation point. Then, it accelerates along the upstream side of the particle, due to the favourable pressure gradient, and decelerates along the downstream side where, depending on the flow regime, it may separate giving rise to a recirculating region. The flow regime substantially changes with $\rho_p/\rho_f$, in agreement with the large variation of the particle Reynolds number shown in figure \ref{fig:Rep_pdf}. For small densities, $\rho_p/\rho_f=1.29$, the mean flow separates and a small recirculating region arises, see e.g.~the regions with $\aver{u}_{cp,\ell}<0$ and $\aver{w}_{cp,\ell}>0$ just downstream the particle. However, no vortex shedding is detected in this case. In fact, both $\aver{u'u'}_{cp,\ell}$ and $\aver{w'w'}_{cp,\ell}$ are relatively small, and do not show the typical localised peaks in the wake region, as shown for example in figure 13 of \cite{constantinescu-squires-2004} and in figure 8 of \cite{mittal-2000}. For $\rho_p/\rho_f=4.98$, instead, the mean flow recirculation is not detected downstream the particle, but a localised peak of $\aver{u'u'}_{cp,\ell}$ is observed along the particle side at $(\xi,\zeta) \approx (0,1.5R)$. Although the absence of a localised peak of $\aver{w'w'}_{cp,\ell}$ indicates that the classical vortex shedding does not occur, this peak of $\aver{u'u'}_{cp,\ell}$ suggests that some unsteadiness has arisen. Interestingly, the maps of the fluctuations are compatible with the flow pattern shown in figure 9 of \cite{mittal-2000}, that discusses the flow past a sphere with an unsteadiness driven by the incoming flow fluctuations, rather than by a global instability of the wake. For $\rho_p/\rho_f \ge 17.45$, instead, the maps of the mean flow and the variances are compatible with an unstable wake, and with an alternate shedding of vortices. Indeed, along the downstream side of the particle a mean recirculating region is observed, together with localised peaks of $\aver{u'u'}_{cp,\ell}$ and $\aver{w'w'}_{cp,\ell}$. When the particle inertia increases, the mean recirculating region enlarges and the intensity of the fluctuations increases, revealing a stronger vortex shedding. In passing, note that, unlike in the flow past a sphere in free-stream \citep{constantinescu-squires-2004}, here the peak of $\aver{u'u'}_{cp,\ell}$ is much larger than that of $\aver{w'w'}_{cp,\ell}$: in this case the vortex shedding is substantially different, being modulated by the particle motion and the fluid velocity fluctuations.

\begin{figure}
\centering
\includegraphics[trim={0 240, 0, 0},clip,width=0.49\textwidth]{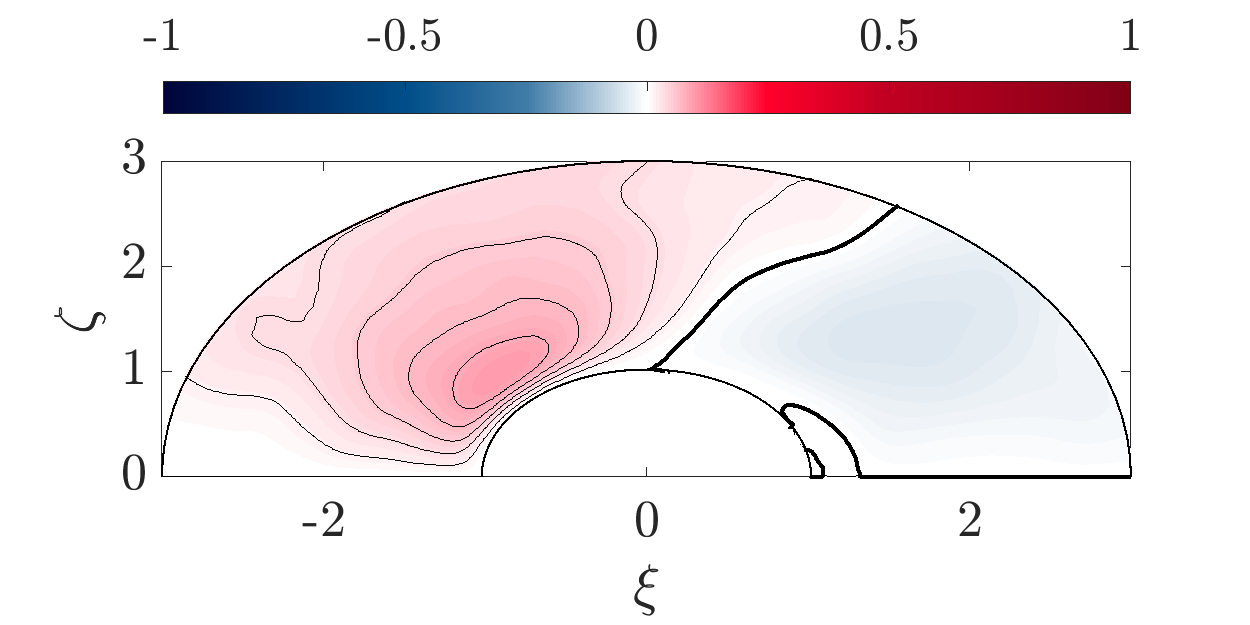}
\includegraphics[trim={0 240, 0, 0},clip,width=0.49\textwidth]{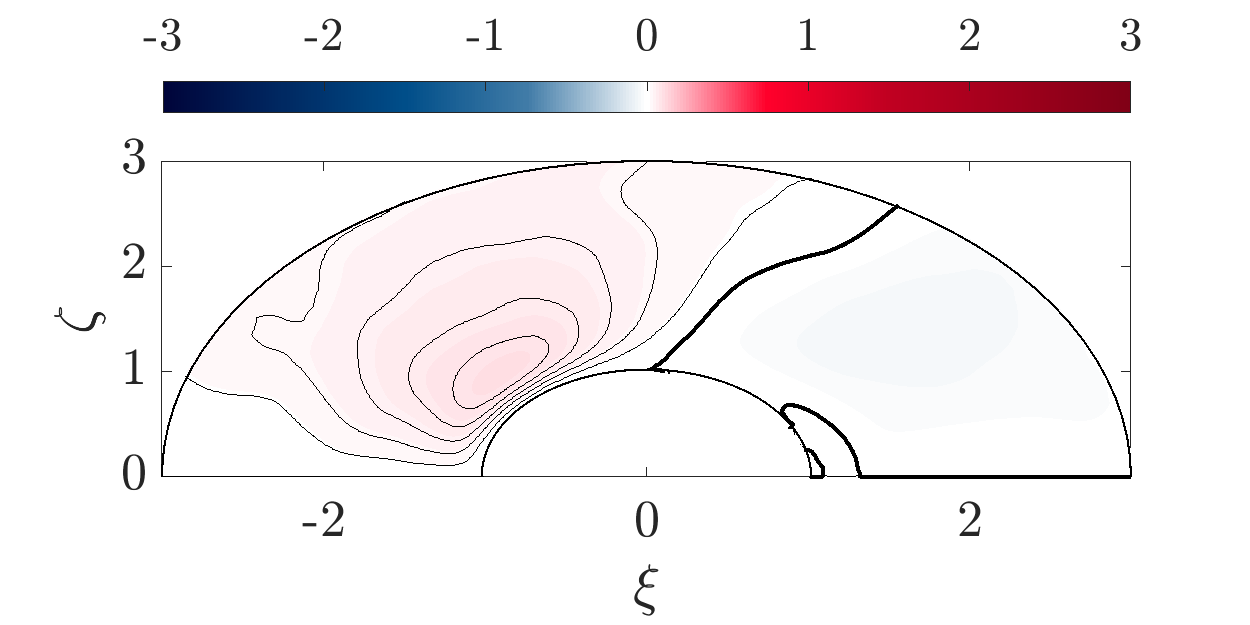}
\includegraphics[width=0.49\textwidth]{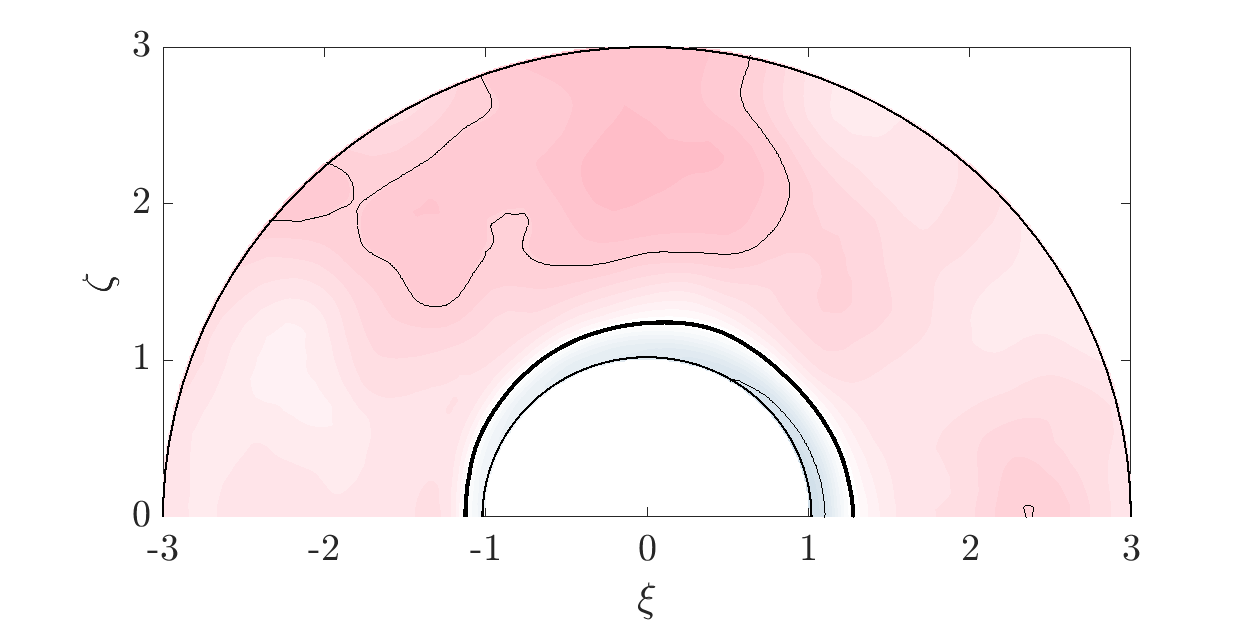}
\includegraphics[width=0.49\textwidth]{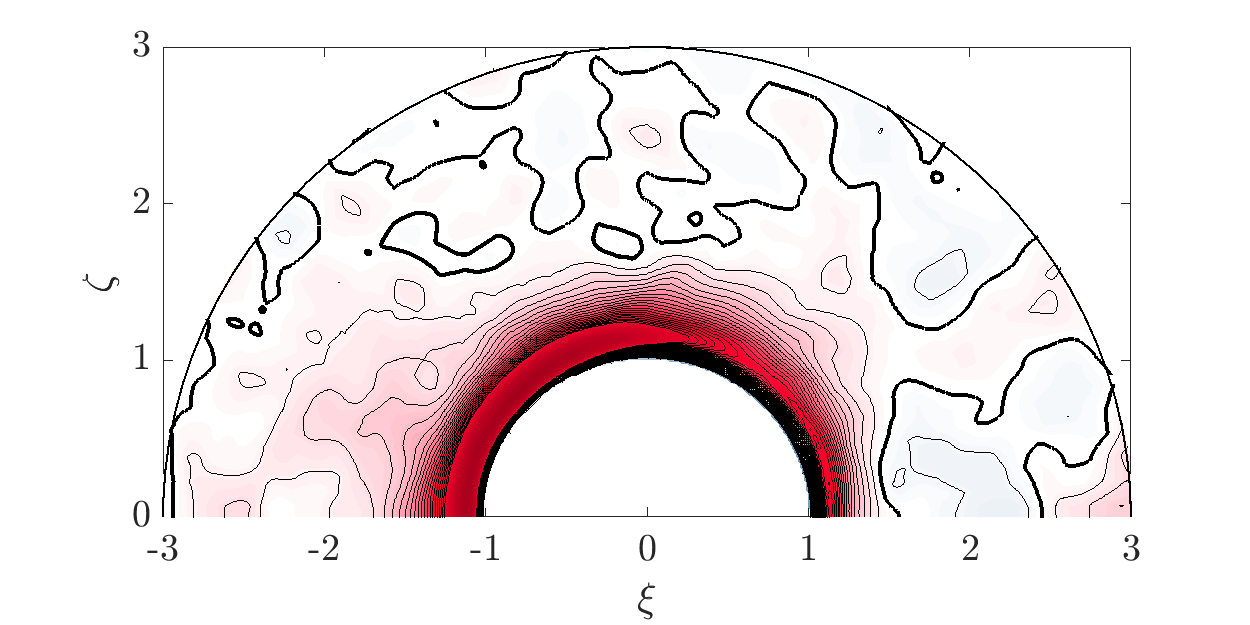}
\includegraphics[width=0.49\textwidth]{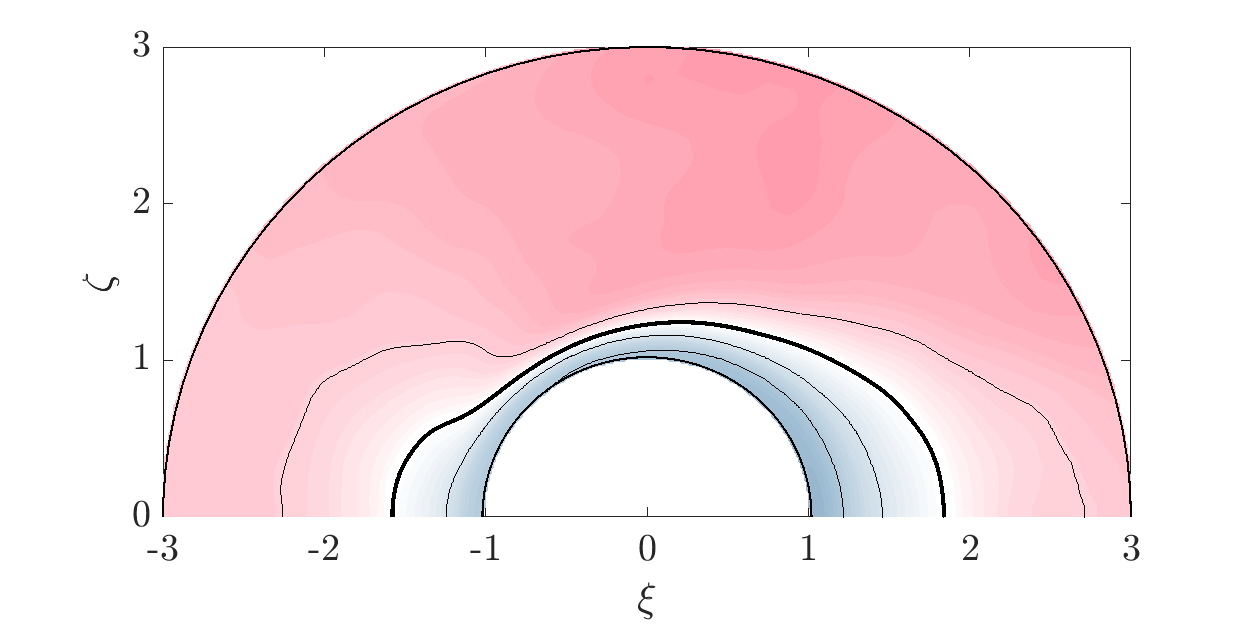}
\includegraphics[width=0.49\textwidth]{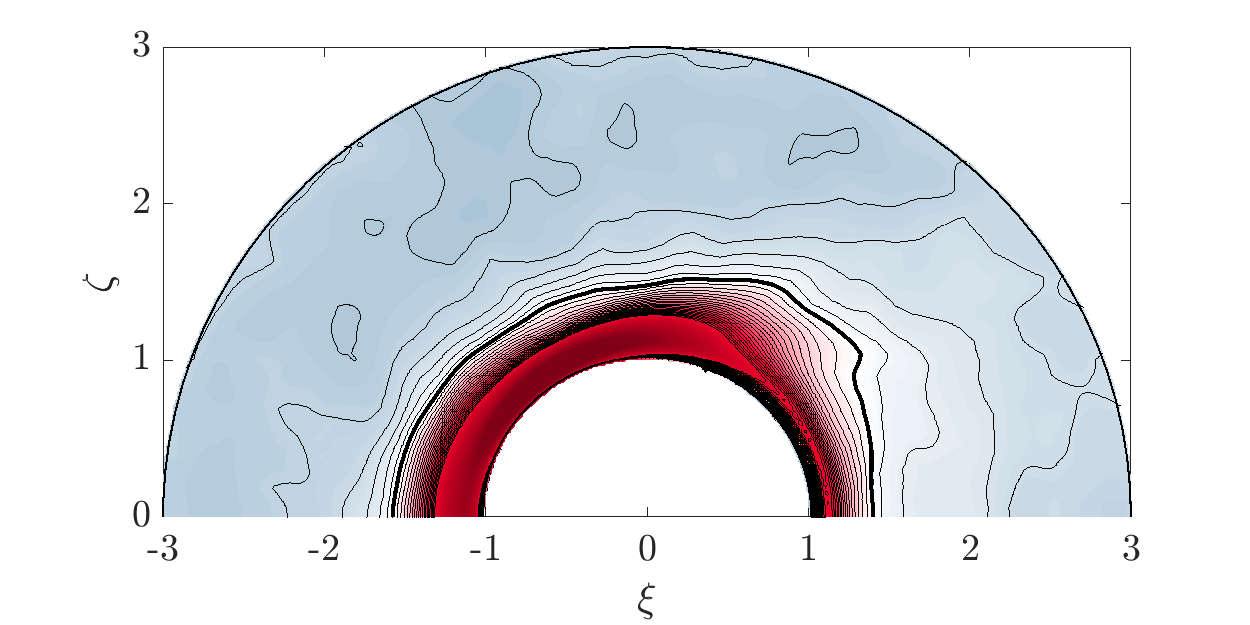}
\includegraphics[width=0.49\textwidth]{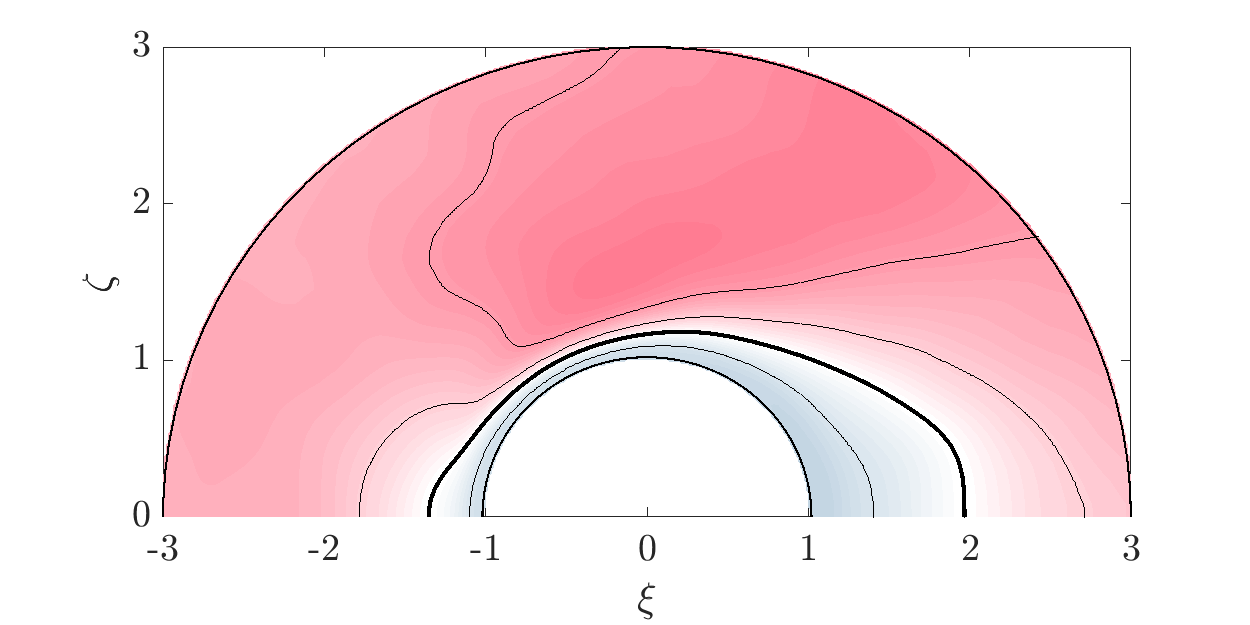}
\includegraphics[width=0.49\textwidth]{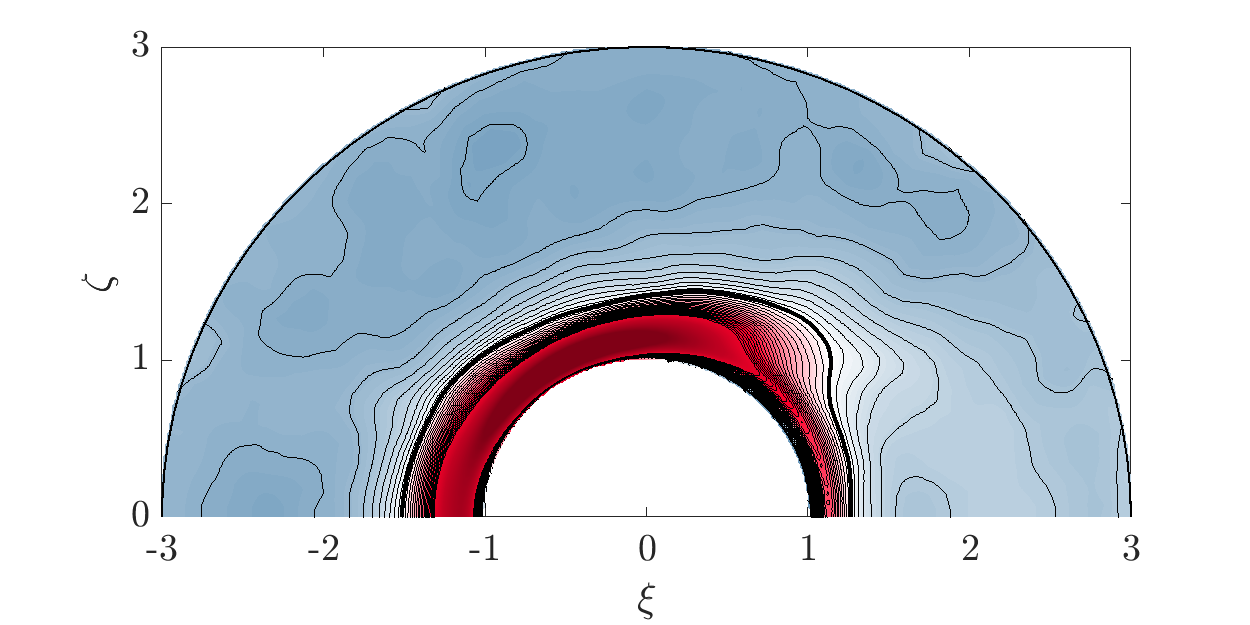}
\includegraphics[width=0.49\textwidth]{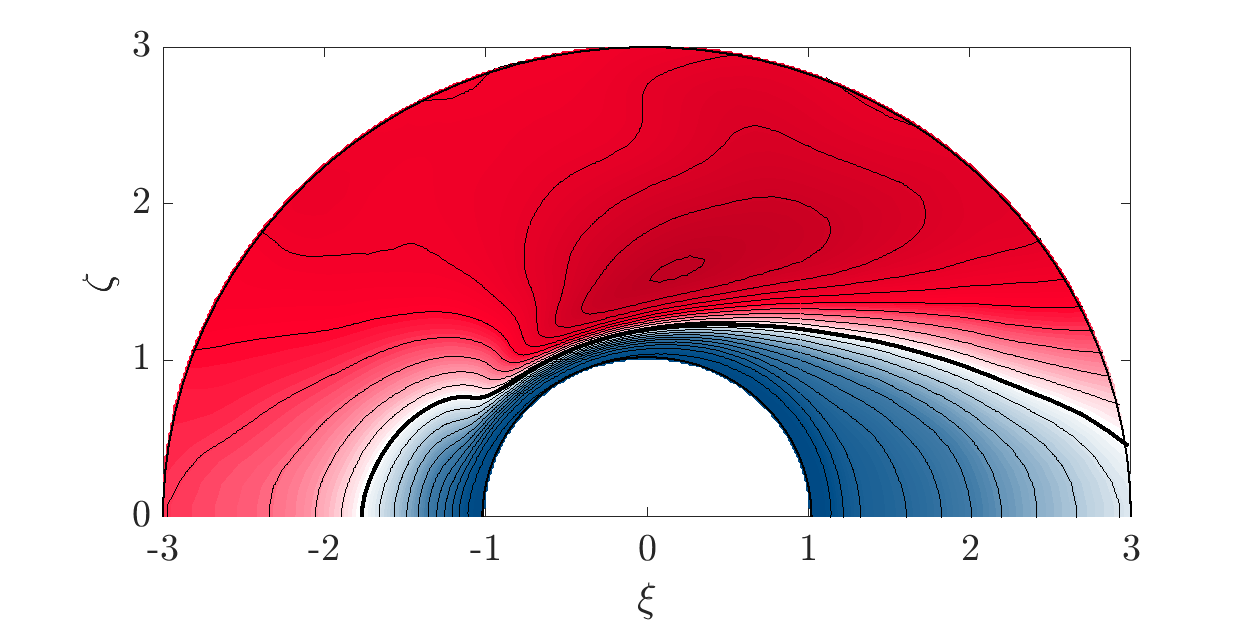}
\includegraphics[width=0.49\textwidth]{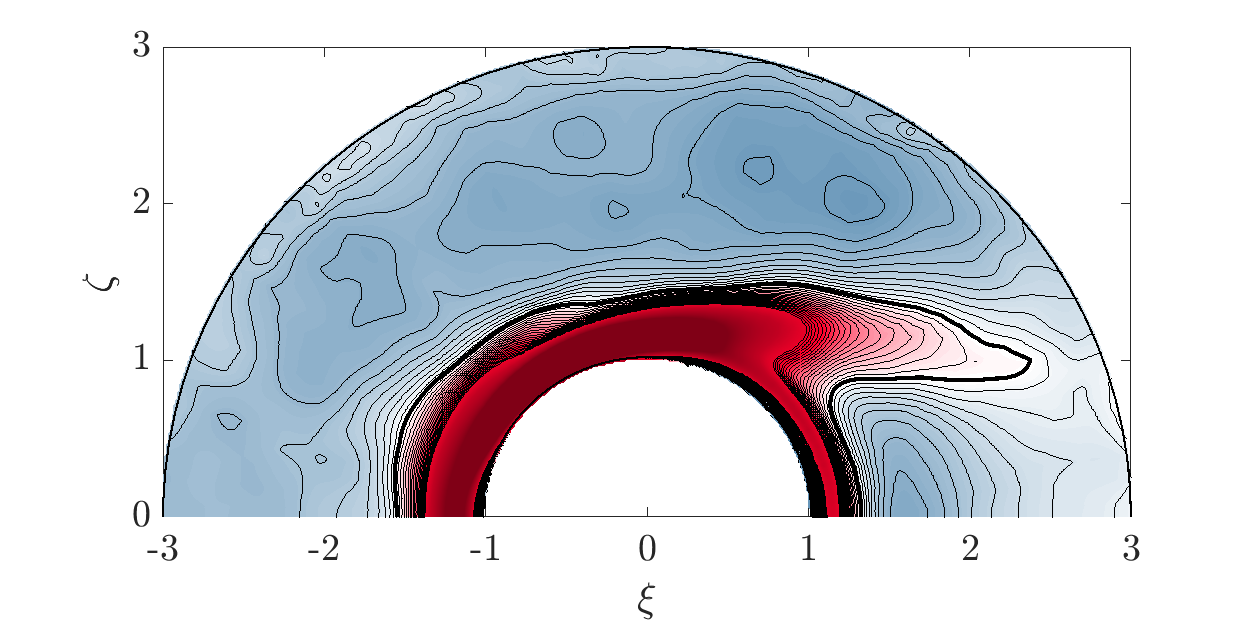}
\caption{Conditionally averaged energy and dissipation, for the case with $D/\eta=32$. From top to bottom: $\rho_p/\rho_f=1.29,4.98,17.45$ and $104.7$. Left: $\aver{e_k}_{cp}/\overline{\aver{E}}-1$. Right: $\aver{\epsilon}_{cp}/\overline{\aver{\epsilon}}-1$. Blue colour denotes region with values smaller that the box average value, while red colour indicates region with values larger than the box average value.}
\label{fig:ParLocEnerDiss}
\end{figure}
We now move to the distribution of the fluid kinetic energy; see the left panels in figure \ref{fig:ParLocEnerDiss}. 
In agreement with figure \ref{fig:parLocEner}, the flow energy is reduced at the particle surface, due to the no-slip and no-penetration boundary conditions, while it becomes larger than the box-average value when moving away. For small densities, $\rho_p/\rho_f=1.29$, the energy distribution is almost symmetric, in agreement with the above discussed lack of vortex shedding in the wake. When $\rho_p/\rho_f$ increases, instead, the low energy region with $\aver{e_k}_{cp}/\overline{\aver{E}}<1$ becomes asymmetric and extends mainly downstream the particle, in agreement with the visualisations of previous authors \citep[see for example][]{tanaka-eaton-2010}. This is consistent with the distribution of the mean flow and of the velocity fluctuations shown in figure \ref{fig:ParLocVarVel}. For large $\rho_p/\rho_f$ a region with $\aver{e_k}_{cp}/\overline{\aver{E}}>1$ arises over the lateral sides of the particle, $(\xi,\zeta) \approx (0, 1.5D)$, being the trace of the larger mean flow acceleration and larger fluctuating energy discussed above. 

The right panels of figure \ref{fig:ParLocEnerDiss} consider the fluid dissipation rate. In agreement with figure \ref{fig:parLocDiss}, a region of high dissipation rate is observed in the immediate neighbourhood of the entire particle surface, being associated with the large velocity gradients that develop in the particle boundary layer. The intensity of $\aver{\epsilon}_{cp}/\overline{\aver{\epsilon}}$ in this region increases with $\rho_p/\rho_f$, consistently with the increase of the particle Reynolds number and with the progressive decorrelation of the particle velocity and the local flow. For small $\rho_p/\rho_f$, $\aver{\epsilon}_{cp}/\overline{\aver{\epsilon}} \approx 1$ outside this region, meaning that the influence of light particles on the flow dissipation rate is relatively limited in space. For larger $\rho_p/\rho_f$, instead, $\aver{\epsilon}_{cp}/\overline{\aver{\epsilon}} < 1$ outside the particle boundary layer, indicating a wider influence of the particle. Aside from this region, two further regions of relatively high dissipation rate are observed. The first is observed for all $\rho_p/\rho_f$ and is placed upstream the particle; here $\aver{\epsilon}_{cp}/\overline{\aver{\epsilon}}$ is maximum. In fact, the largest velocity gradients are attained close to the front stagnation point, where the flow deceleration is strong. Note that this region of large dissipation has been also observed by \cite{brandle-etal-2016}, when considering particles with $1 \le \rho_p/\rho_f \le 4$ in homogeneous isotropic turbulence at a lower Reynolds number $Re_\lambda \approx 70$. The second region is observed only for large $\rho_p/\rho_f$, and is associated with the separated flow in the wake region. Note that it is barely visible for $\rho_p/\rho_f=17.45$, while clearly visible for $\rho_p/\rho_f=104.7$. By means of the vortex shedding, indeed, heavy particles modulate the flow dissipation rate over a wider portion of the wake region.


\section{Particle dynamics}
\label{sec:particles}

\subsection{Particles velocity and trajectories}
\label{sec:traj}

\begin{figure}
\centering
\includegraphics[width=0.49\textwidth]{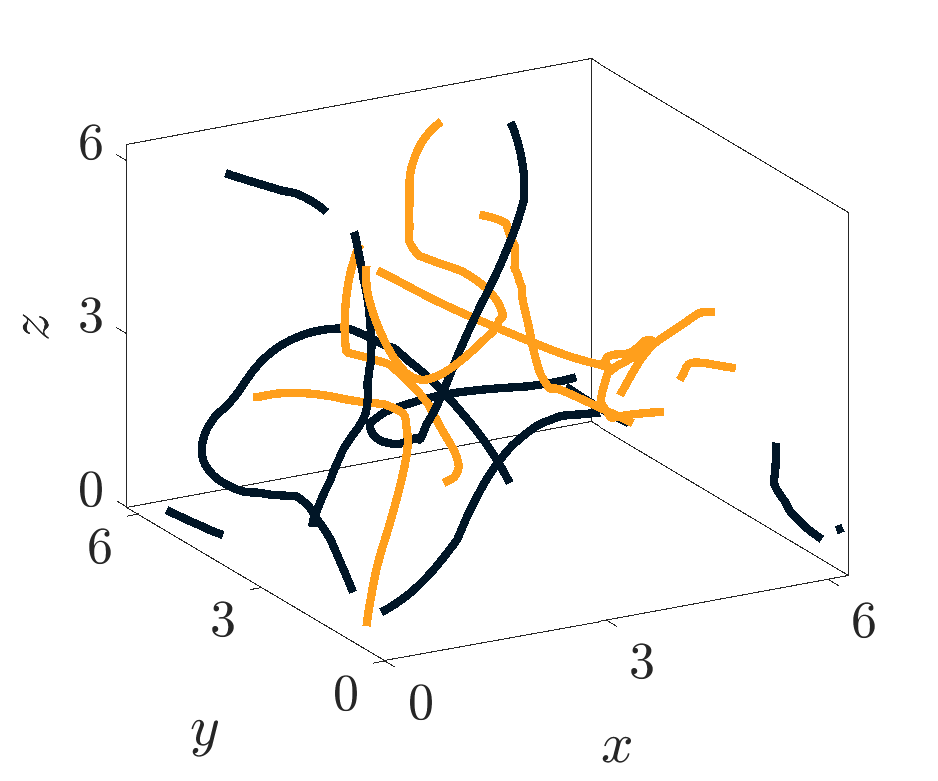}
\includegraphics[width=0.49\textwidth]{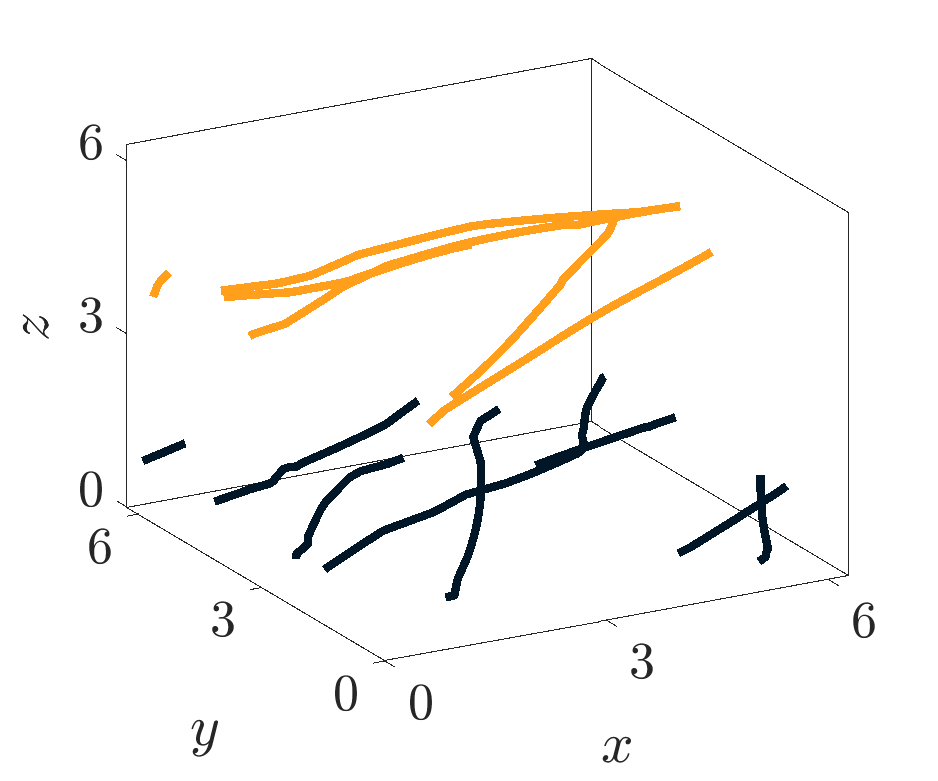}
\caption{Particle dynamics for $D/\eta=32$. Two trajectories for (left) $M=0.3$ and (right) $M=0.6$, representative of regime A and regime B introduced in section \ref{sec:meanflow}. Adapted from \cite{chiarini-etal-2023}. The yellow and black colours are used to ease the visualisation of the two trajectories in the left panel.}
\label{fig:traj}
\end{figure}

\begin{figure}
\centering
\includegraphics[width=1.0\textwidth]{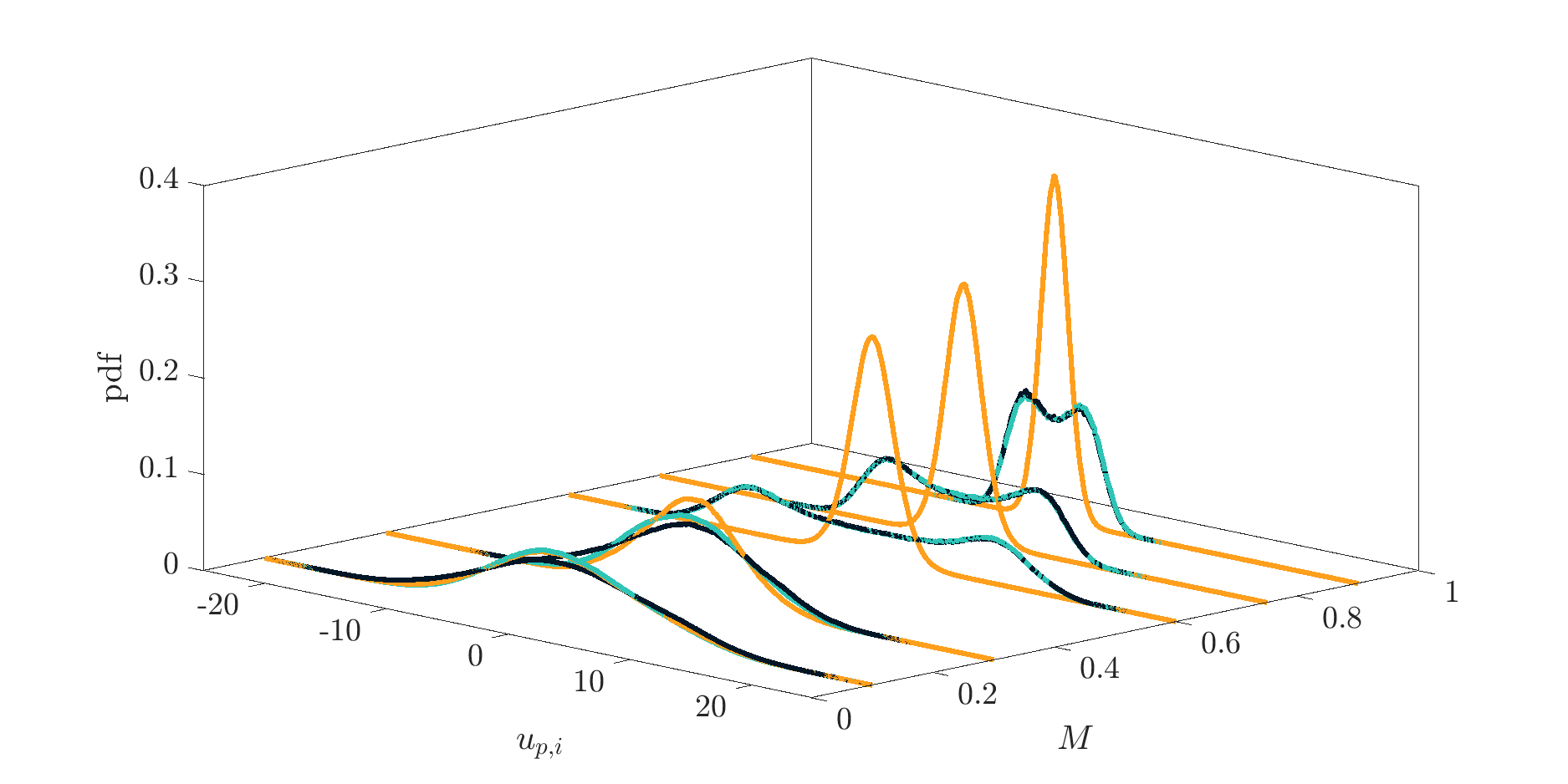}
\includegraphics[width=1.0\textwidth]{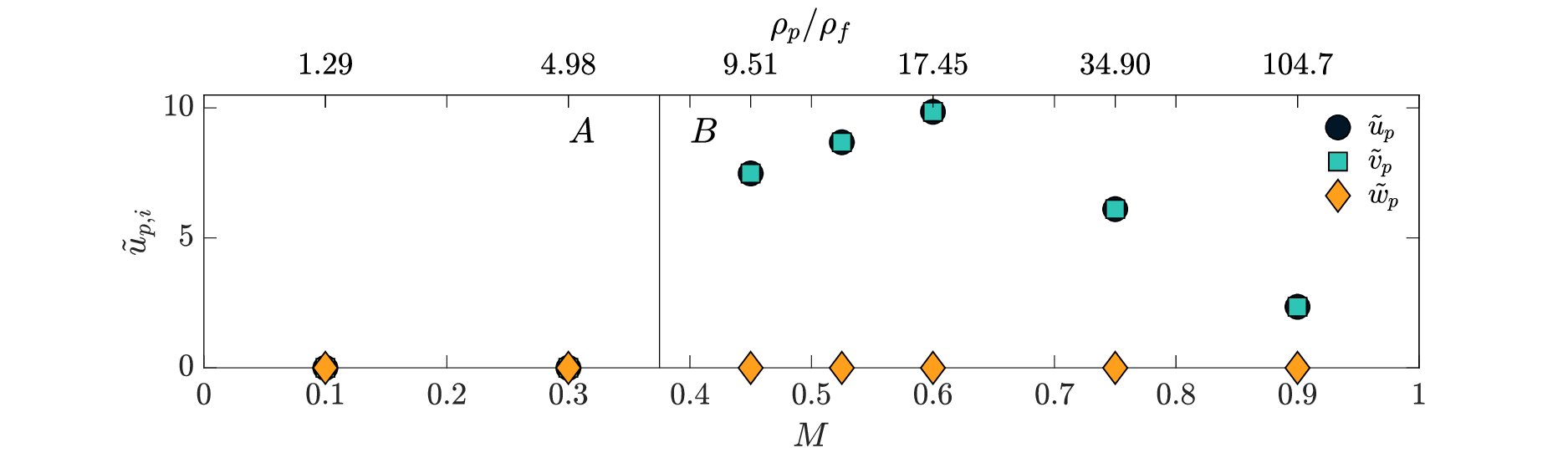}
\caption{(top) Probability density function of the particle velocity components for $D/\eta = 32$ and $0.1 \le M \le 0.9$. Black is for $u_p$, green for $v_p$ and orange for $w_p$. (bottom) Dependence of the modes of the three velocity components of the particles $\hat{u}_p$, $\hat{v}_p$ and $\hat{w}_p$ on the particle density.}
\label{fig:velp_pdf}
\end{figure}

In this section we briefly characterise the dynamics of the particles in the A and B flow regimes mentioned in section \ref{sec:meanflow}; we refer the reader to \cite{chiarini-etal-2023} for a detailed discussion. Figure \ref{fig:traj} shows typical trajectories for the $D/\eta=32$ particle-laden cases with $M=0.3$ (left) and $M=0.6$ (right), which are representative of the motion of the particles in the two regimes. Figure \ref{fig:velp_pdf}, instead, quantitatively characterises the velocity of the particles for $D/\eta=32$ and $0.1 \le M \le 0.9$.

In regime A ($D/\eta > 64$ and $M \le 0.3$) particles follow, at least partially, the cellular motion imposed by the external ABC forcing, and exhibit complex trajectories in agreement with the chaotic Lagrangian structure of the ABC flow \citep{dombre-etal-1986}; see the left panel of figure \ref{fig:traj}. In this case, particles do not select a preferential direction, and progressively span the complete computational domain. Accordingly, the probability density functions of the three velocity components almost collapse, and show a symmetric unimodal distribution with the modes being $ \hat{u}_p = \hat{v}_p = \hat{w}_p = 0$. When $M$ and/or $D$ are increased, the three distributions narrow (i.e. the variances decrease) due to the larger inertia, and the particle velocity experiences smaller fluctuations.

In regime B ($D/\eta \le 64$ and $M \ge 0.45$) particles show a completely different behaviour. Due to the large inertia, they deviate from the cellular flow path imposed by the ABC forcing, and move along almost straight trajectories showing anomalous transport \citep{chiarini-etal-2023}, see the right panel of figure \ref{fig:traj}. Recall that, in regime B, we denote with $z$ the direction orthogonal to the plane where the trajectories of the particles lay, which corresponds to the direction aligned with the attenuated mean flow velocity component (see section \ref{sec:meanflow}). Compared to regime A, here the fluctuations of the in-plane $u_p$ and $v_p$ particle velocity components are strongly enhanced, while the fluctuations of the out-of-plane $w_p$ component are attenuated, see figure \ref{fig:velp_pdf}. The direction of the trajectories in the $x-y$ plane changes with $z$, as the mean flow and particles velocity retain the sinusoidal dependence on $z$ inherited by the ABC forcing (see also section \ref{sec:meanflow}). Accordingly, the probability density function of $w_p$ exhibits a very narrow symmetric unimodal distribution centred in $\hat{w}_p=0$ (see figure \ref{fig:velp_pdf}), while the distributions of $u_p$ and $v_p$ show a symmetric bimodal distribution with the modes $\pm \hat{u}_p = \pm \hat{v}_p$ that change with the particle size and density.

\subsection{Clustering}
\label{sec:clustering}

After the characterisation of the dynamics of a single particle, in this section we focus on their collective motion and investigate how it changes with the particle size and density. We look at the local concentration of the suspensions to inspect the presence of clusters. Different metrics have been proposed over the years for detecting clusters \citep{brandt-coletti-2022}. Vorono\"{i} tesselletion is a computationally efficient tool for the analysis of the spatial arrangement of the dispersed phase in the carrier flow, and has been used by several authors \citep[see for example][]{monchaux-etal-2010,monchaux-etal-2012,uhlmann-chouippe-2017,petersen-etal-2019}. The position of each particle is determined by its centre, and the computational box is partitioned such that each grid point is associated with the closest particle. The Vorono\"{i} cell of each particle, therefore, consists of the ensemble of grid cells that are closer to it. The inverse of the volume of a Vorono\"{i} cell provides a measure of the local concentration: a smaller Vorono\"{i} cell corresponds to a denser particle arrangement, while a larger Vorono\"{i} cell to a more sparse distribution. \cite{ferenc-neda-2007} have shown that when point particles are randomly drawn in a box, the probability density function of the corresponding Vorono\"{i} cell volumes exhibits a gamma distribution with parameters that can be evaluated analytically. However, for finite-size particles this is not the case, as they can not overlap. This means that the parameters describing a random distribution of finite-size particles need to be computed separately for different particle sizes, volume fractions and box sizes. For particle suspensions that exhibit clustering, the variance of the Vorono\"{i} cell volumes is larger than that computed for the corresponding random distribution \citep{monchaux-etal-2012,uhlmann-doychev-2014,uhlmann-chouippe-2017}.

\begin{figure}
\centering
\includegraphics[width=0.7\textwidth]{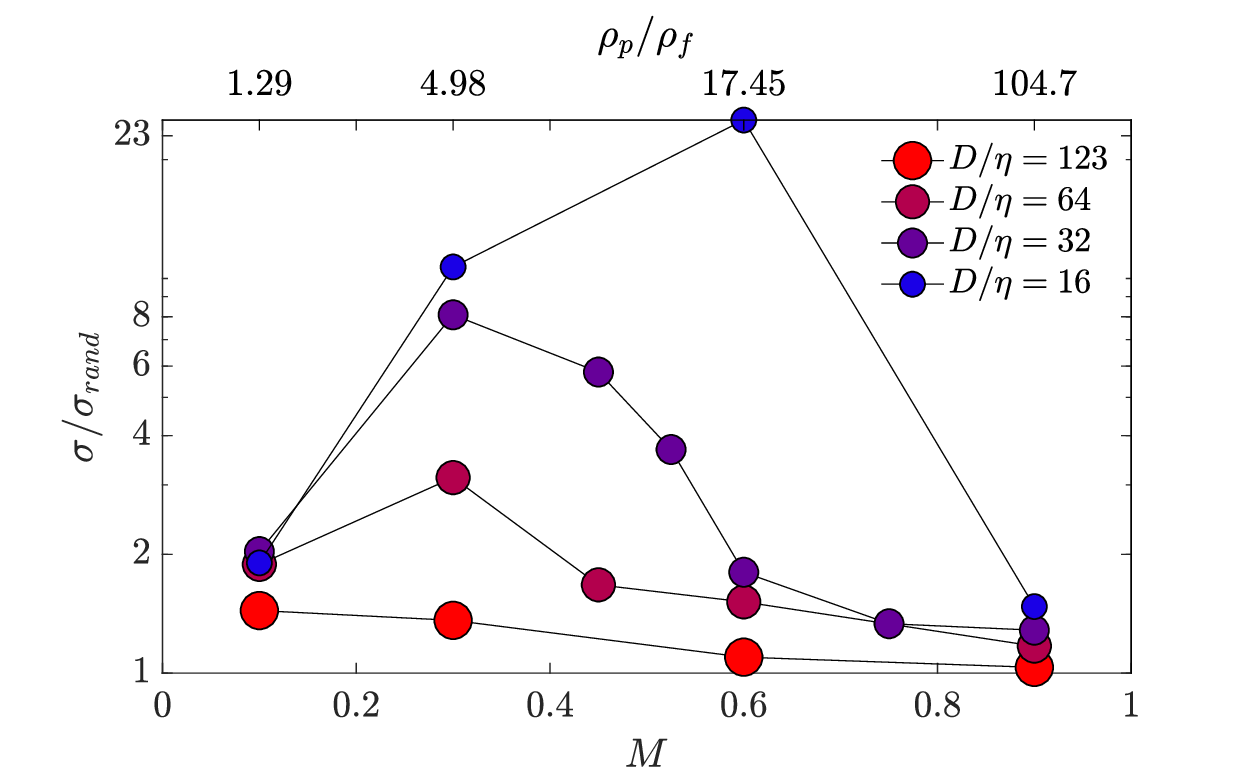}
\caption{Time average of the variance of the volume of the Vorono\"{i} cells for different particles sizes and mass fractions.}
\label{fig:Voronoi_sigma}
\end{figure}
Figure \ref{fig:Voronoi_sigma} shows the time average of the variance of the volume of the Vorono\"{i} cells for different particles sizes and mass fractions. It is clearly visible that for all mass fractions smaller particles exhibit stronger clustering. Due to their lower inertia, indeed, smaller particles are more able to follow the fluid motion, and this may promote their clustering due to the particle preferential sampling (see section \ref{sec:PrefSamp}). In contrast, the influence of the mass fraction on the clustering is not monotonous, and changes with the particle size. For large particles, $D/\eta=123$, heavier particles exhibit weaker clustering; for $M=0.9$ we compute $\sigma/\sigma_{rand} \approx 1$ indicating the almost absence of clusters. For $D/\eta \le 64$, instead, the level of clustering is maximum for intermediate mass fractions; $\sigma/\sigma_{rand}$ is maximum at $M=0.3$ for $D/\eta=32$ and $D/\eta=64$, and at $M=0.6$ for $D/\eta=16$. As shown in the following (see figure \ref{fig:Voronoi_clusters} and relative discussion), this non monotonous dependence of the level of clustering on $M$ is related to the different trajectories of the particles in the A and B regimes discussed in section \ref{sec:traj}.
The low level of clustering observed for $M=0.1$ recalls the results of \cite{fiabane-etal-2012} that report no clustering for neutrally buoyant particles with size $D/\eta \approx 20$. The present results, however, give evidence that clusters are present in non-dilute suspension of relatively small and heavy finite-size particles. 

\cite{monchaux-etal-2010} used the Vorono\"{i} tessellation to provide an objective definition of a cluster.
\begin{figure}
\centering
\includegraphics[width=0.7\textwidth]{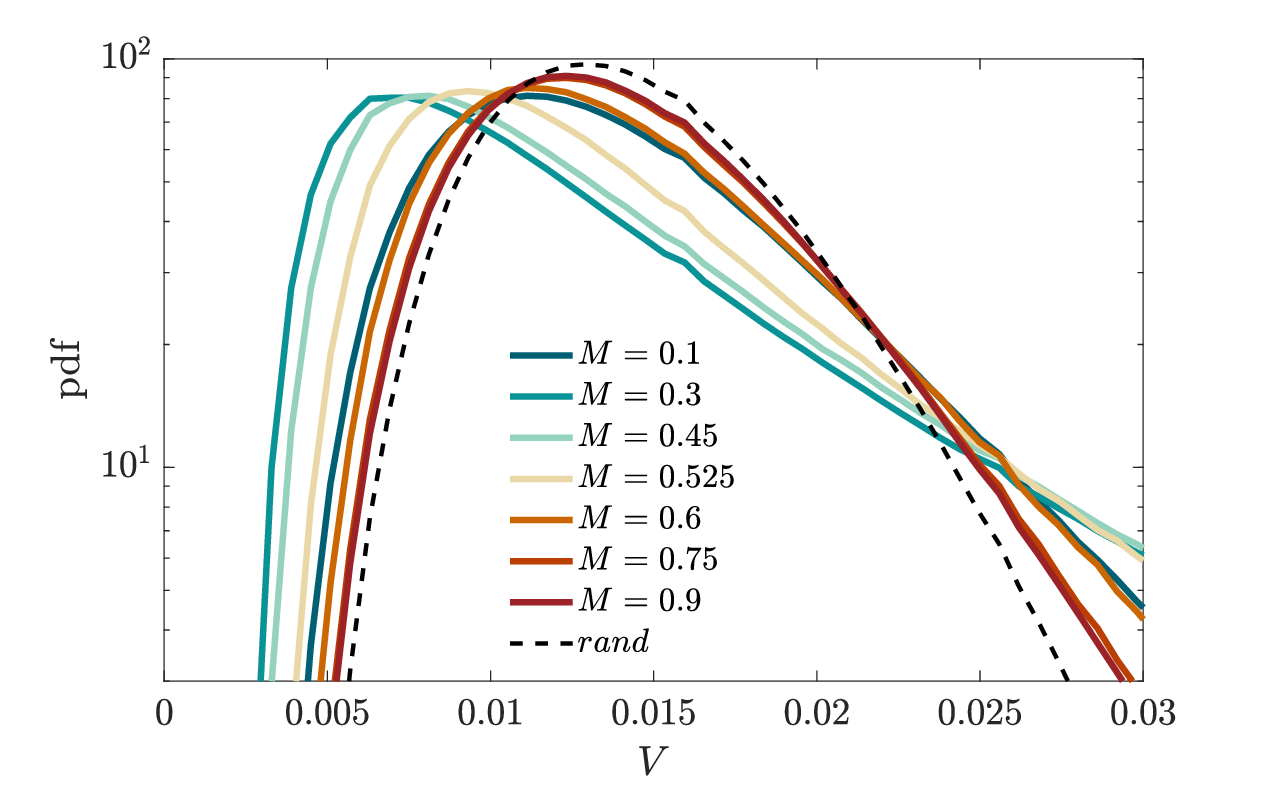}
\caption{Probability density function of the Vorono\"{i} cell volumes for $D/\eta=32$ and $0.1 \le M \le 0.9$ compared to the random distribution.}
\label{fig:Voronoi_pdf}
\end{figure}
As shown in figure \ref{fig:Voronoi_pdf} for $D/\eta=32$, the fact that the variance of the actual simulation is larger than the corresponding random arrangement of particles leads to two cross-overs between the two associated probability density functions of the Vorono\"{i} cells volume. These intersection points can be used to identify clusters and void regions. \cite{monchaux-etal-2010} proposed that all particles associated with a Vorono\"{i} cell with volume smaller than the lower cross-over point are part of a cluster, while all particles associated with a Vorono\"{i} cell with volume larger than the larger cross-over point are part of void regions. Moreover, particles with Vorono\"{i} cells smaller than the lower cross-over point that share at least one vertex, are part of the same cluster.
\begin{figure}
\centering
\includegraphics[width=0.49\textwidth]{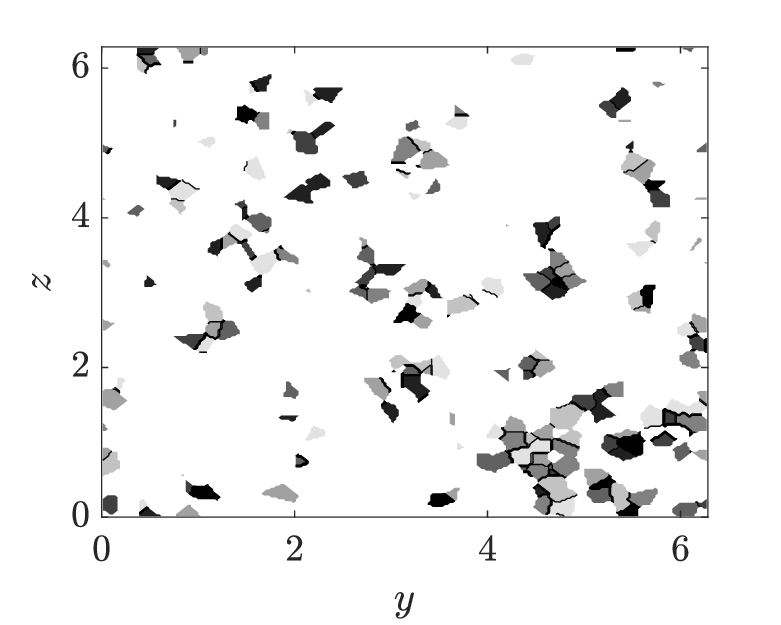}
\includegraphics[width=0.49\textwidth]{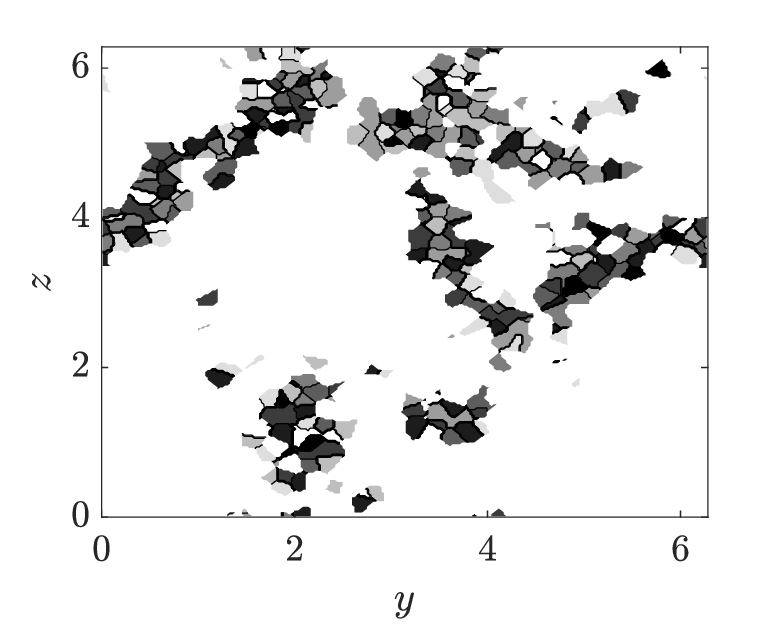}
\includegraphics[width=0.49\textwidth]{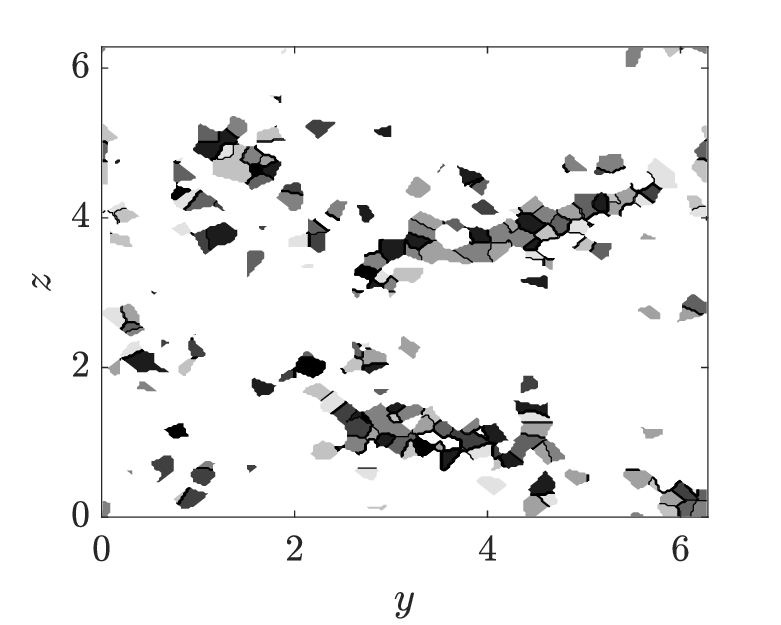}
\includegraphics[width=0.49\textwidth]{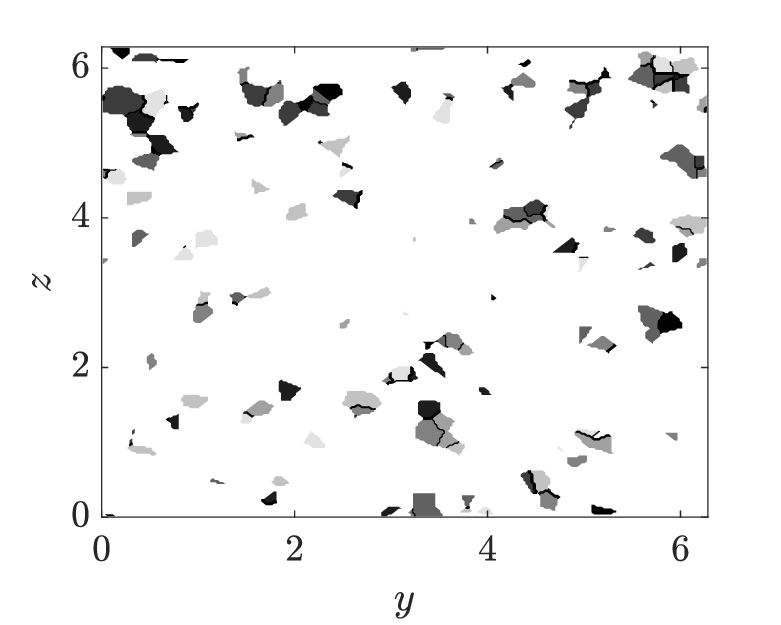}
\caption{Visualisation of the Vorono\"{i} clusters for $D/\eta=32$ and (top left) $M=0.1$, (top right) $M=0.3$, (bottom left) $M=0.6$ and (bottom right) $M=0.9$.}
\label{fig:Voronoi_clusters}
\end{figure}
Following this approach, figure \ref{fig:Voronoi_clusters} provides a qualitative view of the clustering for $D/\eta=32$; note that qualitatively the same results have been found also for the other particle sizes. The figure shows the clusters found in an instantaneous snapshot in the $x=L/2$ plane for $0.1 \le M \le 0.9$. Different colours indicate Vorono\"{i} cells associated with different particles. Accordingly with the previous discussion, the level of clustering is strong for $M=0.3$, while rather weak for $M=0.1$ and $M=0.9$. The spatial arrangement of the clusters changes with the trajectories of the particles (see section \ref{sec:traj}), explaining the non monotonous dependence of $\sigma/\sigma_{rand}$ on $M$ for $D/\eta \le 64$. For $M=0.3$, indeed, the cluster arrangement recalls the cellular shape inherited by the ABC forcing, and does not reveal a preferential direction. For $M=0.6$, instead, clusters are stretched and aligned in the $y$ direction, accordingly with the almost two-dimensional motion of heavy particles previously discussed. The low level of clustering for $M=0.9$ shows that heavy particles move along almost straight lines in a isolated manner.

An alternative way for characterising clustering is the radial distribution function $g(r)$ \citep{saw-etal-2008,salazar-etal-2008} defined as
\begin{equation}
  g(r) = \frac{N_s(r)/\Delta V_i(r)}{N_p/V}.
\end{equation}
Here, $N_s(r)$ is the number of particle pairs separated by a distance within $r-\Delta r/2$ and $r + \Delta r/2$, $\Delta V_i$ is the volume of the discrete shell located at $r$, $N_p=N(N-1)/2$ is the total number of particle pairs in the sample, and $V$ is the total volume of the system. This quantity describes the probability of having a pair of particles at a given mutual distance, and its magnitude characterises the strength of clustering at scale $r$. In a perfectly uniform distribution of point particles (where the overlap between particles is possible) $g(r)=1$ for all $r$. 
\begin{figure}
\centering
\includegraphics[width=0.49\textwidth]{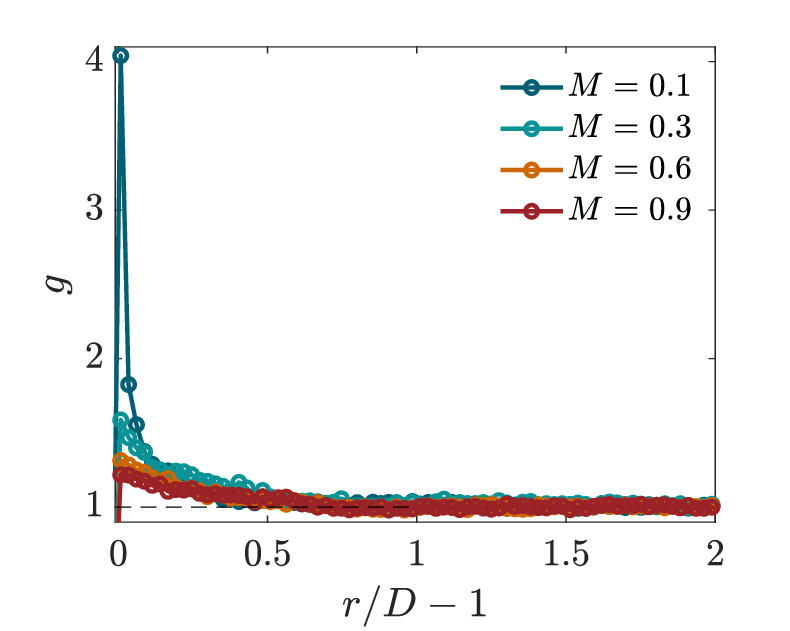}
\includegraphics[width=0.49\textwidth]{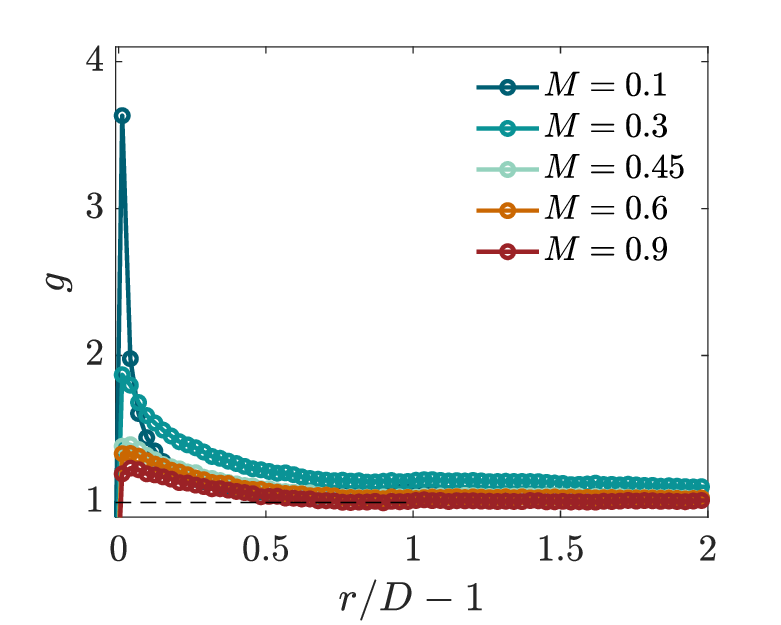}
\includegraphics[width=0.49\textwidth]{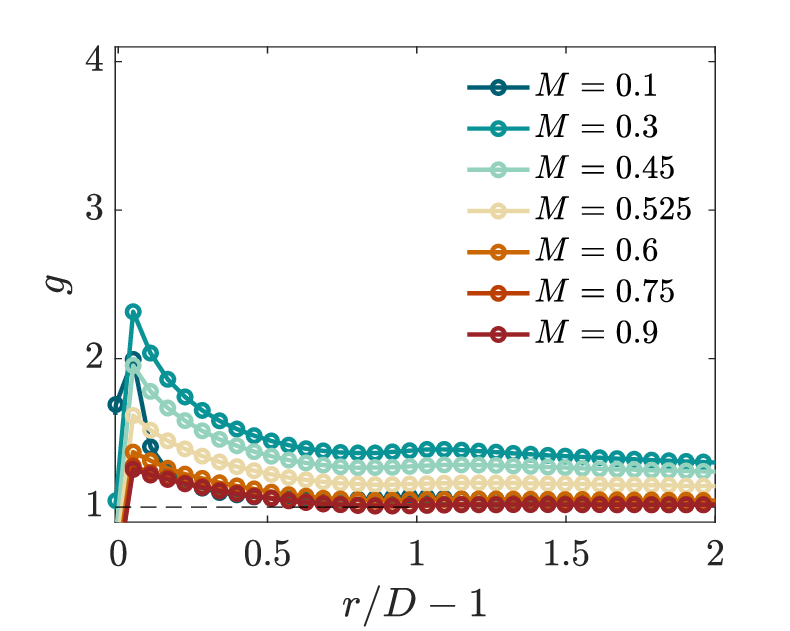}
\includegraphics[width=0.49\textwidth]{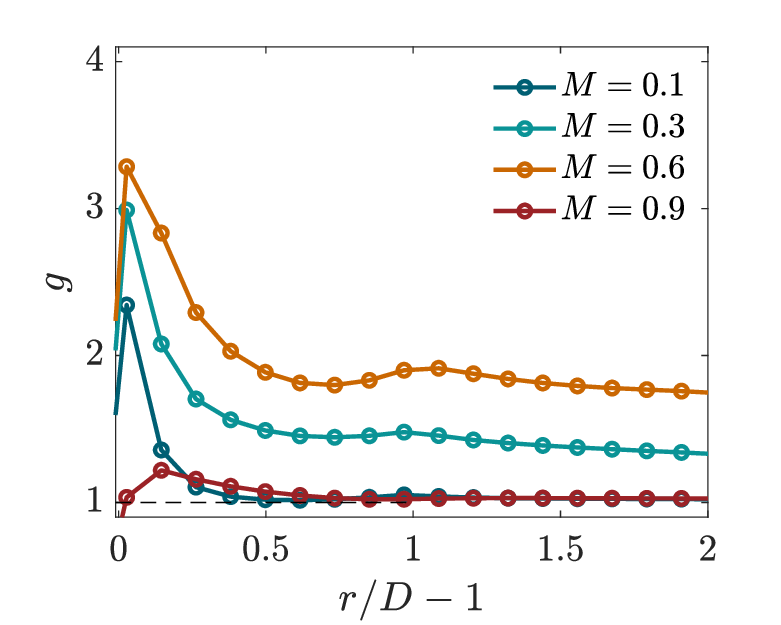}
\caption{Dependence of the radial distribution function on $M$ for (top left) $D/\eta=123$, (top right) $D/\eta=64$, (bottom left) $D/\eta=32$ and (bottom right) $D/\eta=16$.}
\label{fig:rdfMeffect}
\end{figure}

Figure \ref{fig:rdfMeffect} shows the dependence of the radial distribution function $g(r)$ on the mass fraction. For large particles, $D/\eta \ge 64$, it turns out that the accumulation at small length scales $r \approx D$ is maximum for light particles, while progressively decreases with $M$. At large length scales, instead, the accumulation is maximum for $M=0.3$, with the minimum found for $M=0.9$. The overall scenario agrees with the analysis of the Vorono\"{i} tessellation, where the maximum level of clustering was found for $M=0.3$. Note that besides the dominant peak at $r = D$, a second less evident peak is observed for $r \approx 2D$. They are respectively the statistical trace of the first and second coordination shells around the test particle. For smaller particles $D/\eta=32$ and $D/\eta=16$, the accumulation is largest at all length scales for $M=0.3$ and $M=0.6$, respectively. This agrees with the large increase of the variance of the Vorono\"{i} volumes seen in figure \ref{fig:Voronoi_sigma}. Overall, the radial distribution function confirms that for $D/\eta \le 64$ the level of clustering increases with $M$ for small mass fraction, but decreases for larger $M$, when due to their larger inertia particles move along almost two-dimensional and straight trajectories.

\begin{figure}
\centering
\includegraphics[width=0.49\textwidth]{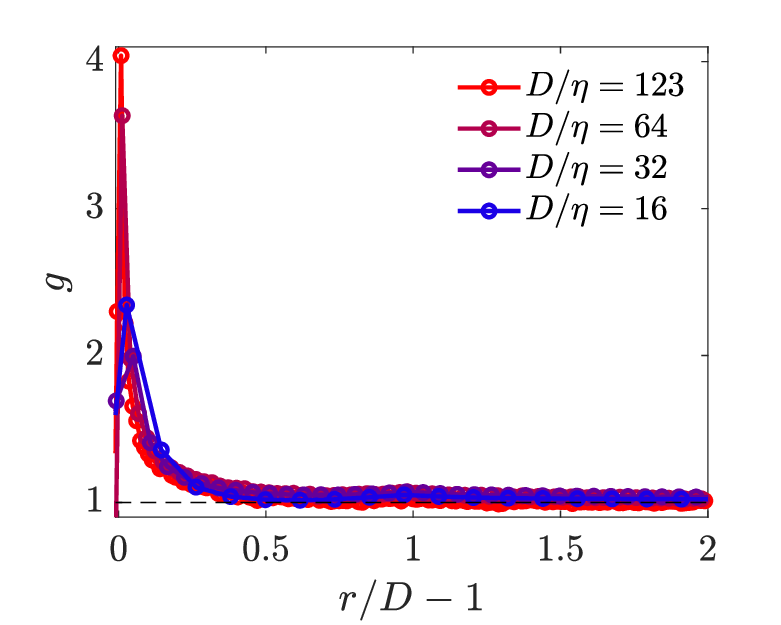}
\includegraphics[width=0.49\textwidth]{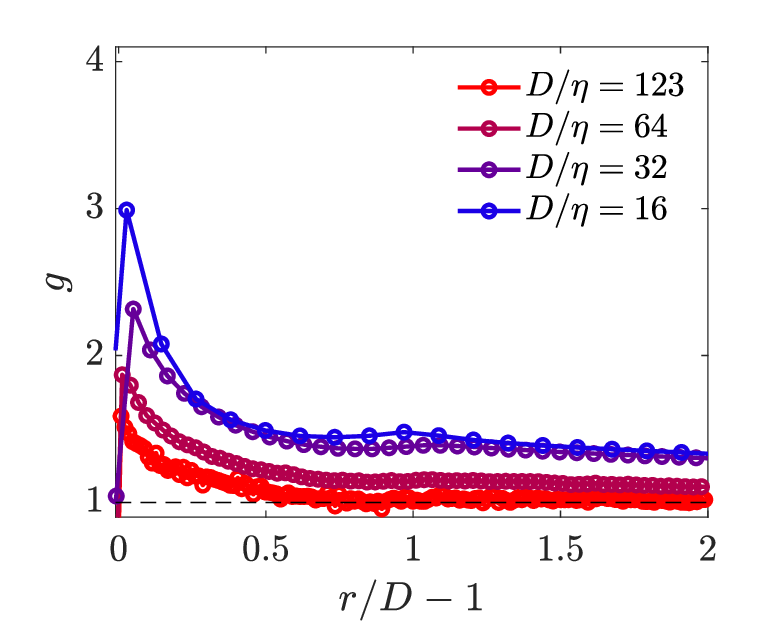}
\includegraphics[width=0.49\textwidth]{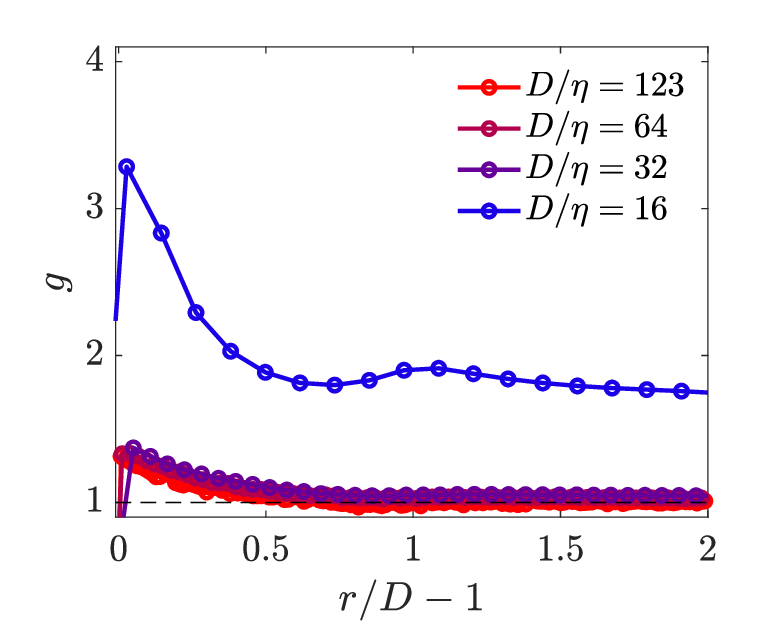}
\includegraphics[width=0.49\textwidth]{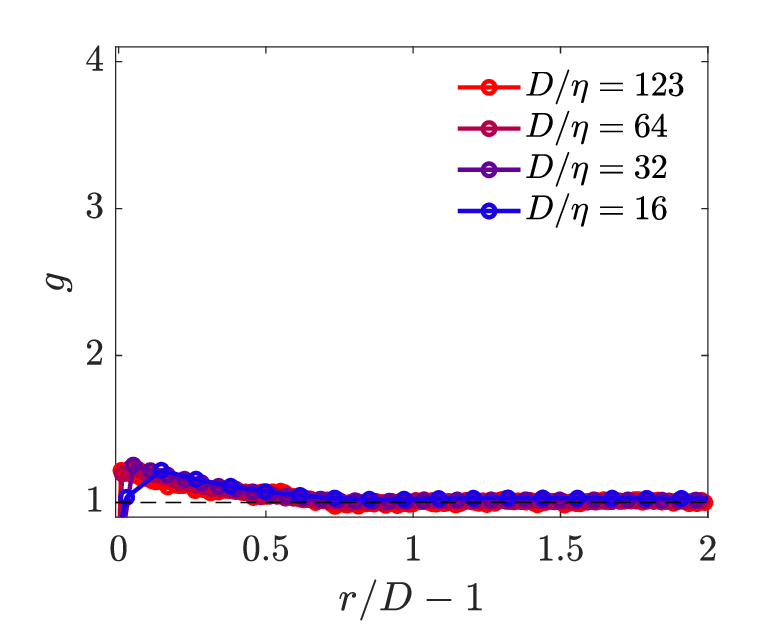}
\caption{Dependence of the radial distribution function on the particles size for (top left) $M=0.1$, (top right) $M=0.3$, (bottom left) $M=0.6$ and (bottom right) $M=0.9$.}
\label{fig:rdfDeffect}
\end{figure}

Figure \ref{fig:rdfDeffect} details the effect of the particles size on the radial distribution function. It is clearly visible that for all mass fractions the level of accumulation increases at all length scales when the particle size decreases. The same trend has been found by \cite{uhlmann-chouippe-2017} when considering finite-size particles with $D/\eta=5$ and $D/\eta=11$ and $\rho_p/\rho_f = 1.5$, while the opposite trend, i.e. an increase of the accumulation with the particle size, has been reported by \cite{salazar-etal-2008} for particles smaller than the Kolmogorov length scale $D < \eta$. 

\begin{figure}
\centering
\includegraphics[width=0.7\textwidth]{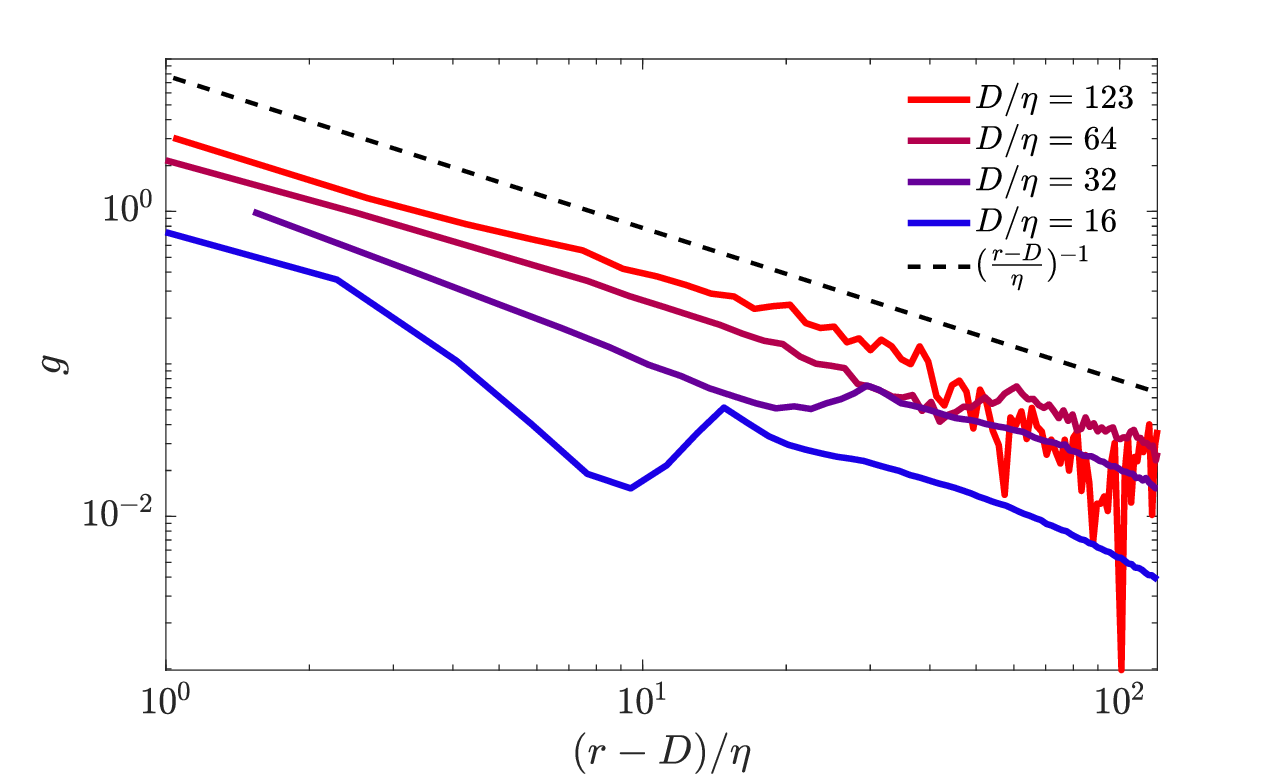}
\caption{Radial distribution function for $M=0.1$ and $16 \le D/\eta \le 123$.}
\label{fig:rdf_log}
\end{figure}
We have also observed that for small mass fractions, $M = 0.1$, the decay of the radial distribution function with the distance $r$ from the test particle approximately follows a power law, which indicates a self-similar spatial distribution \citep{saw-etal-2008,uhlmann-chouippe-2017,petersen-etal-2019}. Despite theoretical arguments support this formulation for dissipative separations $r/\eta<1$ and for small $St$ only, several authors provided numerical and experimental evidence that the power-law form continues to hold also for larger $r$ \citep{saw-etal-2008,petersen-etal-2019}. For point particles, \cite{bragg-ireland-collins-2015} observed that the clustering mechanisms operating in the inertial range are analogous to those operating in the dissipation range \citep{bragg-collins-2014}. They argued that when $St_r \ll 1$ (where $St_r = \tau_p/\tau_r$ is the Stokes number based on
$\tau_r$, i.e. the eddy turnover time at scale $r$ defined as $\tau_r = \overline{\aver{\epsilon}}^{-1/3}r^{2/3}$), preferential sampling of the coarse-grained fluid velocity gradient tensor at scale $r$ generates the inward drift and clustering. When, instead, $St_r > O(1)$ the non local, path-history mechanism contributes to the clustering, breaking thus the self-similarity. In figure \ref{fig:rdf_log}, we consider $M = 0.1$ and $16 \le D/\eta \le 123$, for which the Stokes numbers are between $ 0.1 \lessapprox St \lessapprox 10$. Although the results for $D/\eta = 16$ suggest that self-similarity is broken, for larger $D/\eta$ we measure a power law of approximately $g(r) \sim  (r/\eta)^{-1}$ up to $r/\eta \approx O(10^2)$.

\begin{figure}
\centering
\includegraphics[width=0.9\textwidth]{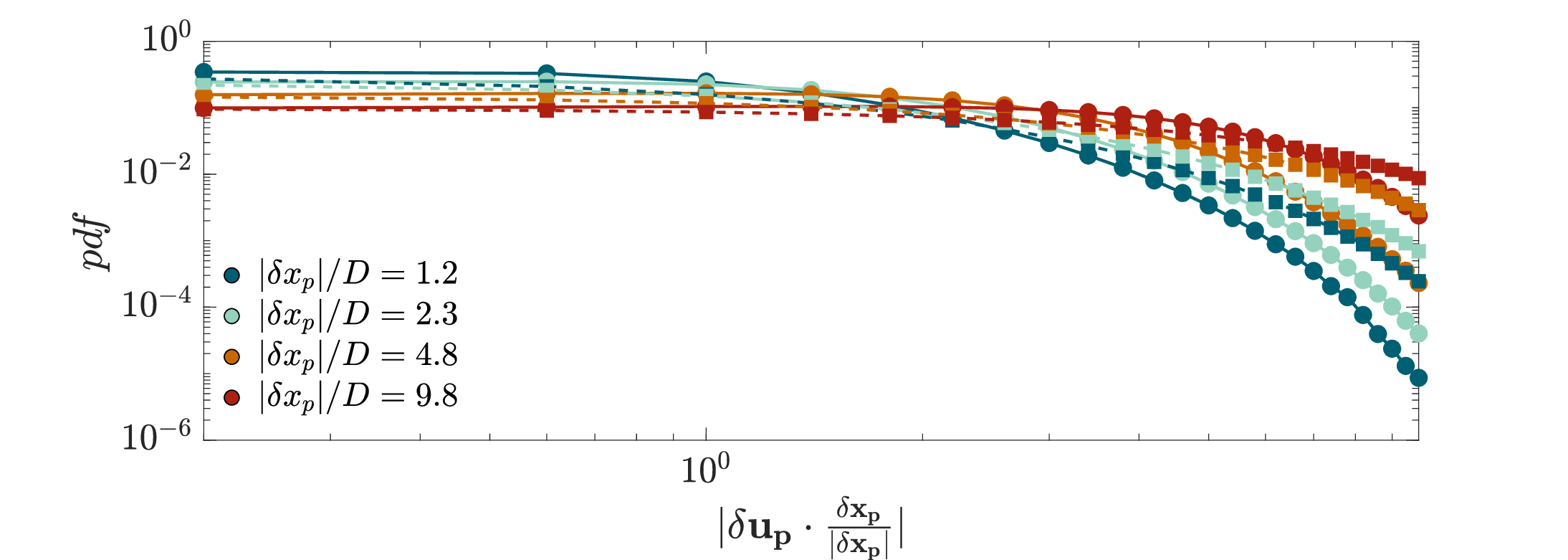}
\includegraphics[width=0.9\textwidth]{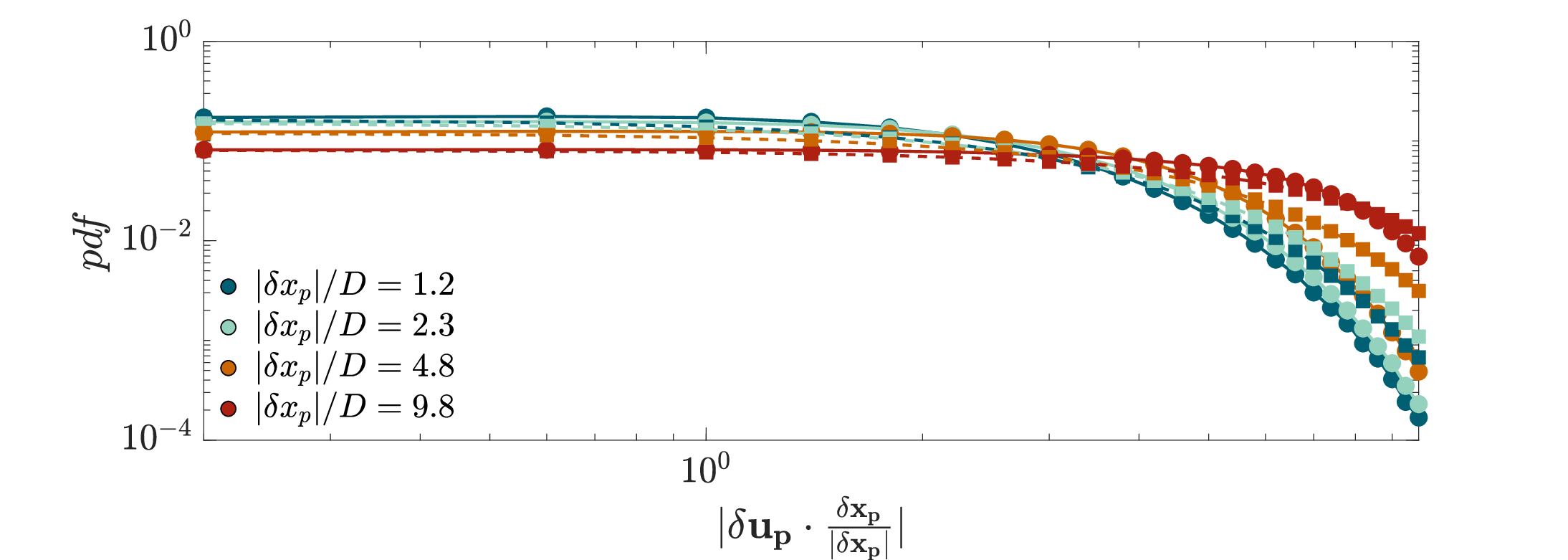}
\includegraphics[width=0.9\textwidth]{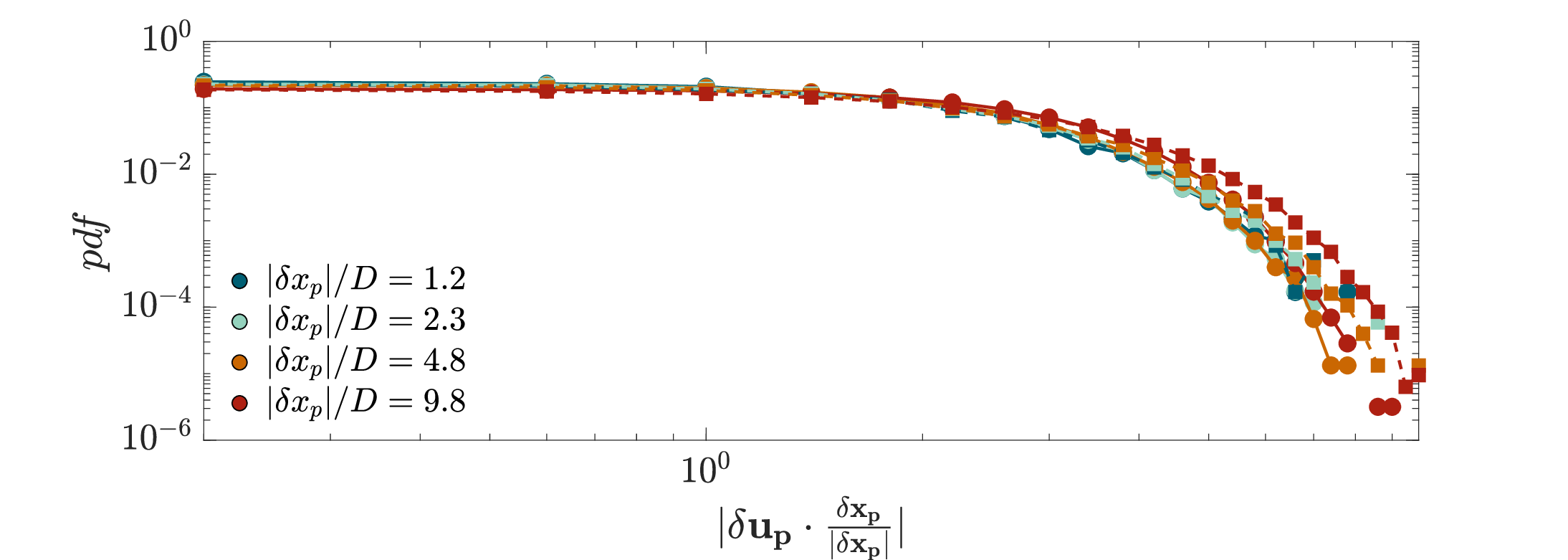}
\caption{Distribution of the radial particle-particle relative velocity for $D/\eta=32$. From top to bottom $M=0.3$, $M=0.6$ and $M=0.9$. $\delta \bm{u}_p$ is the relative velocity between two particles, whereas $\delta \bm{x}_p$ is their separation vector. The solid lines with circles refer to $\delta \bm{u}_p \cdot \delta \bm{x}_p/|\delta \bm{x}_p|>0$, i.e. to diverging particle. The dashed lines with squares refer to $\delta \bm{u}_p \cdot \delta \bm{x}_p/|\delta \bm{x}_p|<0$, i.e. to converging particles. Different colours correspond to different particle-particle distances.}
\label{fig:part-part-relvel}
\end{figure}
It is instructive to investigate the distribution of the particle-particle relative velocity $\delta \bm{u}_p$. In the context of point-particles, indeed, several theories regarding the clustering mechanisms rely on the exact equation for the radial distribution function $g(r)$, in which the distribution of $\delta \bm{u}_p$ plays an important role; see for example \cite{gustavsson-mehlig-2011,bragg-collins-2014,bragg-ireland-collins-2015}. Figure \ref{fig:part-part-relvel} shows the distribution of $\delta \bm{u}_p \cdot \delta \bm{x}_p/|\delta \bm{x}_p|$, i.e. the component of the relative velocity between two particles $\delta \bm{u}_p$ projected along their separation vector $\delta \bm{x}_p$. When $\delta \bm{u}_p \cdot \delta \bm{x}_p/|\delta \bm{x}_p|>0$ the two particles depart, while when $\delta \bm{u}_p \cdot \delta \bm{x}_p/|\delta \bm{x}_p|<0$ they get closer. For conciseness, figure \ref{fig:part-part-relvel} considers only the $D/\eta=32$ case, but the following discussion holds also for the other particle sizes. Note that for all $|\delta \bm{x}_p|/D$, the average value of the distribution is zero, due to the statistically steady state condition considered in this work. However, the distribution of $\delta \bm{u}_p \cdot \delta \bm{x}_p/|\delta \bm{x}_p|$ is not symmetric, in agreement with the above discussed tendency of particles to form clusters. For all $|\delta \bm{x}_p|/D$, indeed, the distributions are left skewed: the mode is slightly positive, but the negative tails are longer. When increasing $|\delta \bm{x}_p|/D$ the distribution becomes progressively more flat, in agreement with a less correlated motion of the particles. Figure \ref{fig:part-part-relvel} shows that the dependence of the distribution on $M$ agrees with what found for $g(r)$. For $M=0.9$, indeed, the positive and negative tails of the distribution almost overlap, consistently with the low value of clustering detected in figure \ref{fig:rdfMeffect}. A similar effect is found when varying the particle size (not shown): when fixing $M$, an increase of $D$ leads to a more symmetric distribution, in agreement with the above discussed decrease of the level of clustering.

\subsection{Preferential particle location}
\label{sec:PrefSamp}

In the previous section we have shown that the solid phase is not randomly distributed in space and that, depending on the size and density of the particles, the suspension features a mild level of clustering. In this section we investigate the probability density functions of the acceleration and vorticity vectors in the region surrounding the particles, to speculate whether the two principal mechanisms that are known to govern the particle preferential location for sub-Kolmogorov particles may play a role also in the case of particles with size that lays in the inertial range. We consider the centrifuge \citep{maxey-1987} and the sweep-stick \citep{goto-vassilicos-2008} mechanisms. The centrifuge mechanism is based on the hypothesis that (i) the particle inertia is sufficiently large, (ii) the particle paths deviate from the fluid path, and (iii) the excess inertia does not place the particles in the ballistic regime. In this case, particles are centrifuged out from regions of high vorticity (vortex cores), preferring regions of large strain.
The sweep-stick mechanism, instead, links the particle locations with the fluid acceleration $\bm{a}_f$. In the framework of the one-way coupling point-particle model, following the work by \cite{chen-goto-vassilicos-2006}, \cite{goto-vassilicos-2008} showed that the particles accumulate in points that satisfy the following criterion $\bm{e}_1 \cdot \bm{a}_f = 0$ and $\lambda_1=0$, where $\lambda_1$ is the largest eigenvalue of the symmetric part of the acceleration gradient tensor, and $\bm{e}_1$ is the associated eigenvector. This results from the observation that the slip velocity between a point-particle and the fluid is proportional to the fluid acceleration, i.e. $\bm{u}_p - \bm{u}_f \approx - \tau_p \bm{a}_f$. Later, \cite{coleman-vassilicos-2009} found that point particles in homogeneous and isotropic turbulence accumulate preferentially in the vicinity of zero-acceleration points, having observed that this simpler criterion is more strongly correlated with the location of their point-particles. 
It is worth mentioning that, by analysing the governing equation for the radial distribution function, \cite{bragg-ireland-collins-2015} suggested that the clustering mechanisms in the inertial and dissipative ranges are analogous. For any $r$ which is less than the integral length scale of the flow, they propose that the clustering mechanism for small $St$ is associated with centrifuging out effect of eddies of that scale. For $St > O(1)$, instead, they found that non local mechanisms contribute to the clustering, generating a statistical asymmetry of the path history of approaching and separating particle pairs. However, despite the universality of the clustering mechanism across the range of scales disagrees with the presence of different mechanisms, they observe that in the inertial range for $St \ll 1$ the mechanism they propose is essentially equivalent to the sweep-stick mechanism.

We investigate whether the centrifuge and sweep-stick mechanisms play a role in determining the preferential location of particles with size in the inertial range in non-dilute suspensions. This analysis follows the work by \cite{uhlmann-chouippe-2017}, that investigated the preferential sampling of light and small particles ($D/\eta = 5-11$ and $\rho_p/\rho_f=1.5$) in more dilute suspensions ($\Phi_V=0.005$), and at smaller Reynolds numbers ($Re_\lambda \approx 100$). The idea is to compute the probability density functions of the modulus of the acceleration and vorticity vectors in shells around the particles, and to compare them with the probability density functions of the same quantities evaluated for the complete fluid phase. The difference between the probability density functions may reveal whether there is a link with these mechanisms or not. In case the sweep-stick mechanism plays a role in the particle locations, we expect an increase of the probability of small $|\bm{a}_f|$ events in the region surrounding the particles. Similarly, an increase of the probability of small $|\bm{\omega}_f|$ events in the neighbourhood of the particles may suggest a link with the centrifuge mechanism. Before going on, it is worth stressing that the interpretation of these results in our case is not straightforward. In fact, unlike what considered by the point-particle approach for sub-Kolmogorov particles in dilute suspensions, finite-size particle substantially influence the surrounding flow (see section \ref{sec:near-par}), and can potentially collide. Therefore, the differences between the global and local probability density functions are due to a combination of three effects: (i) the preferential particle location, (ii) the influence of the particle on the surrounding flow and (iii) the influence of the particle-particle interaction on the surrounding fluid phase. A further consideration regards the radius of the shell $R_{sh}$ around the particles that is used for the local probability density functions. If $R_{sh}$ is too small, only the local perturbations of the particles on the surrounded flow are considered. If $R_{sh}$ is too large, spurious contributions of the fluid phase that do not influence the particle location would be taken into account. Following section \ref{sec:near-par}, we choose $R_{sh} = 1.5 R_p$, but we have verified that small variations of $R_{sh}$ qualitatively do not influence the results. Clearly, for larger $R_{sh}$ the differences between the local and global probability density functions progressively decrease. 

\begin{figure}
\centering
\includegraphics[width=0.49\textwidth]{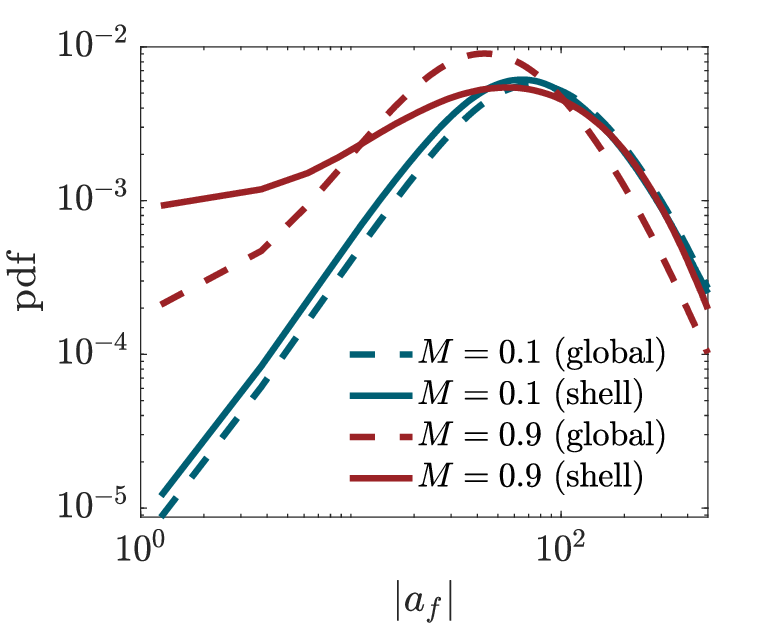}
\includegraphics[width=0.49\textwidth]{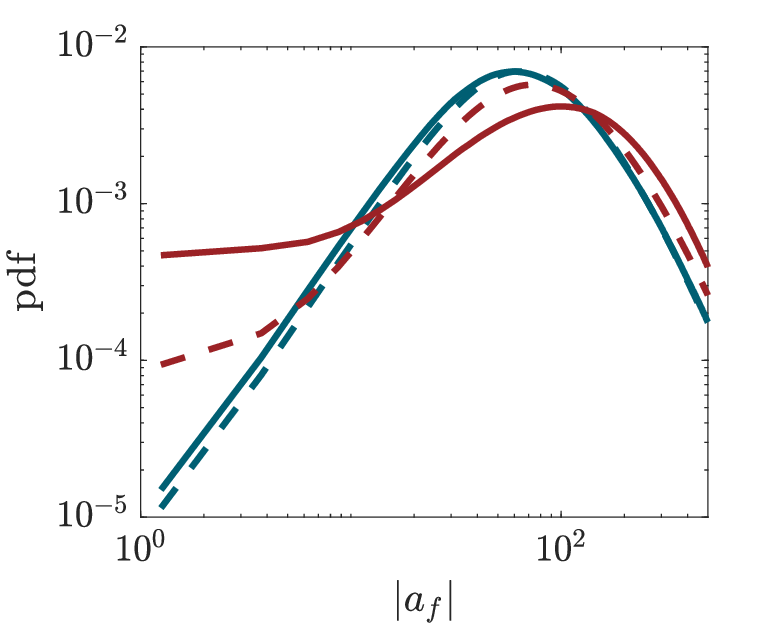}
\caption{Local and global probability density functions of the modulus of the fluid acceleration $|\bm{a}_f|$ for (left) $D/\eta=64$ and (right) $D/\eta=123$. Solid line is for the shell around the particle, while dashed line is for the global fluid phase.}
\label{fig:pdf_shell_a}
\end{figure}

Figure \ref{fig:pdf_shell_a} plots the local and global probability density functions of the modulus of the fluid acceleration $|\bm{a}_f|$ for $D/\eta=64$ and $D/\eta=123$ as representative cases.
We start considering small mass fractions $M=0.1$ ($\rho_p/\rho_f = 1.29$). In this case, for all $D$ considered, the probability of large acceleration in the shell decreases compared to the rest of the fluid phase, while the probability of small values increases suggesting a link with the sweep-stick mechanism. Note, however, that for this $M$ the difference between the local and global probability density functions is relatively small, in agreement with the low level of clustering above discussed. This result is in line with the results of \cite{uhlmann-chouippe-2017}, that for particles with $D/\eta=5-11$ and $\rho_p/\rho_f = 1.5$ found a non negligible correlation between the particle position and the sticky points. Therefore, despite the theory developed by \cite{goto-vassilicos-2008} rigorously holds only in the limit of the point-particle approach, our results indicate that a link between the preferential location of light particles with size within the inertial range and low fluid acceleration areas may exist.

When larger $M$ are considered the scenario changes and the correlation between the particle location and regions with low fluid acceleration is less clear. The peak of the global probability density functions moves towards smaller $|\bm{a}_f|$, in agreement with a decrease of the mean fluid acceleration. In the shell surrounding the particles, instead, the probability density function becomes progressively more flat as $M$ increases. Comparing the two probability density functions, the probability of both low and large $|\bm{a}_f|$ values increases in the neighbourhood of the particles, while the probability of intermediate values decreases. The increased probability of large values of $|\bm{a}_f|$ with $M$ in the region surrounding the particles is consistent with the increase of the particle inertia and of the relative velocity between the fluid phase and the particle, that results into an increase of the momentum exchange between the fluid and solid phases.

\begin{figure}
\centering
\includegraphics[width=0.7\textwidth]{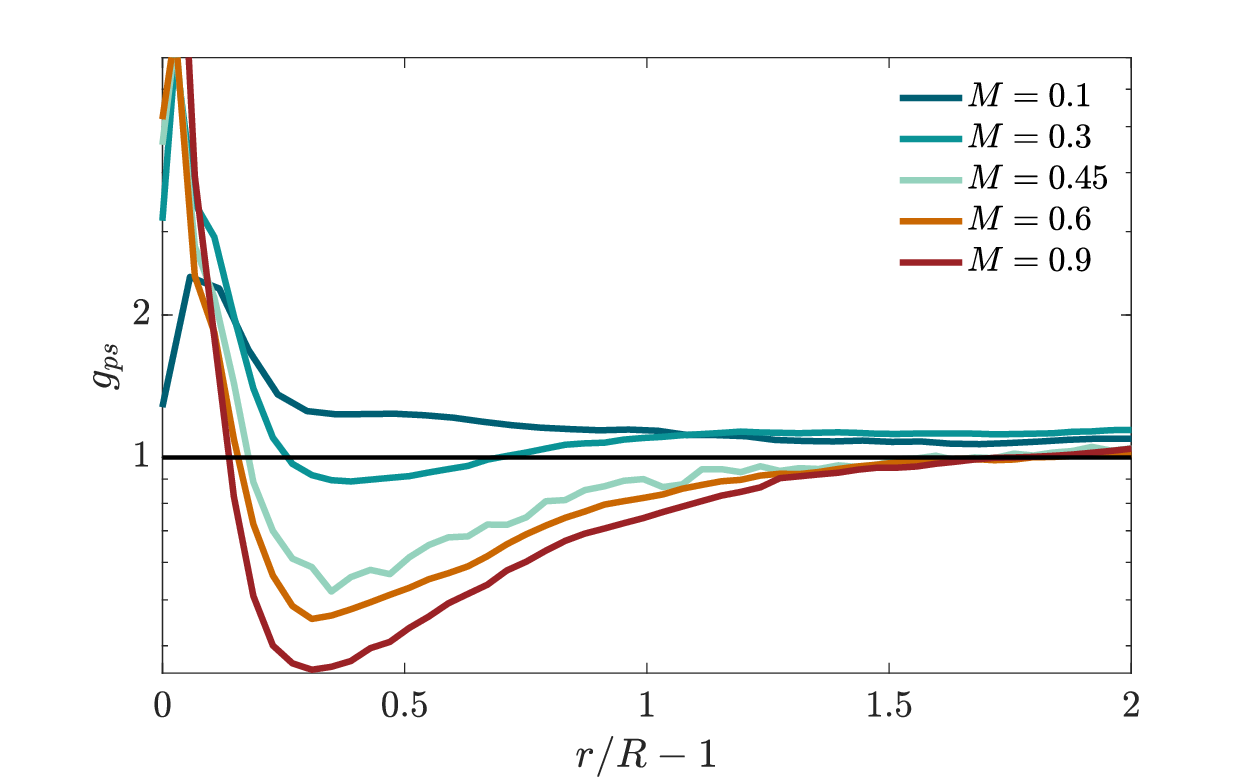}
\caption{Radial distribution function for the distance between particles and point with fluid acceleration modulus below $a_{th}=0.05 |\bm{a}|_{rms}$ for $D/\eta=64$.}
\label{fig:rdf_part_sticky}
\end{figure}

To investigate whether the sweep-stick mechanism actually plays a role for light and/or heavy particles, we compute the probability $g_{ps}(r)$ for each particle to have at a certain distance $r$ a point where the fluid acceleration is smaller than a threshold $a_{th}$. When $g_{ps}(r)> 1$ the probability is larger compared to a random distribution, whereas when $g_{ps}(r)<1$ it is lower. In figure \ref{fig:rdf_part_sticky}, we show the dependence of $g_{ps}(r)$ on $M$ for $D/\eta=64$. Following the work by \cite{uhlmann-chouippe-2017} we have set $a_{th}=0.05 |\bm{a}|_{rms}$, and we have verified that the results do not change significantly by slightly decreasing or increasing this value. 
For all mass fractions there is a large probability of having fluid points with low acceleration immediately close to the particles. This is a direct effect of inertial particles on the surrounding flow, and is associated with the no-slip and no-penetration condition on the surface. Due to their inertia, particles face a lower acceleration with respect to the fluid phase and, therefore, the same holds for the boundary layer surrounding them. Note that the increase of $g_{ps}$ with the mass fraction for small $r$ agrees with figure \ref{fig:pdf_shell_a}, that shows that the probability density function of $|\bm{a}|_f$ in the shell surrounding the particles progressively flattens when $M$ increases. Moving away from the particle boundary layer, i.e. at larger $r$, the distribution of $g_{ps}(r)$ changes with the mass fraction. For $M=0.1$ ($\rho_p/\rho_f=1.29$), $g_{ps}$ remains larger than $1$, with a value of approximately $g \approx 1.2$, up to $r \approx 3R$, while for $M>0.3$ ($\rho_p/\rho_f > 4.98$) $g_{ps}<1$ for all $r$. This clarifies the above discussion: for small mass fractions a link between the particle location and regions of low fluid acceleration is possible also for particles with size within the inertial range. This can be explained by the fact that a particle of size $D$ behaves to leading order like a point particle with respect to eddies of scale $\ell \gg D$. Hence, from a qualitative point of view, particles of size $D$ can exhibit sweep-stick like phenomena when the scale of the flow that drives their motion is of size $\ell \ge O(D)$. For larger $M$, instead, this is not the case. Indeed, the increase of the probability of small $|\bm{a}|$ values seen in figure \ref{fig:pdf_shell_a} is only due to the increase of the probability of negative values in the boundary layer attached to the particle surface, while $g_{ps}$ is less than one for $0.14 \lessapprox r/R_p-1 \lessapprox 1.4$, indicating a lower probability of having points with $|\bm{a}|<a_{th}$ compared to a random distribution.

We now move to the centrifuge mechanism.
\begin{figure}
\centering
\includegraphics[width=0.49\textwidth]{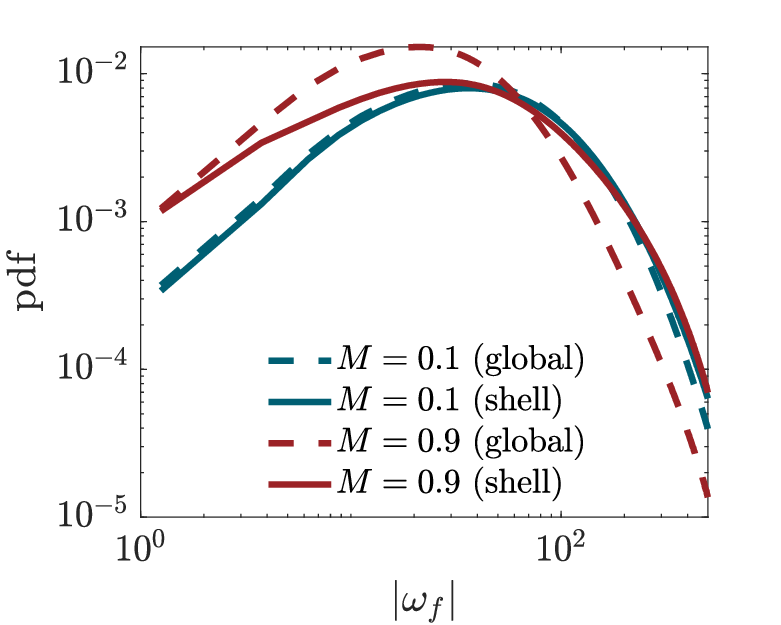}
\includegraphics[width=0.49\textwidth]{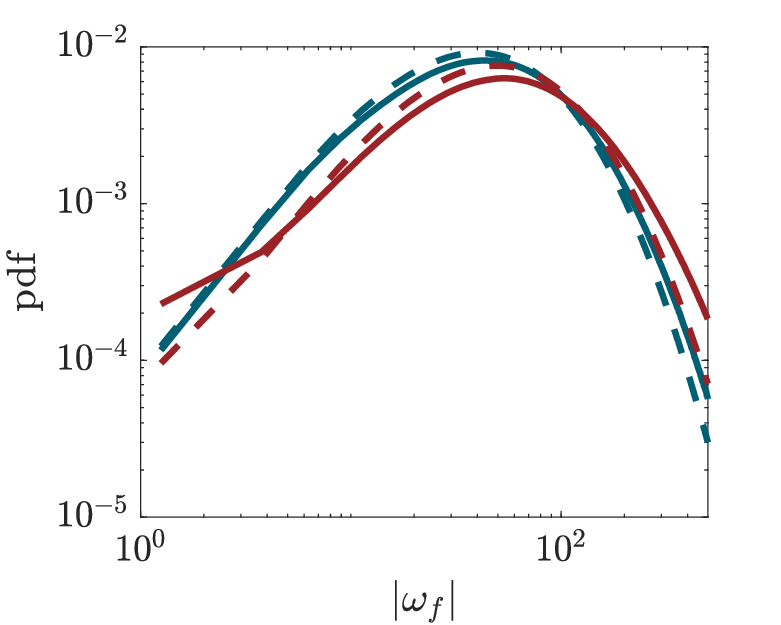}
\caption{Local and global probability density functions of the modulus of the vorticity $|\bm{\omega}_f|$ for (left) $D/\eta=64$ and (right) $D/\eta=123$. Solid line is for the shell around the particle, while dashed line is for the global fluid phase.}
\label{fig:pdf_shell_o}
\end{figure}
For all mass fractions and particle sizes, the probability of small vorticity values in the region surrounding the particles is smaller than in the complete box, while the probability of large values increases (see figure \ref{fig:pdf_shell_o}). For larger mass fraction and larger particles we observe that the difference between the two probability density functions progressively increases, in agreement with the increase of the particle Reynolds number and with the occurrence of vortex shedding in the particle wake (see section \ref{sec:near-par}). 
Based on this observation, one may be tempted to exclude that the centrifuge mechanism plays a role in determining the preferential location of finite-size particles. However, it is possible that a particle of size $D$ is centrifuged out by a vortex of size $\ell \gg D$. While doing this, a strong vorticity may generate on the surface of the particle due to the no-slip and no-penetration boundary conditions, potentially leading to the above effect on the local probability density function.


\section{Conclusion}
\label{sec:conclusions}

In this work we have investigated the fluid-solid interaction of suspensions of solid spherical particles in triperiodic turbulence, at the relatively large micro-scale Reynolds number of $Re_\lambda \approx 400$. The study is based on direct numerical simulations coupled with an immersed boundary method, to properly resolve the flow around each particle. The fluid-solid interaction is studied in the two-dimensional parameter space of the particle size and density. The particle size is varied between $16 \le D/\eta \le 123$, while the solid-to-fluid density ratio is varied between $1.29 \le \rho_p/\rho_f \le 104.7$, corresponding to a variation of the mass fraction between $ 0.1 \le M \le 0.9$. Non-dilute suspensions with a volume fraction of $\Phi_V = 0.079$ are considered. Turbulence is sustained with the Arnold-Beltrami-Childress (ABC) cellular forcing \citep{podvigina-pouquet-1994}, that generates a three-dimensional and inhomogeneous shear at the largest scales of the flow.

In the considered portion of the parameter space, the solid phase modulates the largest scales of the flow in a way that changes with the size and the density of the particles. Depending on the ratio between the particle size and the length-scale of the inhomogeneous mean shear induced by the forcing, the solid phase may modulate the largest scales of the flow towards an anisotropic and almost two-dimensional state \citep{chiarini-etal-2023}. The smaller scales, instead, are homogeneous and isotropic for all considered cases, and their modulation does not depend on the external forcing. For these scales, the independence of the results on the external forcing has been assessed using the forcing introduced by \cite{eswaran-pope-1988}. The influence of the solid phase on the energy spectrum indicates that the mechanism driving the energy transfer across scales changes with the particle size and density. By means of the scale-by-scale energy budget, we have shown that the two mechanisms that drive the energy transfer, i.e. the classical energy cascade described by the Kolmogorov theory and the energy transfer associated with the fluid-solid coupling, play a different role depending on the size and density of the particles. For large and light particles ($D/\eta \ge 64$ and $M \le 0.3$), the flow modulation is weak and the classical energy cascade dominates, being only marginally altered by the solid phase. For large and heavier particles ($D\eta \ge 64$ and $M \ge 0.6$), the two mechanisms coexist: the fluid-solid coupling term drives the energy transfer at scales larger than the particle size, while the classical energy cascade is recovered at smaller scales. For small and heavy particles ($D/\eta \le 32$ and $M>0.6$), instead, the classical energy cascade is subdominant, and the overall energy transfer across scales is driven by the fluid-solid coupling term, which links directly the largest scales where energy is injected, and the smallest scales where energy is dissipated. By means of the extended similarity \citep{benzi-1993}, we have shown that the solid phase increases the flow intermittency in a way that depends on the relative velocity between the particles and the surrounding flow.

By means of a conditional average of the flow in a shell surrounding the particles, we have shown that the near-particle flow pattern largely changes with $D$ and $\rho_p/\rho_f$, with interesting implications for modelling. Light particles ($\rho_p/\rho_f < 5$) partially follow the flow, and their influence on the near-flow region is limited within the boundary layer that develops over their surface. In this case, vortex shedding does not occur, and the energy drained by the particle-fluid relative velocity is mainly dissipated at the particle surface. When the particle density increases ($\rho_p/\rho_f \ge 5$), instead, the influence of the particles on the surrounding flow becomes more intense, and the particle wake becomes unsteady. Here, part of the energy drained by the particle is dissipated away from the particle, and this requires a more sophisticated approach for modelling. For intermediate particle sizes, our results suggest that an actual vortex shedding occurs only for heavy particles with $\rho_p/\rho_f \ge 17$, and that for lighter particles with $\rho_p/\rho_f \approx 5$ the wake unsteadiness is essentially due to the relative motion between the particles and the fluid phase \citep{mittal-2000}.

We have also quantified the particle clustering with the aid of the Vorono\"{i} tessellation \citep{monchaux-etal-2010} and the radial distribution function \citep{saw-etal-2008}. Both methods show that, for all mass fractions, the level of clustering increases when the particle size decreases. The dependence of the clustering on the particle density, instead, is not monotonic and changes with $D$. For large particles ($D/\eta=123$) the level of clustering decreases with the particle density, being almost negligible for the largest mass fractions. For smaller particles ($D/\eta \le 64$), instead, the level of clustering is not monotonic, being maximum for intermediate $\rho_p/\rho_f$. Following the work by \cite{uhlmann-chouippe-2017}, we have then explored the possibility that at the considered parameters the preferential location of the particles might be determined by similar effects as for point-particles. We have investigated whether the centrifuge \citep{maxey-1987} and the sweep-stick \citep{goto-vassilicos-2008} mechanisms play a role in determining the preferential location of light and heavy particles, with size that lays in the inertial range. Overall, our results indicate that a link between the preferential location of the particles and areas of very small fluid acceleration is possible, but only for light particles with $\rho_p/\rho_f = 1.3$; however, particular care is needed when looking at these results because, unlike in the limit of point particles, here the particles are altering the flow in their surroundings, thus different effects are mixed in the analysis. 

\section*{Acknowledgments} 
The authors acknowledge the computer time provided by the Scientific Computing \& Data Analysis section of the Core Facilities at OIST and the computational resources of the supercomputer Fugaku provided by RIKEN through the HPCI System Research Project (Project IDs: hp210229 and hp210269).
  
\section*{Funding} 
The research was supported by the Okinawa Institute of Science and Technology Graduate University (OIST) with subsidy funding from the Cabinet Office, Government of Japan.
  
\section*{Declaration of Interests} 
The authors report no conflict of interest.

\section*{Author ORCIDs}

Alessandro Chiarini, https://orcid.org/0000-0001-7746-2850; \\
Marco Edoardo Rosti,  https://orcid.org/0000-0002-9004-2292.

\appendix
\section{Dependence of the results on the external forcing}
\label{sec:Pope}

In this section the dependence of the results on the external forcing is assessed. We show that the flow modulation for $\kappa/\kappa_L>1$ does not depend on the inhomogeneous mean shear induced by the ABC forcing at $\kappa/\kappa_L=1$. In this respect, we have carried out additional simulations using the forcing introduced by \cite{eswaran-pope-1988}, that does not generate an inhomoegeneous mean shear at the largest scales, and looked at the flow modulation induced by the addition of particles. The particle size has been fixed at a value of $D/L = 0.0207$, the volume fraction has been kept constant at $\Phi_V = 0.079$, and the mass fraction has been varied between $0.1 \le M \le 0.9$. For these additional simulations the Reynolds number has been set to a smaller value of $Re_\lambda \approx 260$ for the single phase case, meaning that $D/\eta \approx 18$. The grid resolution is kept constant with $N_{point}=1024$ points along each direction.

\begin{figure}
\centering
\includegraphics[width=0.9\textwidth]{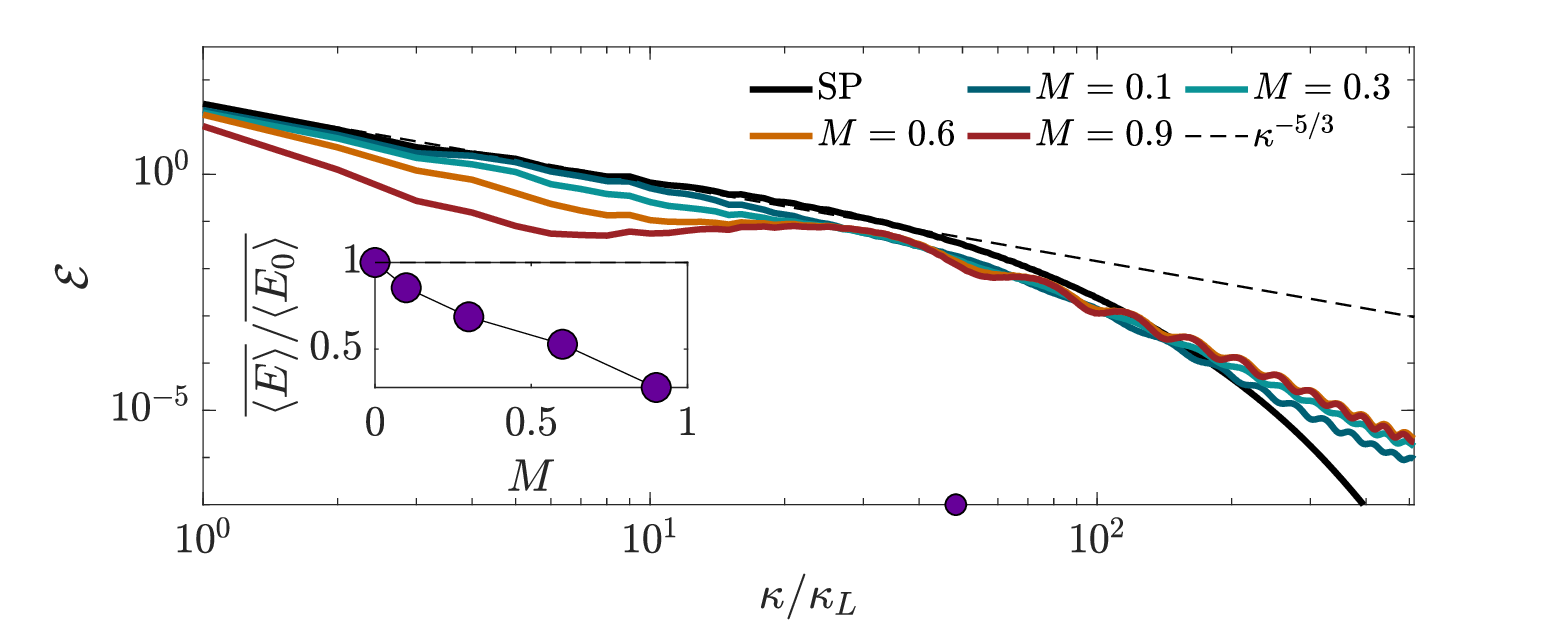}
\caption{Dependence of the energy spectrum $\mathcal{E}$ on the mass fraction for the case forced with the forcing introduced by \cite{eswaran-pope-1988}.}
\label{fig:SpecPope}
\end{figure}
Due to the absence of the inhomogeneous mean shear, when using the \cite{eswaran-pope-1988}'s forcing, particles modulate all scales in an isotropic way for all $M$, and the energy enhancement observed in figures \ref{fig:energy} and \ref{fig:energyM} for regime B does not occur. This is shown in the inset of figure \ref{fig:SpecPope} where $\overline{\aver{E}}/\overline{\aver{E_0}}$ is plotted as a function of $M$.
We now look at the scale-by-scale energy spectrum. Figure \ref{fig:SpecPope} plots $\mathcal{E}$ for these additional simulations, to be compared with figure \ref{fig:spec_M}. For $\kappa/\kappa_L>1$ the way the particles modulate the carrier flow does not change with the external forcing. Indeed, figure \ref{fig:SpecPope} confirms the conclusions drawn with the ABC forcing. Particles drain energy at scales larger than $D$ (see the energy depletion for $\kappa < \kappa_{p,1}$) and release it at smaller scales by means of their wake (see the energy enhancement for $\kappa > \kappa_{p,2}$), with heavier particles being more effective. When comparing the energy spectra obtained with the different forcings, the only qualitative difference regards the largest scale. Indeed, when using the ABC forcing, for $0.45 \le M \le 0.75$ $\mathcal{E}(\kappa/\kappa_L=1)$ becomes larger than the single-phase value, due to the above described mean flow enhancement.
\begin{figure}
\centering
\includegraphics[trim={0 10 0 0},clip,width=0.65\textwidth]{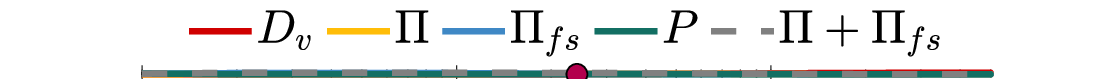}
\includegraphics[width=0.49\textwidth]{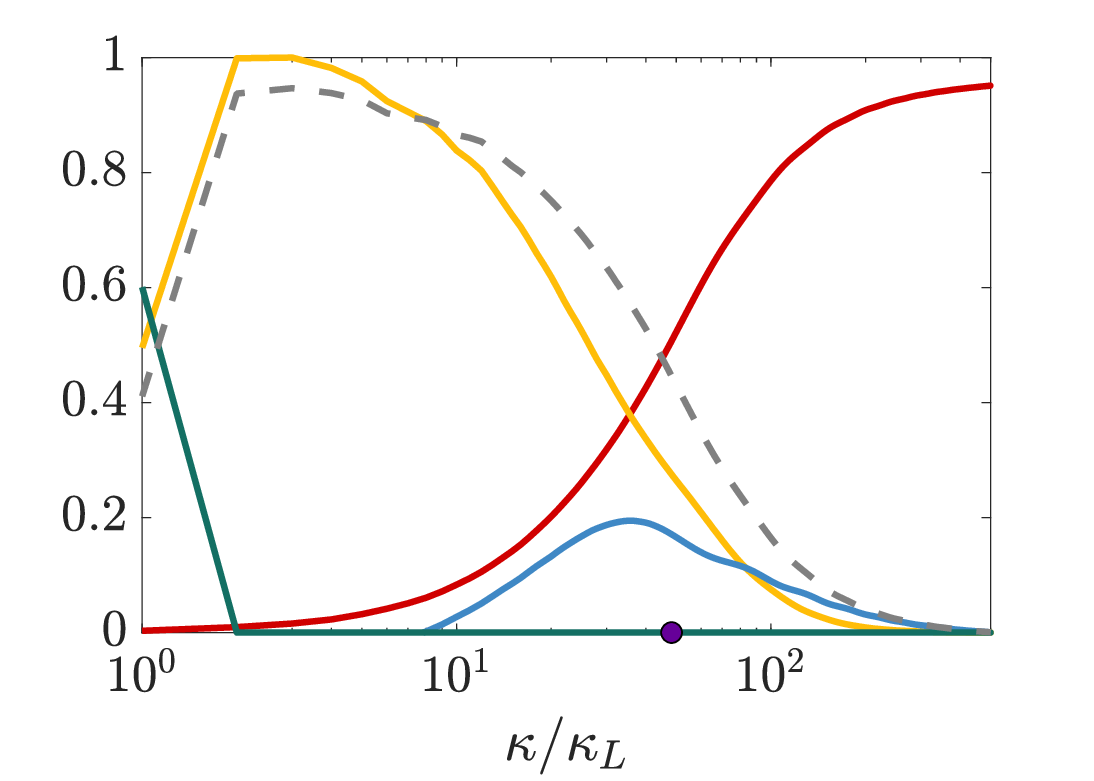}
\includegraphics[width=0.49\textwidth]{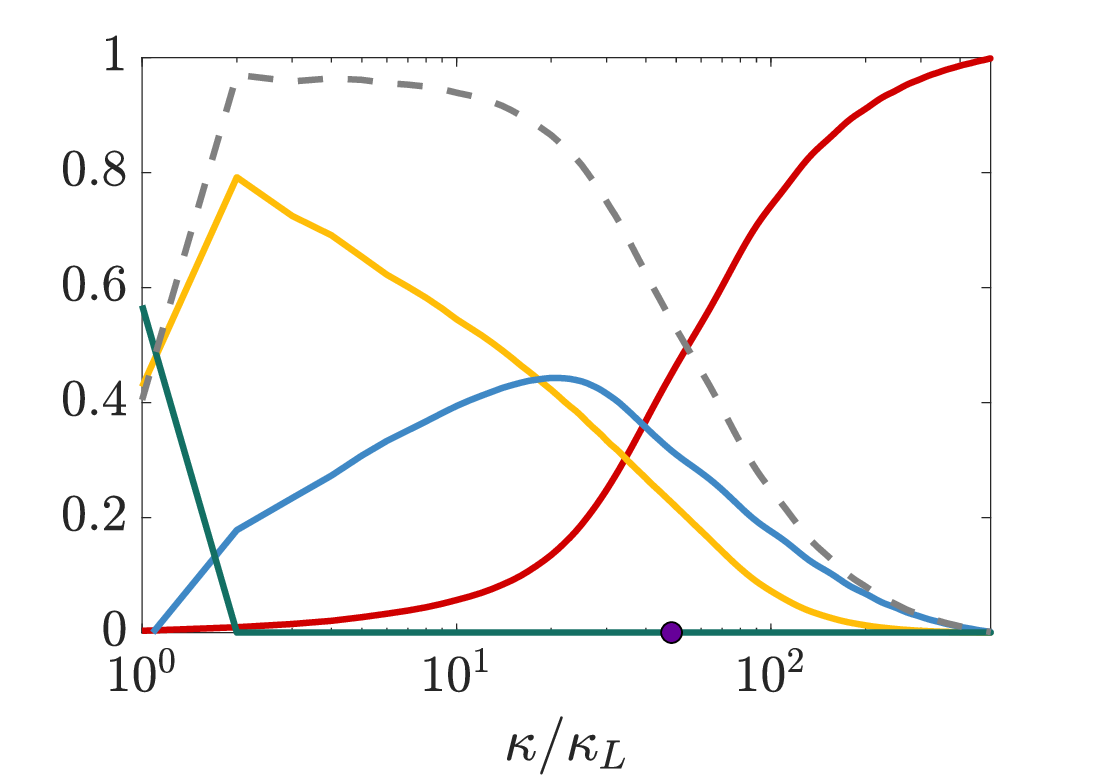}
\includegraphics[width=0.49\textwidth]{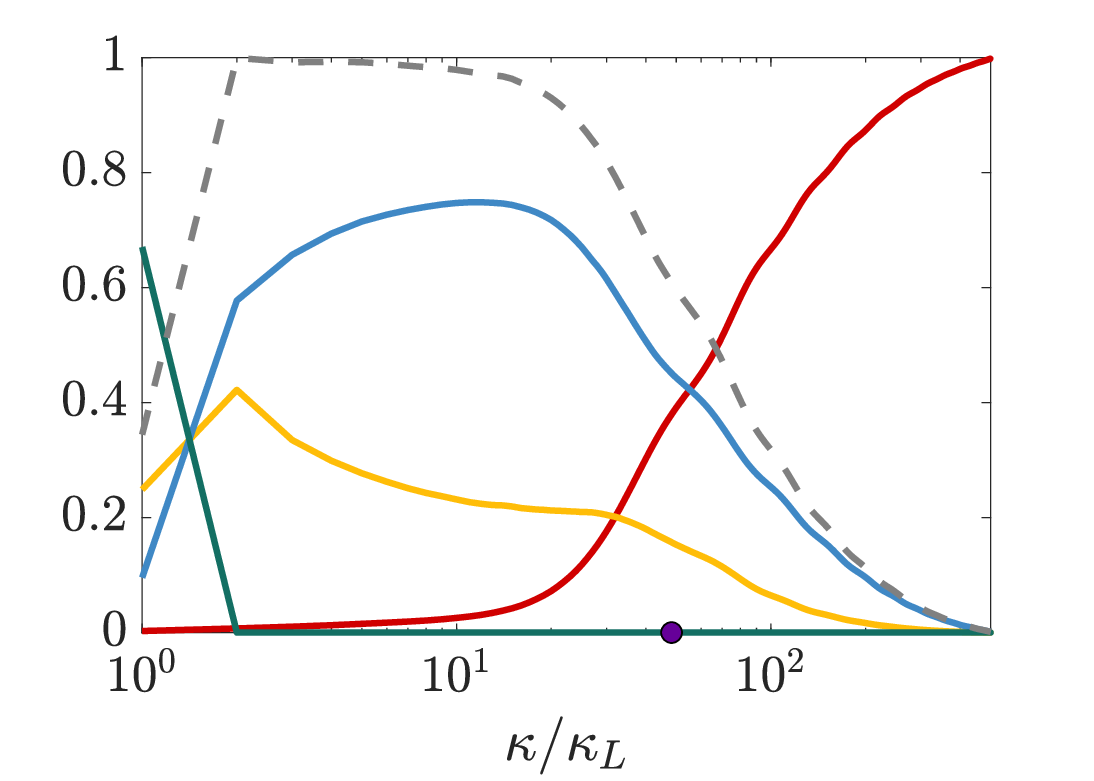}
\includegraphics[width=0.49\textwidth]{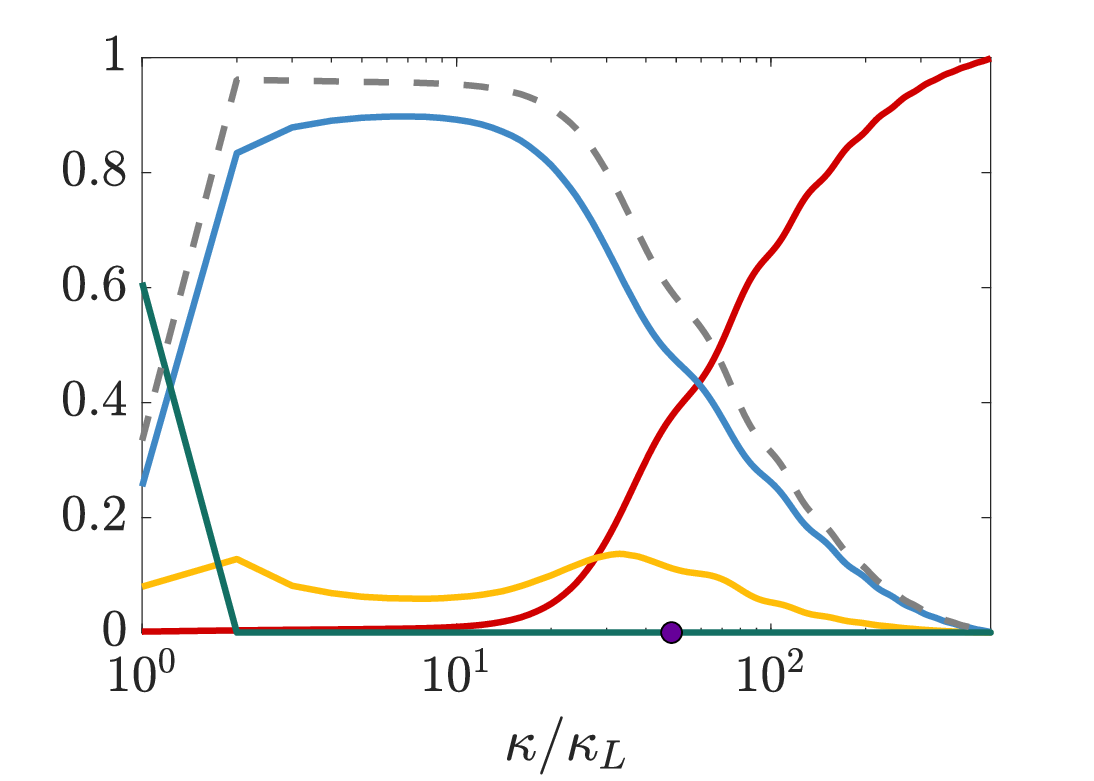}
\caption{Scale-by-scale energy budget for the case forced with the forcing introduced by \cite{eswaran-pope-1988}. Top left: $M=0.1$ (top left). Top right: $M=0.3$. Bottom left: $M=0.6$. Bottom right: $M=0.9$.}
\label{fig:budPope}
\end{figure}
In addition, in figure \ref{fig:budPope} we report the scale-by-scale energy transfer balance computed with these additional simulations, to be compared with figures \ref{fig:bud_gamma5}, \ref{fig:bud_gamma17} and \ref{fig:bud_gamma100}. Again, the results are in agreement with what found using the ABC forcing. For small mass fractions the non linear term $\Pi$ dominates at all scales, and the fluid-solid coupling term only marginally influences the scale energy transfer. For large $M$, instead, $\Pi_{fs}$ dominates at all scales and the classical energy cascade is substantially annihilated.

\section{Sensitivity to the grid resolution}
\label{sec:resolution}

\begin{figure}
\centering
\includegraphics[width=0.9\textwidth]{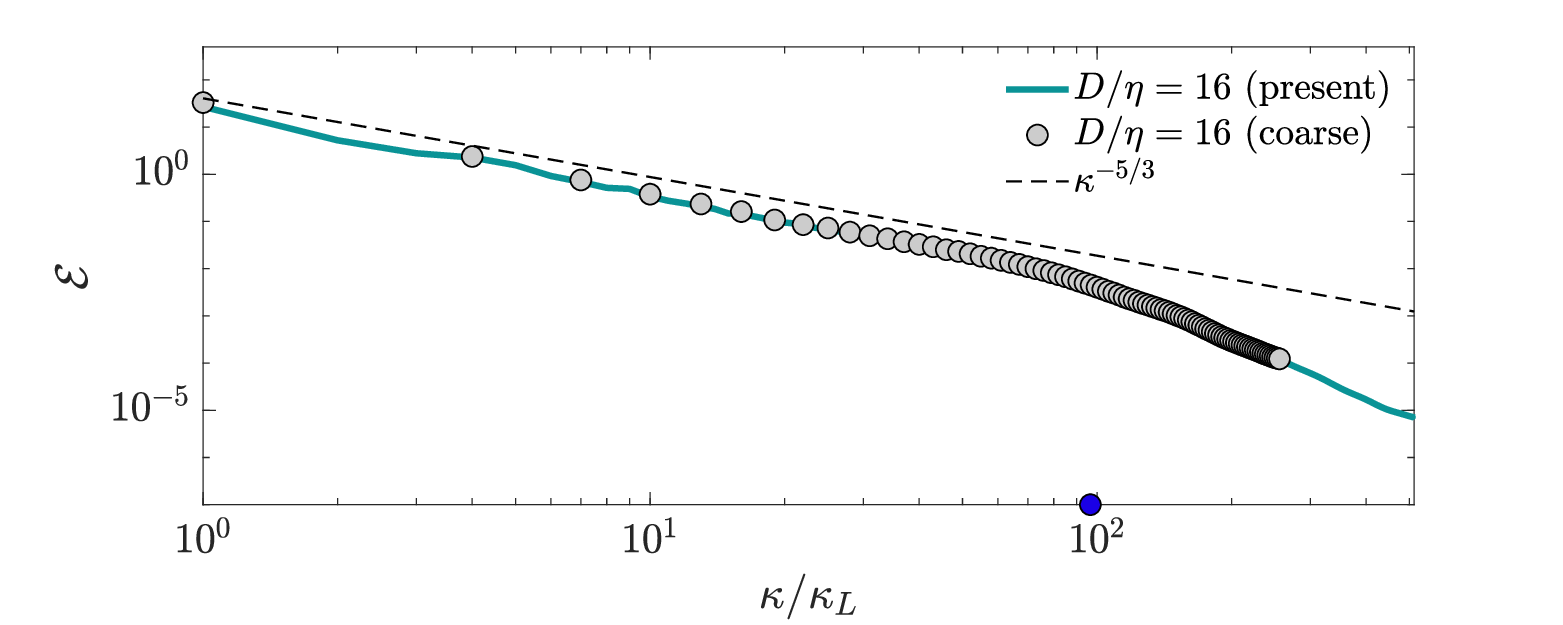}
\caption{Dependence of the scale-by-scale energy spectrum for $D/\eta = 16$ and $M = 0.3$. The line is for the standard grid with $N_{point}=1024$ along each direction. The circles refer to the coarser grid with $N_{point}= 512$ along each direction.}
\label{fig:spec-coarse}
\end{figure}
In this section the adequacy of the grid resolution is investigated. An additional simulation has been carried out for $D/\eta=16$, i.e. the smallest particle considered, on a coarser grid. The mass fraction has been set to $M=0.3$, and turbulence is sustained by means of the ABC forcing. In this coarser grid the fluid domain is discretised using $N_{point}=512$ points in the three directions, leading to $\eta/\Delta x \approx 2$. In doing this, the number of points across each particle decreases from $16$ (in the standard grid) to $8$. Figure \ref{fig:spec-coarse} shows that the energy spectra obtained with the standard and coarse grids collapse pretty well, indicating that the resolution of the standard grid is adequate.

\bibliographystyle{jfm}

\end{document}